\documentclass[a4paper,11pt]{article}
\pdfoutput=1
\usepackage{jcappub}
\usepackage{amsmath}
\usepackage{graphicx}
\usepackage{amssymb}
\usepackage{sidecap}
\usepackage{color}
\usepackage{natbib,hyperref}

\def\dalemb#1#2{{\vbox{\hrule height.#2pt
        \hbox{\vrule width.#2pt height#1pt \kern#1pt \vrule width.#2pt}
        \hrule height.#2pt}}}

\def\ba{\begin{eqnarray}}
\def\ea{\end{eqnarray}}
\def\be{\begin{equation}}
\def\ee{\end{equation}}
\newcommand{\vc}[1]{\boldsymbol{#1}}

\def\Var{{\rm Var}}
\def\gtorder{\mathrel{\raise.3ex\hbox{$>$}\mkern-14mu
             \lower0.6ex\hbox{$\sim$}}}
\def\ltorder{\mathrel{\raise.3ex\hbox{$<$}\mkern-14mu
             \lower0.6ex\hbox{$\sim$}}}

\title
{The binned bispectrum estimator: 
template-based and non-parametric CMB non-Gaussianity searches}

\author[a,b,c]{Martin Bucher}
\author[d,a]{Benjamin Racine}
\author[e,a]{Bartjan van Tent}

\affiliation[a]{APC, AstroParticule et Cosmologie, Universit\'e Paris Diderot,\\
CNRS/IN2P3, CEA/Irfu, Observatoire de Paris, Sorbonne Paris Cit\'e,\\ 
10 rue Alice Domon et L\'eonie Duquet, 75205 Paris Cedex 13, France}
\affiliation[b]{Astrophysics and Cosmology Research Unit,\\ School of 
Mathematical Sciences, University of KwaZulu-Natal,\\ Westville Campus, 
Private Bag X54001, Durban 4000, South Africa}
\affiliation[c]{Kavli Institute for Theoretical Physics China, Chinese Academy of Sciences, \\P. O. Box 2735, Beijing 100190, P. R. China}
\affiliation[d]{Institute of Theoretical Astrophysics, University of Oslo, \\P.O. Box 1029 Blindern, NO-0315 Oslo, Norway}
\affiliation[e]{Laboratoire de Physique Th\'eorique (UMR 8627),\\ 
CNRS, Univ. Paris-Sud, Universit\'e Paris-Saclay,\\ B\^atiment 210,
91405 Orsay Cedex, France}

\emailAdd{bucher@apc.univ-paris7.fr}
\emailAdd{benjar@uio.no}
\emailAdd{vantent@th.u-psud.fr}

\abstract{We describe the details of the binned bispectrum estimator as used 
for the official 2013 and 2015 analyses of the temperature and
polarization CMB maps from the ESA Planck satellite. The defining
aspect of this estimator is the determination of a map bispectrum
(3-point correlation function) that has been binned in harmonic space. 
For a parametric determination of the non-Gaussianity in the map 
(the so-called $f_\mathrm{NL}$ parameters), one takes the inner product of 
this binned bispectrum with theoretically motivated templates.
However, as a complementary approach one can also smooth the binned
bispectrum using a variable smoothing scale in order to suppress noise
and make coherent features stand out above the noise. This allows one
to look in a model-independent way for any statistically significant
bispectral signal. This approach is useful for characterizing
the bispectral shape of the galactic foreground emission, for which a
theoretical prediction of the bispectral anisotropy is lacking, and
for detecting a serendipitous primordial signal, for which a
theoretical template has not yet been put forth. Both the
template-based and the non-parametric approaches are described in this
paper.}

\begin{document}

%\date{1 September 2011; revised 11 September 2011; revised 13 October 2011}

%\pagerange{\pageref{firstpage}--\pageref{lastpage}} \pubyear{2002}

\label{firstpage}

\maketitle

\section{Introduction}

A fundamental question of observational cosmology is whether the
primordial cosmological perturbations were precisely Gaussian, or
whether small departures from exact Gaussianity can be detected at a
statistically significant level and then characterized. Here the
qualification `primordial' is essential because our goal is to probe
the new physics at play in the very early Universe.
However, it is also important to study
the non-Gaussianity that was subsequently imprinted at
late times through known processes, in particular the nonlinear
dynamics of gravitational clustering,
in order to `decontaminate' the primordial non-Gaussianity.
Observations of the cosmic
microwave background (CMB) anisotropies in temperature and
polarization are particularly well-suited to addressing this
fundamental question, as they provide a clean probe of the initial
conditions because most (but not all) of the CMB anisotropy was
imprinted well before nonlinear effects became important.

Non-Gaussianity is a vast subject because Gaussian stochastic
processes are the exception rather than the rule, comprising a set of
measure zero within the space of all possible stochastic processes for
generating an initial state. Given the highly exceptional nature of
the stochastic processes that are exactly Gaussian, there is a great
need to identify departures from Gaussianity that are in some sense
theoretically well-motivated. Bispectral non-Gaussianity may be
regarded as the leading-order correction to exact Gaussianity in some
sort of a perturbative expansion.

Early investigations of the statistics of large-scale structure of the
universe, mainly on grounds of simplicity, assumed Gaussian statistics
before there was a well-motivated theoretical explanation for why the
primordial cosmological perturbations should be very nearly
Gaussian. However in the mid-1980s the calculation of the cosmological
perturbations generated from quantum vacuum fluctuations during
inflation provided a theoretical justification for using Gaussian
statistics to model the large-scale structure. 
Indeed Gaussianity was proclaimed as being one of the core predictions of 
inflation.

Inflation however does not predict exact Gaussianity no matter what
model of inflation is assumed. It cannot be modeled by a free
field theory because at a minimum the gravitational sector is
nonlinear. Additional nonlinearity will of course also arise from
other sources, such as for example from the nonlinearity of the
inflationary potential.
The departures from Gaussianity predicted within the framework of
single-field inflation were calculated by Maldacena~\cite{Maldacena:2002vr} 
and by Acquaviva et al.~\cite{Acquaviva:2002ud}.
Many other inflation models have been introduced in the literature
that can produce non-negligible non-Gaussianity. For example, models
where multiple scalar fields play a role during inflation, where 
isocurvature perturbations are generated, or where inflation starts 
in an excited vacuum state. In some string-based models, as well as in 
some modified gravity or effective-field theories, the kinetic part 
of the inflaton Lagrangian can be non-standard, leading to novel bispectral
signatures. Deviations from the pure slow-roll phase in the inflaton
potential can also produce oscillations in the bispectrum.
See e.g.,~\cite{2010AdAst2010E..73L} or \cite{Ade:2013ydc,Ade:2015ava} for
a review.

Under the assumption of statistical isotropy, the bispectrum of the
map of a scalar quantity reduces to a function of three multipole
numbers $B_{\ell_1 \ell_2 \ell_3}$, where the bispectrum is
symmetric under permutations and vanishes unless the $\ell$-triplet satisfies
the triangle inequality \cite{Luo:1993xx}. If we include polarization, which 
in turn can be decomposed into $E$ and $B$ components, then the bispectrum needs
to be generalized to $B^{p_1 p_2 p_3}_{\ell_1 \ell_2 \ell_3}$, where
$p_1,p_2,p_3=T,E$ (we will not consider $B$-polarization in this paper).
If we insist on exploiting the highest possible
spectral resolution of the CMB maps (not necessarily the best idea),
then the number of reduced bispectral coefficients that can be
measured is huge, scaling with $\ell_\mathrm{max}$ as $\ell_\mathrm{max}^3$, and
the individual coefficients are too contaminated by noise to be useful
in detecting bispectral Gaussianity. A major and unavoidable
contribution to this noise arises from cosmic variance --- that is, from
the departures from zero of $B^{p_1 p_2 p_3}_{\ell_1 \ell_2 \ell_3}$
that would occur even if the underlying stochastic process were exactly
Gaussian. While Gaussianity requires that the expectation value of the
bispectral coefficients, calculable only in the limit of an infinite
number of sky realizations, vanishes, the value calculated for any single sky
realization will include fluctuations about this expectation value.
For this reason, in order to make any meaningful detection of
bispectral non-Gaussianity in the data, it is necessary to combine, in
one way or another, many measured bispectral coefficients in order to
make the signal stand out over the noise.

There are basically two situations to be considered. If we have a
simple parametric model for the expected pattern of bispectral
non-Gaussianity (generally parameterized by an amplitude called
$f_\mathrm{NL}$), then an optimal estimator can be constructed by
summing the observed bispectral coefficients over
$\ell_1,\ell_2,\ell_3$ using inverse variance weighting. Another
situation to be considered involves non-parametric reconstruction of the
bispectrum, where we do not have a specific template in mind, but want
to smooth the bispectrum in order to reduce the noise and see whether
there is a broad signal that stands out over the noise at a statistically
significant level. This latter approach is particularly relevant for
studying the bispectral properties of foregrounds, for which a theory
of the expected shape of the bispectral non-Gaussianity is lacking.

Combining the bispectral coefficients is not only required from the
physical point of view (to obtain statistically significant results), but
also computationally: computing $\mathcal{O}(\ell_\mathrm{max}^3)$ bispectral
coefficients for each map is not feasible in practice. A natural
solution, motivated by the second case mentioned above as well as the
observation that many of the templates of the first case are very smooth,
is to bin the bispectrum in harmonic space. This is the basis of the
binned bispectrum estimator, which was first described in \cite{Bucher:2009nm},
and which is the subject of this paper (see also \cite{Casaponsa:2013mja} for
an independent investigation of the binned bispectrum estimator, and
\cite{Santos:2002df} for a first rudimentary flat-sky estimator based on a 
binned bispectrum applied to the MAXIMA data).
The binned bispectrum estimator has established itself as one of the
three main bispectrum estimators used successfully for the official
analysis of the Planck data, both in 2013~\cite{Ade:2013ydc} and in
2015~\cite{Ade:2015ava}. The other two are the KSW 
estimator~\cite{Komatsu:2003iq,Yadav:2007rk,Yadav:2007ny} and the modal
estimator~\cite{Fergusson:2009nv,Fergusson:2010dm,Fergusson:2014gea},
and we will now briefly describe the main methodological differences between
these three estimators. In addition, other bispectrum estimators exist,
based on wavelets (e.g.~\cite{Curto:2010si}), 
needlets (e.g.~\cite{2008EJSta...2..332L}), 
and Minkowski functionals (e.g.~\cite{Ducout:2012it})
(see \cite{Ade:2013ydc} for more complete references).

The KSW estimator (separable template fitting) is based
on the observation that if the primordial bispectrum template is
separable as a function of $k_1,k_2,k_3$ (or alternatively the CMB bispectrum
template is separable as a function of $\ell_1,\ell_2,\ell_3$ modulo a possible
overall integral over $r$, the radial distance towards the surface of last
scattering), then the terms in the optimal estimator for
$f_\mathrm{NL}$ can be reordered as a product of terms depending only on
$k_1,\ell_1$, terms depending only on $k_2,\ell_2$, and terms depending only
on $k_3,\ell_3$ (within an overall integral over $r$). This significantly
reduces the computational cost (by effectively replacing a three-dimensional
integral and sum by the product of three one-dimensional integrals and sums),
at the cost of losing the ability for full bispectrum reconstruction. The KSW
estimator is fast, but only works for separable templates and can only be used
for the first case mentioned above (template fitting).\footnote{The
skew-$C_\ell$ extension~\cite{Munshi:2009ik} of the KSW estimator allows the
determination of a so-called bispectrum-related power spectrum, which contains
the contribution to $f_\mathrm{NL}$ (for a given shape) of all triangles with
one side equal to $\ell$.}

The modal estimator builds on the idea of the KSW estimator by first
expanding the theoretical bispectrum templates and the bispectrum of the
map in a basis of separable templates, the so-called modes. (For the Planck
2015 analysis two pipelines were used, one with a basis of 600 polynomials,
and the other with 2000.) The coefficients of the individual modes are then
computed using the KSW technique. In this way one can in principle treat
any bispectrum template, separable or not, as well as reconstruct the full
bispectrum of the map. These advantages come at the cost of often needing
a large number of modes for sufficient convergence, which can become
computationally heavy.

The binned bispectrum estimator does not use the KSW technique and keeps
the full three-dimensional sum. The required computational reduction comes
from reducing the number of terms in the sum by binning the bispectrum in
harmonic space, as will be discussed in detail in this paper.
In this way one can do both template fitting (with templates that do not need
to be separable) and full bispectrum reconstruction as mentioned above.
Moreover, the estimator is very fast when applied to a map, has a convenient
modular structure (which means for example that one can analyze an additional
template without having to rerun the map), and gives the dependence of
$f_\mathrm{NL}$ on $\ell$ as a free bonus. The possible drawback is that the 
method works only for bispectra that are relatively smooth (or have rapid 
oscillations only in a limited $\ell$-range) in order for a limited number 
of bins (about 50--60 in practice) to suffice.

The basic output of the binned bispectrum estimator is a binned, or
coarse-grained, pseudo-bispectrum. Here `pseudo' indicates that
full-sky spherical harmonic transforms have been applied to a masked
sky, so that the recovered $a_{\ell m}$ coefficients are in fact a
convolution of the real CMB multipole coefficients with the multipole
coefficients of the mask. How one corrects for the artefacts of the
mask is discussed in detail in Section~\ref{masking}. Below we shall
almost always assume the presence of a mask but will omit the
qualification `pseudo'. The coarse-grained pseudo-bispectrum can be
combined with a
library of theoretical templates by means of an inner product that
generates optimally matched filters. This approach was described
in~\cite{Bucher:2009nm}, where it was shown that with a modest number
of bins, the loss of information compared to an unbinned analysis is
negligible. One can thus determine the so-called $f_\mathrm{NL}$
parameter for various templates, but one can also construct other
estimators, for example to look for the acoustic peaks in the
bispectrum (see~\cite{Bucher:2009nm}).

Section~\ref{binnedbispsec} describes the binned
bispectrum and how it is calculated from a sky map.
Because there have been many refinements of
the binned estimator (such as its generalization to include
polarization) since the previous paper, we will discuss
the parametric analysis in detail in Section~\ref{fNLsec}.
As mentioned above, Section~\ref{masking} describes how to deal with
the masked sky. The implementation details of the code are discussed in
Section~\ref{implementationsec}.
The binned bispectrum can also be used to carry out a non-parametric,
or model-independent, analysis. In such an analysis the binned
bispectrum can be smoothed to search for a serendipitous statistically
significant signal of bispectral non-Gaussianity in the CMB for which
templates have not yet been proposed, or to characterize the
bispectral properties of foregrounds without a well-motivated
theoretical template. The construction of the full smoothed
bispectrum is treated in Section~\ref{smoothingsec}. The smoothing
complicates the statistical analysis of the significance of any 
non-Gaussian features because it introduces correlations between
neighbouring bins. In Section~\ref{statanalsec} we develop a method
to address this complication and provide an illustration by applying it
to the bispectrum of a realistic map containing a Gaussian CMB and radio
point sources.
Some conclusions are presented in Section~\ref{conclusionsec}. 
Appendix~\ref{App:AppendixA} provides both a further discussion of
the templates of Section~\ref{fNLsec} and a further illustration of the 
techniques of Section~\ref{smoothingsec} by presenting
two-dimensional cross-sections of the smoothed template bispectra.

\section{Binned bispectrum estimator}
\label{binnedbispsec}

\subsection{Binned bispectrum}
\label{sec_binned_bispectrum}

In this paper we define the bispectrum by means of the 
expectation value
\be
B_{\ell_1 \ell_2 \ell_3}^{p_1 p_2 p_3}  =
\left\langle 
\int d\hat \Omega \,
M_{\ell _1}^{p_1}(\hat \Omega ) M_{\ell _2}^{p_2}(\hat \Omega ) M_{\ell _3}^{p_3}(\hat \Omega )
\right\rangle
\label{Bobsaa}
\ee
where $M_\ell^p(\hat \Omega )$ is a map (as a function of position 
$\hat \Omega$ on the celestial sphere) where all but the $(2\ell +1)$ 
components having 
multipole number $\ell$ in the spherical harmonic decomposition have been 
filtered out, so that 
\be
M_\ell^p (\hat \Omega) = \sum_{m=-\ell}^{+\ell} a_{\ell m}^p Y_{\ell m}(\hat \Omega).
\label{Texpans}
\ee
The label $p$ refers in general to either temperature ($T$), 
$E$-polarization ($E$), or $B$-polarization ($B$), although in this paper
we will only consider the former two.
Inserting the expansion (\ref{Texpans}) into (\ref{Bobsaa}) and using
the Gaunt integral
$\int d\hat \Omega \, Y_{\ell_1 m_1} Y_{\ell_2 m_2} Y_{\ell_3 m_3}
= \sqrt{N_\triangle} \left( \begin{smallmatrix}
\ell _1 &\ell _2 &\ell _3\cr
m_1 & m_2 & m_3\cr
\end{smallmatrix} \right)$,
we find that 
\be
B_{\ell_1 \ell_2 \ell_3}^{p_1 p_2 p_3}  =
\sqrt{N_\triangle^{\ell_1 \ell_2 \ell_3}}
\sum_{m_1,m_2,m_3} \begin{pmatrix}
\ell _1 &\ell _2 &\ell _3\cr
m_1 & m_2 & m_3\cr
\end{pmatrix} 
\langle a_{\ell_1 m_1}^{p_1} a_{\ell_2 m_2}^{p_2} a_{\ell_3 m_3}^{p_3} \rangle,
\label{Bangleav}
\ee
where we have defined
\be
N_\triangle^{\ell_1 \ell_2 \ell_3} \equiv
\frac{(2\ell_1+1)(2\ell_2+1)(2\ell_3+1)}{4\pi }
\begin{pmatrix}
\ell _1 &\ell _2 &\ell _3\cr
0&0 &0\cr
\end{pmatrix}^2 ,
\label{numtriangles}
\ee
which 
may be interpreted 
as the number of possible $(\ell_1,\ell_2,\ell_3)$
triangles on the celestial sphere. (This quantity is sometimes denoted as 
$h_{\ell_1 \ell_2 \ell_3}^2$ in the literature.)
The quantity $\langle a_{\ell_1 m_1}^{p_1} a_{\ell_2 m_2}^{p_2} a_{\ell_3 m_3}^{p_3} \rangle$
is the CMB angular bispectrum, 
while the quantity 
$B_{\ell_1 \ell_2 \ell_3}^{p_1 p_2 p_3} / N_\triangle^{\ell_1 \ell_2 \ell_3}$ is known as the
reduced bispectrum.

From these equations we can derive two 
mathematical properties of the bispectrum. Firstly,
the bispectrum is symmetric under the simultaneous interchange of its 
three multipole numbers $\ell_1,\ell_2,\ell_3$ and its three polarization
indices $p_1,p_2,p_3.$ 
This means that it is sufficient 
to consider only the subspace $\ell _1\le \ell _2\le \ell _3$. It should be 
noted, however, that once we have both temperature and polarization, imposing 
this condition means that we no longer have the freedom to rearrange the
polarization indices, so that for example the $TTE$, $TET$, and $ETT$ 
combinations correspond to three distinct bispectra.
Secondly, because of the presence of the Wigner $3j$-symbol with all $m$'s
equal to zero in 
(\ref{numtriangles}), both the parity condition ($\ell_1+\ell_2+\ell_3$ even)
and the triangle inequality (consisting of 
$|\ell_1-\ell_2| \le \ell_3 \le \ell_1+\ell_2$ and permutations thereof) must 
be satisfied. Otherwise the bispectrum coefficient vanishes.
Note that the parity condition is a consequence of our starting point
(\ref{Bobsaa}). One could also define the angle-averaged bispectrum
as in (\ref{Bangleav}) but without the $\sqrt{N_\triangle}$ factor in front.
That expression would still include parity-odd modes as well in principle. 
The parity condition is a selection rule that results under the assumption 
that the underlying stochastic process is invariant under spatial 
inversion, which is the case for example for most scalar field models of 
inflation. In that case the odd-parity bispectrum can only be noise
from cosmic variance and thus is not worth analyzing. In this paper
we do not consider parity-odd combinations involving $B$-polarization
or chiral models where parity is not a good symmetry (like in the
standard electroweak model). See for example \cite{Kamionkowski:2010rb,
Shiraishi:2014roa, Shiraishi:2014ila} for studies of parity-odd bispectra.

To compute the observed bispectrum with the 
maximum possible resolution, we would simply evaluate
the integral over the sky of triple products of 
maximally filtered observed sky maps
\be
B_{\ell_1 \ell_2 \ell_3}^{p_1 p_2 p_3, \mathrm{obs}}=
\int d\hat \Omega \,
M_{\ell _1}^{p_1, \mathrm{obs}}(\hat \Omega) M_{\ell _2}^{p_2, \mathrm{obs}}(\hat \Omega) 
M_{\ell _3}^{p_3, \mathrm{obs}}(\hat \Omega) 
\label{Bobsab}
\ee
for each distinct triplet satisfying the above selection rules.
(In practice this integral is evaluated as 
a sum over pixels.)
The total number of triplets would 
be $\mathcal{O}(10^{7})$ for a WMAP or $\mathcal{O}(10^{9})$ 
for a Planck temperature map. 

But we can also use broader filters for the integral in 
(\ref{Bobsab}), as we shall see later with very little 
loss of information because a modest resolution in $\ell$ suffices for 
many physically motivated templates for which the predicted 
$B_{\ell_1 \ell_2 \ell_3}^{p_1 p_2 p_3}$ varies slowly with its $\ell$ arguments.
We end up having to compute only $\mathcal{O}(10^{4})$ 
bin triplets, leading to an enormous reduction in the 
computational resources required. We divide the $\ell$-range 
$[\ell_\mathrm{min}, \ell_\mathrm{max}]$ into subintervals denoted by 
$\Delta_i= [\ell_i,\ell_{i+1}-1]$ where $i=0,\ldots ,(N_\mathrm{bins}-1)$
and $\ell_{N_\mathrm{bins}} = \ell_\mathrm{max}+1$, so that 
the filtered maps are
\be
M_i^p(\Omega) = 
\sum_{\ell\in\Delta_i} \sum_{m=-\ell}^{+\ell}
a_{\ell m}^p Y_{\ell m}(\hat\Omega),
\label{Tmapbinned}
\ee
and we use these instead of $M_\ell^p$ in the expression for the bispectrum
(\ref{Bobsab}). The binned bispectrum is 
\be
B_{i_1 i_2 i_3}^{p_1 p_2 p_3, \mathrm{obs}}= \frac{1}{\Xi_{i_1 i_2 i_3}}
\int d\hat\Omega \,
M_{i_1}^{p_1, \mathrm{obs}}(\hat\Omega) M_{i_2}^{p_2, \mathrm{obs}}(\hat\Omega) 
M_{i_3}^{p_3, \mathrm{obs}}(\hat\Omega)
\label{Bobsbinned}
\ee
where $\Xi_{i_1 i_2 i_3}$ is the number of $\ell$ triplets within 
the $({i_1, i_2, i_3})$ bin triplet
satisfying the triangle inequality and parity condition selection rule. 
Because of this normalization factor,
$B_{i_1 i_2 i_3}^{p_1 p_2 p_3}$ may be considered an average over 
all valid $B_{\ell_1 \ell_2 \ell_3}^{p_1 p_2 p_3}$ inside the bin triplet.

\subsection{Variance}
\label{sec_variance}

We start by considering only the temperature bispectrum.
The covariance of the 
bispectra $B_{\ell_1 \ell_2 \ell_3}$ and $B_{\ell_4 \ell_5 \ell_6}$ equals
the average of the product minus the product of the 
averages. Under the assumption of weak non-Gaussianity 
the calculation simplifies significantly. In that case one can neglect
the average value of the bispectra, and the average of the product,
\begin{align}
\langle B_{\ell_1 \ell_2 \ell_3} B_{\ell_4 \ell_5 \ell_6} \rangle
= & \sqrt{N_\triangle^{\ell_1 \ell_2 \ell_3} N_\triangle^{\ell_4 \ell_5 \ell_6}}
 \nonumber\\
& \times
 \sum_{\tiny\begin{array}{c@{}c@{}c} m_1, & m_2, & m_3, \\ m_4, & m_5, & m_6\\ 
\end{array}} \!\!\!\! 
\begin{pmatrix}
\ell _1 &\ell _2 &\ell _3\cr
m_1 & m_2 & m_3\cr
\end{pmatrix} 
\begin{pmatrix}
\ell _4 &\ell _5 &\ell _6\cr
m_4 & m_5 & m_6\cr
\end{pmatrix} 
\langle a_{\ell_1 m_1} a_{\ell_2 m_2} a_{\ell_3 m_3}
a_{\ell_4 m_4}^* a_{\ell_5 m_5}^* a_{\ell_6 m_6}^*  \rangle
\end{align}
(using the fact that $B$ is real so that $B=B^*$),
can be rewritten as the product of three power spectra 
$C_\ell \equiv \langle a_{\ell m} a_{\ell m}^* \rangle$ using Wick's
theorem
\begin{align}
\langle a_{\ell_1 m_1} & a_{\ell_2 m_2} a_{\ell_3 m_3}
a_{\ell_4 m_4}^*  a_{\ell_5 m_5}^* a_{\ell_6 m_6}^*  \rangle
= C_{\ell_1} C_{\ell_2} C_{\ell_3} [  \delta_{\ell_1 \ell_4} \delta_{\ell_2 \ell_5} 
\delta_{\ell_3 \ell_6} \delta_{m_1 m_4} \delta_{m_2 m_5} \delta_{m_3 m_6}
 \nonumber\\
& + (14)(26)(35) + (15)(24)(36) + (15)(26)(34) + (16)(24)(35) + (16)(25)(34) ],
\label{6ptfunc}
\end{align}
using obvious shorthand to denote the other permutations of 
$\delta$-functions. Due to the $\delta$-functions, the covariance matrix is
diagonal, so we need to consider only the (diagonal) variance of 
$B_{\ell_1 \ell_2 \ell_3}$. We use the identity
\be
\sum_{m_1 m_2 m_3} \begin{pmatrix} \ell _1 &\ell _2 &\ell _3\cr
m_1 & m_2 & m_3\cr \end{pmatrix}
\begin{pmatrix} \ell _1 &\ell _2 &\ell _3\cr
m_1 & m_2 & m_3\cr \end{pmatrix} = 1
\ee
and the fact that for even parity of $\ell_1+\ell_2+\ell_3$ the columns of the
Wigner $3j$-symbol can be permuted to obtain 
\be
\Var(B_{\ell_1 \ell_2 \ell_3}) = 
g_{\ell_1 \ell_2 \ell_3} N_\triangle^{\ell_1 \ell_2 \ell_3}
C_{\ell_1} C_{\ell_2} C_{\ell_3}
\equiv V_{\ell_1 \ell_2 \ell_3}
\ee
with $g_{\ell_1 \ell_2 \ell_3}$ equal to 6, 2, or 1, depending on whether 3,
2, or no $\ell$'s are equal, respectively, and $N_\triangle$ defined
as in (\ref{numtriangles}). 
Similarly the variance of the binned bispectrum 
$B_{i_1 i_2 i_3} = ({\Xi_{i_1 i_2 i_3}})^{-1}
\sum_{\ell_1\in\Delta_1} \sum_{\ell_2\in\Delta_2} \sum_{\ell_3\in\Delta_3}
B_{\ell_1 \ell_2 \ell_3}$ is given by
\be
\Var(B_{i_1 i_2 i_3}) = 
\frac{g_{i_1 i_2 i_3}}{(\Xi_{i_1 i_2 i_3})^2} 
\sum_{\ell_1\in\Delta_1} \sum_{\ell_2\in\Delta_2} \sum_{\ell_3\in\Delta_3}
N_\triangle^{\ell_1 \ell_2 \ell_3}
C_{\ell_1} C_{\ell_2} C_{\ell_3}
\equiv V_{i_1 i_2 i_3}
\label{idealVariance}
\ee
with $g_{i_1 i_2 i_3}$ equal to 6, 2, or 1, depending on whether 3,
2, or no $i$'s are equal, respectively. The $\delta$-functions in 
(\ref{6ptfunc}) lead here to conditions of equality on the bins, since due
to the sum over all $\ell$'s inside a bin, $\delta_{\ell_a \ell_b}$ will always
give 1 if $\ell_a$ and $\ell_b$ are in the same bin, and 0 if not.

With the noise and beam smoothing present in a real experiment, 
(\ref{idealVariance}) becomes
\be
V_{i_1 i_2 i_3} =
\frac{g_{i_1 i_2 i_3}}{(\Xi_{i_1 i_2 i_3})^2} 
 \sum_{\ell_1\in\Delta_1} \sum_{\ell_2\in\Delta_2} \sum_{\ell_3\in\Delta_3}
N_\triangle^{\ell_1 \ell_2 \ell_3} 
(b_{\ell_1}^2 C_{\ell_1} + N_{\ell_1})
(b_{\ell_2}^2 C_{\ell_2} + N_{\ell_2}) 
(b_{\ell_3}^2 C_{\ell_3} + N_{\ell_3})
\ee
where $b_\ell$ is the beam transfer function and $N_\ell$ the instrument 
noise power spectrum. This expression is exact only for 
an axisymmetric beam and isotropic noise; otherwise it is an approximation
(because the beam and noise properties would include off-diagonal matrix 
elements). For a Gaussian beam, the beam transfer function is typically 
specified by the full width at half maximum $\theta_\mathrm{FWHM}$ (in radians),
so that
$
b_\ell = \exp \left[ - \frac{1}{2} \ell(\ell+1) \theta^2_\mathrm{FWHM} 
/ (8 \ln 2) \right].
$
A pixel window function $w_\ell$ to account for pixelization effects
is combined with the beam transfer function according to 
$b_\ell \rightarrow w_\ell b_\ell$.

For bispectral elements including both $T$ and $E$, the
variance is replaced by the covariance matrix in polarization space, whose
expression without binning is
\be
\mathrm{Covar}(B_{\ell_1 \ell_2 \ell_3}^{p_1 p_2 p_3}, B_{\ell_1 \ell_2 \ell_3}^{p_4 p_5 p_6}) 
= g_{\ell_1 \ell_2 \ell_3} N_\triangle^{\ell_1 \ell_2 \ell_3}
(\tilde{C}_{\ell_1})^{p_1 p_4} (\tilde{C}_{\ell_2})^{p_2 p_5}
(\tilde{C}_{\ell_3})^{p_3 p_6} \equiv V_{\ell_1 \ell_2 \ell_3}^{p_1 p_2 p_3 p_4 p_5 p_6},
\label{covarexact_polar}
\ee
where
\be
\tilde{C}_\ell = \left( \begin{array}{cc}
(b_\ell^T)^2 C_\ell^{TT} + N_\ell^T & b_\ell^T b_\ell^E C_\ell^{TE}\\
b_\ell^T b_\ell^E C_\ell^{TE} & (b_\ell^E)^2 C_\ell^{EE} + N_\ell^E
\end{array} \right).
\label{Cmatrix_polar}
\ee
Here noise uncorrelated in $T$ and $E$ has been assumed.
Similarly, for the binned case 
\begin{align}
\mathrm{Covar}(B_{i_1 i_2 i_3}^{p_1 p_2 p_3}, B_{i_1 i_2 i_3}^{p_4 p_5 p_6}) 
& =  \frac{g_{i_1 i_2 i_3}}{(\Xi_{i_1 i_2 i_3})^2} 
\sum_{\ell_1\in\Delta_1} \sum_{\ell_2\in\Delta_2} \sum_{\ell_3\in\Delta_3}
N_\triangle^{\ell_1 \ell_2 \ell_3}
(\tilde{C}_{\ell_1})^{p_1 p_4} (\tilde{C}_{\ell_2})^{p_2 p_5}
(\tilde{C}_{\ell_3})^{p_3 p_6} \nonumber\\ & \equiv V_{i_1 i_2 i_3}^{p_1 p_2 p_3 p_4 p_5 p_6}.
\label{covarbinned_polar}
\end{align}

Some subtleties arise in the derivation of equation 
(\ref{covarexact_polar}). The covariance matrix is in principle
an $8 \times 8$ matrix, given that there are 8 independent polarized bispectra
$TTT$, $TTE$, $TET$, $TEE$, $ETT$, $ETE$, $EET$, and $EEE$. As mentioned 
before, note that for
example $TTE$ and $TET$ are not the same: each polarization index $p_i$ is 
coupled to a multipole index $\ell_i$, and cannot be exchanged due to the 
restriction $\ell_1 \leq \ell_2 \leq \ell_3$ that we will always impose in 
order to reduce computation time. A naive 
calculation of this $8 \times 8$ matrix appears to lead to a more complicated
expression in the case of equal $\ell$'s that is not proportional to 
$g_{\ell_1 \ell_2 \ell_3}$. However, one should treat the cases
where two or three $\ell$'s are equal separately. For example,
when $\ell_2 = \ell_3$, one {\em can} exchange the last two polarization 
indices and one finds that $TTE = TET$ and $ETE = EET$. Hence in that case there
are only 6 independent bispectra, and the covariance matrix is $6\times 6$.
Similarly, when all three $\ell$'s are equal, 
the covariance matrix is $4\times 4$.

However, it turns out that as far as computing $f_\mathrm{NL}$ is concerned,
when evaluating the sum in (\ref{innerprodexact_polar}),
properly treating the special cases where $\ell$'s are equal
by reducing the dimension of the covariance matrix and bispectrum vector,
the final result is identical to the following calculation: taking the
covariance matrix to be the $8\times 8$ matrix as computed in the case
of all $\ell$'s unequal, multiplying it by $g_{\ell_1 \ell_2 \ell_3}$, and 
then computing the sum in (\ref{innerprodexact_polar})
directly without treating the cases of equal $\ell$'s separately.
This second computation is much more convenient from a practical point
of view. Finally it can be shown that the latter expression of the 
covariance matrix can be rewritten as the separable product involving only
$2\times 2$ matrices in (\ref{covarexact_polar}).

Similarly it can be shown that the variances of the combinations
$B^{T2E} \equiv TTE+TET+ETT$ and $B^{TE2} \equiv TEE+ETE+EET$ used for the 
smoothed bispectrum (see Section~\ref{smoothingsec})
are also recovered correctly when using (\ref{covarexact_polar}) or
(\ref{covarbinned_polar}). Here one should use of course that Var($B^{T2E}$) = 
Var($TTE$) + Var($TET$) + Var($ETT$) + 2 Covar($TTE,TET$) + 2 Covar($TTE,ETT$)
+ 2 Covar($TET,ETT$), and similarly for Var($B^{TE2}$).
So in the end, while one should remember the caveats regarding 
(\ref{covarexact_polar}) and (\ref{covarbinned_polar}) in the case of equal
$\ell$'s or $i$'s, for the practical purposes of this paper they can be used
without any problem.

\subsection{Linear correction term}
\label{sec_lincorr}

The definition of the bispectrum in (\ref{Bobsaa}) assumes a
rotationally invariant CMB sky and that the bispectral expectation
values have even parity (as a consequence of the parity invariance of
the underlying stochastic process, which we assume here).  Because 
of rotational invariance, the $m$ dependence of the
expectation can be factored out, and the sample reduced bispectral
coefficients provide a lossless compression of the data concerning the
bispectrum. 
However, in a real experiment as opposed to idealized observations of 
the primordial sky, two sources of anisotropy arise that break rotational
invariance and require corrections to the bispectrum estimation to avoid
spurious results.

The first is anisotropic superimposed instrument noise,
due to for example an anisotropic scanning pattern of the satellite.
The second is anisotropy introduced by a mask needed to remove the brightest 
parts of our galaxy and the strongest point sources. These two anisotropic 
`contaminants', unlike for example foreground contaminants,
cannot be removed by cleaning and must be accounted for in the analysis. 
These anisotropic contaminants can mimic a primordial bispectrum signal. For 
example, due to an anisotropic scanning pattern of the experiment, certain 
(large-scale) areas of the sky may have less (small-scale) noise than other
areas. This 
correlation between large and small scales produces a contaminant bispectrum 
that peaks in the squeezed limit (bispectrum configurations with one small
$\ell$ and two large ones). As explained in the next Section, that is also
where the primordial so-called local shape has its main signal. 
Since the CMB and the noise are uncorrelated, the effect will average out
to zero in the central value of the bispectrum over a large number of maps 
(no bias), but it will increase the variance. And while an unbiased estimator 
will find the correct central value when averaged over a large number of maps, 
a larger variance does mean that there is more chance to find a value far from 
the true one when applied to a single map.

These contaminants can be mitigated by subtracting from the cubic 
expression of the observed bispectrum given in (\ref{Bobsab}) or 
(\ref{Bobsbinned}) a linear correction term, as shown in 
\cite{Creminelli:2005hu,Yadav:2007ny}, that is,
\be
B_{i_1 i_2 i_3}^{p_1 p_2 p_3, \mathrm{obs}} \rightarrow \Bigl(
B_{i_1 i_2 i_3}^{p_1 p_2 p_3, \mathrm{obs}} - B_{i_1 i_2 i_3}^{p_1 p_2 p_3, \mathrm{lin}}\Bigr) .
\ee
`Cubic' and `linear' here mean 
cubic and linear in the observed map, respectively.
The linear correction term is
\begin{align}
B_{i_1 i_2 i_3}^{p_1 p_2 p_3, \mathrm{lin}} = \int d\hat\Omega & \left [
M_{i_1}^{p_1, \mathrm{obs}}
\left\langle M_{i_2}^{p_2, G} M_{i_3}^{p_3, G} \right\rangle \right .
 \nonumber\\ & \left. + M_{i_2}^{p_2, \mathrm{obs}} \left\langle M_{i_1}^{p_1, G} M_{i_3}^{p_3, G} \right\rangle 
+ M_{i_3}^{p_3, \mathrm{obs}} \left\langle M_{i_1}^{p_1, G} M_{i_2}^{p_2, G} \right\rangle 
\right ],
\label{Bisp_lincorr}
\end{align}
where the average is over Gaussian CMB maps with the same beam,
(anisotropic) noise, and mask as the observed map.
The effect of the linear correction term on the measured value and variance 
of $f_\mathrm{NL}$ (see Section~\ref{fNLsec}) is illustrated in great detail
in Section~\ref{masking}. With this correction applied as well as an appropriate
treatment of the masked regions that will be discussed in 
Section~\ref{masking}, we will see that we once again have an effectively 
optimal estimator of the bispectrum of the sky.

\section{Parametric bispectrum estimation}
\label{fNLsec}

Once the binned bispectrum of a map has been determined, it can be
compared with the theoretical bispectra predicted by early Universe models. 
In particular the so-called $f_\mathrm{NL}$ parameter
can be determined, which is a measure of the amplitude of the bispectrum,
corresponding roughly to the bispectrum divided by the power
spectrum squared, although this ratio is generally a momentum-dependent
function and not a constant.

\subsection{Bispectrum templates}

\subsubsection{Standard primordial and foreground templates}
\label{sec_templates}

Since the temperature and $E$-polarization fluctuations are assumed to 
originate in density fluctuations produced in the early Universe, for example 
during inflation, the predicted values of the power spectrum and the 
bispectrum of the maps can be expressed in terms of the primordial power 
spectrum $P(k)$ and bispectrum $B(k_1,k_2,k_3)$ of the
gravitational potential $\Phi$ and the radiation transfer functions 
$\Delta_\ell^p(k)$. One finds (see e.g.~\cite{Komatsu:2001rj})
\be
C_\ell^{p_1 p_2} = \frac{2}{\pi} \int_0^\infty k^2 dk \, P(k) \Delta_\ell^{p_1}(k)
\Delta_\ell^{p_2}(k)
\label{Clth}
\ee
and
\begin{align}
B_{\ell_1 \ell_2 \ell_3}^{p_1 p_2 p_3, \mathrm{th}} = 
N_\triangle^{\ell_1 \ell_2 \ell_3}
\left( \frac{2}{\pi} \right)^3 
\int _0^\infty \!\!\! k_1^2 dk_1 \int _0^\infty \!\!\! k_2^2 dk_2
\int _0^\infty \!\!\! k_3^2 dk_3 & \Bigl[
\Delta_{\ell _1}^{p_1}(k_1) \Delta_{\ell _2}^{p_2}(k_2) \Delta_{\ell_3}^{p_3}(k_3)
B(k_1,k_2,k_3)
\nonumber\\
& \times \int _0^\infty \!\!\! r^2dr \, j_{\ell _1}(k_1r) j_{\ell _2}(k_2r) 
j_{\ell _3}(k_3r) \Bigr]
\label{Bth}
\end{align}
where the $j_\ell$ are spherical Bessel functions.
This expression should be multiplied by $f_\mathrm{NL}$ to find the full
bispectrum, but we consider $f_\mathrm{NL}$ an unknown parameter 
to be determined from the data and define the theoretical bispectrum 
template $B^\mathrm{th}$ assuming $f_\mathrm{NL}=1$.
The radiation transfer functions can be computed with freely available computer
codes like CAMB.\footnote{\tt http://camb.info} 
The power spectrum is generally parametrized by its 
amplitude $A$, pivot scale $k_0$, and reduced spectral index\footnote{The 
reduced spectral index is related to the normal one by $\tilde{n}=n-1$, so that
it is a quantity close to zero instead of close to one.} $\tilde{n}$ 
as $P(k) = A (k/k_0)^{-3 + \tilde{n}}$.

Many inflation or other early Universe models predict a primordial bispectrum
that can be approximated by one (or a combination) of only a few distinct 
shapes in momentum space (see e.g.~\cite{Babich:2004gb, Fergusson:2008ra}). 
Hence it makes sense to search for these canonical
shapes. However, it should be kept in mind that these shapes are only 
approximations, and with sufficient 
sensitivity and resolution the difference between slightly 
different templates that all fall within the same approximate category can 
be resolved. Inflation models can also produce shapes that are very different 
from the canonical shapes, for example with localized features or oscillations.
See \cite{Ade:2015ava} for an overview of all the different shapes
that were tested using the Planck 2015 data, as well as more complete
references. 
The purpose of this paper is not to give an exhaustive list of templates,
but to describe the methodology of the binned bispectrum estimator, providing 
only the most common templates as examples. 
In principle any bispectrum shape can be investigated with
the binned bispectrum estimator, as long as it lends itself well to binning
(either by being smooth everywhere, or by having rapid features only in a small
region of $\ell$-space, where the bin density can be increased without the
total number of bins becoming too large).

One often discussed property of templates is separability.
A primordial template is separable if the primordial bispectrum
$B(k_1,k_2,k_3)$ can be written as a product $f_1(k_1)f_2(k_2)f_3(k_3)$, 
or as the sum of a few of such terms. 
Similarly, at the level of the bispectrum of the CMB, separability means that 
it can be written as a (sum of a few) product(s) 
$f_1(\ell_1)f_2(\ell_2)f_3(\ell_3)$.
From (\ref{Bth}) we see that separability of $B(k_1,k_2,k_3)$ implies 
separability of the reduced bispectrum 
$B_{\ell_1 \ell_2 \ell_3}/N_\triangle^{\ell_1 \ell_2 \ell_3}$, except for the overall 
integral over $r$. Separability is a crucial feature for the oldest 
$f_\mathrm{NL}$ estimator (KSW, \cite{Komatsu:2003iq}), 
which can only handle 
separable templates (the integral over $r$ is not a problem), but is of
no importance for the binned bispectrum estimator, which does not require
any separation of factors in the bispectrum templates. 

The most well-known primordial bispectrum type is the so-called local 
bispectrum, 
\be
B^\mathrm{local}(k_1,k_2,k_3) = 
2 [ P(k_1) P(k_2) + P(k_1) P(k_3) + P(k_2) P(k_3) ].
\ee
It is called local because in real space it corresponds to
the local relation
$
\Phi(\vc{x}) = \Phi_G(\vc{x}) + f_\mathrm{NL}^\mathrm{local} 
( \Phi_G^2(\vc{x}) - \langle \Phi_G \rangle^2 )
$
where the subscript $G$ denotes the linear (Gaussian) part.
Squeezed configurations where one $k$ (or $\ell$) is much
smaller than the other two contribute the most to the local bispectrum.
The local bispectrum shape is typically produced in multiple-field
inflation models on superhorizon scales (see e.g.~\cite{Babich:2004gb,
Tzavara:2010ge}), or by other mechanisms that act on superhorizon scales,
such as curvaton models (see e.g.~\cite{Bartolo:2003jx}).

The two other canonical primordial shapes are the equilateral and orthogonal 
templates.
The equilateral bispectrum is dominated by equilateral configurations
where all $k$'s (or $\ell$'s) are approximately equal, and is typically
produced at horizon crossing in inflation models with higher-derivative
or other non-standard kinetic terms (or rather, the equilateral bispectrum 
is a separable approximation to the bispectrum produced in such models, see
\cite{Creminelli:2005hu}). It is given by
\begin{align}
B^\mathrm{equi}(k_1,k_2,k_3) = &
-6 [ P(k_1)P(k_2) + (\mathrm{2\ perms}) ] 
-12 \, P^{2/3}(k_1) P^{2/3}(k_2) P^{2/3}(k_3)
\nonumber\\ &+6 [ P(k_1) P^{2/3}(k_2) P^{1/3}(k_3) + (\mathrm{5\ perms}) ].
\end{align}
The orthogonal bispectrum \cite{Senatore:2009gt} has 
been constructed to be orthogonal to the equilateral shape in such
a way that the bispectrum predicted by generic single-field inflation 
models can be written as a linear combination of the equilateral and
orthogonal shapes. It gets its main contribution from configurations
that are peaked both on equilateral and on flattened triangles
(where two $k$'s are approximately equal and the third is approximately
equal to their sum), with opposite sign, and is given by
\begin{align}
B^\mathrm{ortho}(k_1,k_2,k_3) = &
-18 [ P(k_1)P(k_2) + (\mathrm{2\ perms}) ] 
-48 \, P^{2/3}(k_1) P^{2/3}(k_2) P^{2/3}(k_3)
\nonumber\\ & +18 [ P(k_1) P^{2/3}(k_2) P^{1/3}(k_3) + (\mathrm{5\ perms}) ]. 
\end{align}
It should be noted that the orthogonal shape is not at all orthogonal to
the local shape (as sometimes incorrectly stated in older literature). It
has a large correlation (about 40--50\%) with the local shape at Planck
resolution (see Section~\ref{sec_joint} and Table~\ref{tab_corr_coeff}).

In addition to these three shapes, it is also interesting to look for
non-primordial contaminant bispectra, either to study these foregrounds
or to remove them. In the first place a bispectrum will be produced by 
diffuse extra-galactic point sources. These can generally be divided into
two populations: unclustered and clustered sources. The former are radio and 
late-type infrared galaxies, while the latter are dusty star-forming galaxies 
constituting the cosmic infrared background (CIB). 
Secondly, gravitational lensing of the CMB will produce a bispectrum
that mimics the local shape, because there is a correlation between the 
lenses that produce modifications to the CMB power spectrum on small scales
and the integrated Sachs-Wolfe effect on large scales (both are due to the 
same mass distribution at low redshift).

The unclustered sources
can be assumed to be distributed according to a Poissonian distribution, 
and hence have a white noise power spectrum (i.e., with an amplitude 
independent of $\ell$). Then their bispectrum has a very simple theoretical 
shape~\cite{Komatsu:2001rj}:
\be
B_{\ell_1 \ell_2 \ell_3}^\mathrm{unclust} = 
N_\triangle^{\ell_1 \ell_2 \ell_3} \, b_\mathrm{ps}
\label{Bunclust}
\ee
where $b_\mathrm{ps}$, the amplitude of the unclustered point source bispectrum,
is the parameter that can be determined in the same way as the 
$f_\mathrm{NL}$ parameters for the primordial templates. Like most foregrounds,
but unlike primordial signals, the amplitude depends on the frequency channel,
which allows a multi-frequency experiment like Planck to (partially) clean
these contaminants from its maps.
The above relation is valid both in temperature and in polarization. However,
since not all point sources are polarized, the amplitude $b_\mathrm{ps}$ is not
the same in temperature and polarization, with the difference depending on the
mean polarization fraction of the point sources. Without taking into account
that fraction, it would not make sense to look at the mixed $TTE$ and $TEE$
components of its bispectrum, nor to try to determine $b_\mathrm{ps}$ jointly 
from temperature and polarization maps. In practice for Planck the contribution
from polarized point sources is negligible (see \cite{Ade:2015ava}), so that 
we might as well consider it a temperature-only template.

The clustered point sources (CIB) have a more complicated bispectrum. A simple
template that fits the data well was established in \cite{Lacasa:2013yya}
(see also \cite{Ade:2015ava}):
\be
B_{\ell_1 \ell_2 \ell_3}^\mathrm{CIB} = 
N_\triangle^{\ell_1 \ell_2 \ell_3} \, b_\mathrm{CIB}
\left[ \frac{(1+\ell_1/\ell_\mathrm{break}) (1+\ell_2/\ell_\mathrm{break}) 
(1+\ell_3/\ell_\mathrm{break})}{(1+\ell_0/\ell_\mathrm{break})^3}\right]^q,
\ee
where the index is $q=0.85$, the break is located at $\ell_\mathrm{break}=70$, 
and $\ell_0=320$  is the pivot scale for normalization.
In addition, $b_\mathrm{CIB}$ is the amplitude parameter to be determined.
As for the unclustered point sources, it depends on the frequency.
The CIB is found to be negligibly polarized, so that the above template is
only used in temperature.

The theoretical shape for the lensing-ISW bispectrum was worked out in
\cite{Goldberg:1999xm, Smith:2006ud, Lewis:2011fk} and is given by
\begin{align}
B_{\ell_1 \ell_2 \ell_3}^{p_1 p_2 p_3, \mathrm{lensISW}} = 
N_\triangle^{\ell_1 \ell_2 \ell_3} \Bigl [ &
C_{\ell_2}^{p_2\phi} C_{\ell_3}^{p_1p_3} f_{\ell_1 \ell_2 \ell_3}^{p_1}
+ C_{\ell_3}^{p_3\phi} C_{\ell_2}^{p_1p_2} f_{\ell_1 \ell_3 \ell_2}^{p_1}
+ C_{\ell_1}^{p_1\phi} C_{\ell_3}^{p_2p_3} f_{\ell_2 \ell_1 \ell_3}^{p_2}
\nonumber\\
& + C_{\ell_3}^{p_3\phi} C_{\ell_1}^{p_1p_2} f_{\ell_2 \ell_3 \ell_1}^{p_2} + C_{\ell_1}^{p_1\phi} C_{\ell_2}^{p_2p_3} f_{\ell_3 \ell_1 \ell_2}^{p_3}
+ C_{\ell_2}^{p_2\phi} C_{\ell_1}^{p_1p_3} f_{\ell_3 \ell_2 \ell_1}^{p_3} \Bigr ].
\end{align}
Here $C_\ell^{T\phi}$ and $C_\ell^{E\phi}$ are the temperature/polarization-lensing 
potential cross power spectra, while the CMB power spectra $C_\ell^{TT}$,
$C_\ell^{TE}$, $C_\ell^{EE}$ should be taken to be the {\em lensed}
$TT$, $TE$, $EE$ power spectra. The functions $f_{\ell_1 \ell_2 \ell_3}^{p}$ are 
defined by
\begin{align}
f_{\ell_1 \ell_2 \ell_3}^T & =
\frac{1}{2} \left[ \ell_2 (\ell_2 + 1) + \ell_3 (\ell_3 + 1) - \ell_1 (\ell_1+1)
\right ], \nonumber\\
f_{\ell_1 \ell_2 \ell_3}^E & =
\frac{1}{2} \left[ \ell_2 (\ell_2 + 1) + \ell_3 (\ell_3 + 1) - \ell_1 (\ell_1+1)
\right ]  
\left(\begin{array}{ccc} \ell_1 & \ell_2 & \ell_3 \\ 2 & 0 & -2 \end{array}\right)
\left(\begin{array}{ccc} \ell_1 & \ell_2 & \ell_3 \\ 0 & 0 & 0 \end{array}\right)^{-1},
\end{align}
if $\ell_1+\ell_2+\ell_3$ is even and $\ell_1,\ell_2,\ell_3$ satisfy the
triangle inequality, and zero otherwise. Using some mathematical properties of
the Wigner 3j-symbols we find that, under the same conditions as above,
the ratio of the two Wigner 3j-symbols can be computed explicitly as
\begin{align}
& \left(\begin{array}{ccc} 
\ell_1 & \ell_2 & \ell_3 \\ 2 & 0 & -2 \end{array}\right)
\left(\begin{array}{ccc} 
\ell_1 & \ell_2 & \ell_3 \\ 0 & 0 & 0 \end{array}\right)^{-1} = \nonumber\\
& \Bigl\{ [\ell_2 (\ell_2 + 1) - \ell_1 (\ell_1 + 1) - \ell_3(\ell_3+1)]
[\ell_2 (\ell_2 + 1) - \ell_1 (\ell_1 + 1) - \ell_3(\ell_3+1) + 2] \nonumber\\ 
& - 2\ell_1 (\ell_1 + 1) \ell_3(\ell_3+1) \Bigr\} \:
\Bigl[ 4(\ell_1-1)\ell_1(\ell_1+1)(\ell_1+2)(\ell_3-1)\ell_3(\ell_3+1)(\ell_3+2)
\Bigr]^{-\frac12}.
\end{align}
Note that there is no unknown amplitude parameter in front of this template:
its $f_\mathrm{NL}$ parameter should be unity.
A further discussion of the templates presented in this Section can be found
in Appendix~\ref{App:AppendixA}, where we present two-dimensional
sections of the template bispectra according to the techniques 
described in Section~\ref{smoothingsec}.

\subsubsection{Isocurvature non-Gaussianity}
\label{sec_isocurv}

The generalization of the binned bispectrum $f_\mathrm{NL}$ estimator
to the case where non-Gaussian isocurvature components are present in
addition to the standard adiabatic component was treated in
\cite{Langlois:2011hn,Langlois:2012tm}. For completeness we 
summarize the results here. In fact this boils down to the joint analysis
of a number of additional templates.

We make two simplifying assumptions: we consider only the local shape and
assume the same spectral index for the primordial isocurvature power spectrum
and the isocurvature-adiabatic cross power spectrum as for the adiabatic
power spectrum. In that case the primordial bispectrum can be written as
\be
B^{IJK}(k_1, k_2, k_3) = 
2 f_{\rm NL}^{I, JK}  P(k_2) P(k_3) 
+ 2 f_{\rm NL}^{J, KI}  P(k_1) P(k_3) + 2 f_{\rm NL}^{K, IJ}  P(k_1)P(k_2), 
\ee
where $I,J,K$ label the different modes (adiabatic and isocurvature).
(Note that unlike the expressions before, we here include the 
$f_\mathrm{NL}$ parameter in the expression for the theoretical 
bispectrum.)
The invariance of this expression under the simultaneous interchange of
two of these indices and the corresponding momenta means that 
$f_{\rm NL}^{I, JK} = f_{\rm NL}^{I, KJ}$, explaining the presence of the comma, and
reducing the number of independent $f_\mathrm{NL}$ parameters (from 8 to 6
in the case of two modes).
Inserting this expression into (\ref{Bth}), where 
$\Delta_{\ell _1}^{p_1}(k_1) \Delta_{\ell _2}^{p_2}(k_2) \Delta_{\ell_3}^{p_3}(k_3)$
should be replaced by
$\sum_{I,J,K} \Delta_{\ell _1}^{p_1\, I}(k_1) \Delta_{\ell _2}^{p_2\, J}(k_2) 
\Delta_{\ell_3}^{p_3\, K}(k_3)$,
finally leads to the result
\be
B_{\ell_1 \ell_2 \ell_3}^{p_1 p_2 p_3, \mathrm{th}} = 
\sum_{I,J,K} f_{\rm NL}^{I, JK} B_{\ell_1 \ell_2 \ell_3}^{p_1 p_2 p_3\, I,JK},
\ee
where
\be
B_{\ell_1 \ell_2 \ell_3}^{p_1 p_2 p_3\, I,JK}= 6 \int_0^\infty r^2 dr \, 
\alpha^{p_1\, I}_{(\ell_1}(r)\beta^{p_2\, J}_{\ell_2}(r)\beta^{p_3\, K}_{\ell_3)}(r),
\label{Bth_isocurv}
\ee
with   
\be
\alpha^{p\, I}_{\ell}(r) \equiv \frac{2}{\pi} \int k^2 dk\,  j_\ell(kr) \, 
\Delta^{p\, I}_{\ell}(k),
\qquad\qquad
\beta^{p\, I}_{\ell}(r) \equiv \frac{2}{\pi}  \int k^2 dk \,  j_\ell(kr) 
\, \Delta^{p\, I}_{\ell}(k)\,  P(k).
\ee
Here we use the notation
$(\ell_1 \ell_2 \ell_3)\equiv [\ell_1\ell_2\ell_3+ 5\,  {\rm perms}]/3!$
and it should be kept in mind that the $\ell_i$ and $p_i$ are always kept
together (so the $p_i$ are also permuted in the same way).

We can conclude that including the possibility of isocurvature non-Gaussianity
in our investigations means that we have to replace the single local 
adiabatic bispectrum template by the family of templates (\ref{Bth_isocurv}),
each with their individual $f_\mathrm{NL}$ parameter. In particular, if we 
assume the presence of only a single isocurvature mode in addition to the
adiabatic one (i.e. one of cold dark matter, neutrino density, or neutrino
velocity), we have six local $f_\mathrm{NL}$ parameters to determine instead
of just one, and these should always be estimated jointly (see 
Section~\ref{sec_joint}). Two-dimensional sections of the
isocurvature bispectra can be found in Appendix~\ref{App:AppendixA}.

\subsection{$f_\mathrm{NL}$ estimation}

We start by considering the case where we have only temperature.
In order to estimate $f_\mathrm{NL}$ using a template 
$B_{\ell_1 \ell_2 \ell_3}^\mathrm{th}$, the estimator 
\be
{\hat f}_\mathrm{NL}=
\frac{
\left\langle
B^\mathrm{th,exp},
B^\mathrm{obs}
\right\rangle
}{
\left\langle
B^\mathrm{th,exp},
B^\mathrm{th,exp}
\right\rangle
}
\label{fNL_estimator}
\ee
is constructed using the inner product 
\be 
\langle B^A, B^B \rangle^\mathrm{no\ binning} =
\sum_{\ell _1 \leq \ell _2 \leq \ell_3}
\frac{
B^A_{\ell _1 \ell _2 \ell _3}
B^B_{\ell _1 \ell _2 \ell _3}
}{
V_{\ell _1 \ell _2 \ell _3}
}.
\label{innerprodexact}
\ee
This definition satisfies the mathematical axioms 
of an inner product as long as bin triplets with infinite
variance are excluded from the sum.
The theoretical bispectrum for the experiment is related to the 
theoretically predicted infinite angular resolution
bispectrum by the relation 
$B^\mathrm{th,exp}_{\ell_1 \ell_2 \ell_3}= b_{\ell _1} b_{\ell _2} b_{\ell _3}
B^\mathrm{th,f_\mathrm{NL}=1}_{\ell_1 \ell_2 \ell_3}$.
For the binned estimator the template is first binned as
$
B^\mathrm{th,exp}_{i_1 i_2 i_3} = 
(\sum_{\ell_1\in\Delta_1} \sum_{\ell_2\in\Delta_2} \sum_{\ell_3\in\Delta_3}
B^\mathrm{th,exp}_{\ell_1 \ell_2 \ell_3})/\Xi_{i_1 i_2 i_3}
$
and then the above estimator can be used with the binned version of the 
inner product:
\be
\langle B^A, B^B \rangle^\mathrm{binned} =
\sum_{i_1 \leq i_2 \leq i_3}
\frac{B_{i_1 i_2 i_3}^A B_{i_1 i_2 i_3}^B}{V_{i_1 i_2 i_3}}.
\label{innerprod}
\ee

One sees that the above estimator is of the form 
$
{\hat f}_\mathrm{NL} \propto \sum [(B^\mathrm{th})^2/V] 
[B^\mathrm{obs}/B^\mathrm{th}]
$
(where from now on we drop the explicit ``exp'' label).
Since $V_{i_1 i_2 i_3}$ is the theoretical estimate of the variance of 
$B_{i_1 i_2 i_3}^\mathrm{obs}$ in the approximation of weak non-Gaussianity,
the estimator is inverse variance weighted: $B^\mathrm{obs}/B^\mathrm{th}$
is an estimate of $f_\mathrm{NL}$ based on a single bin triplet, and all these 
estimates are combined, weighted by the inverse of their variance, 
$V/(B^\mathrm{th})^2$.
The proportionality factor 
$1/\langle B^\mathrm{th},B^\mathrm{th}\rangle$ is the
normalization of the weights and gives the theoretical (Gaussian) estimate 
for the variance\footnote{If we have independent quantities
$y_i$ with variances $v_i$ and define the inverse-variance weights as
$w_i = (1/v_i)/(\sum_j 1/v_j)$, then the variance of the weighted mean
$\sum_i w_i y_i$ is $\sum_i w_i^2 v_i = (\sum_i v_i/v_i^2)/(\sum_j 1/v_j)^2
= 1/(\sum_j 1/v_j)$.}
of the total estimator $\hat{f}_\mathrm{NL}$. This is the same 
as saying that $\langle B^\mathrm{th},B^\mathrm{th}\rangle$ is the 
$\chi^2$ or $(S/N)^2$ of the estimator in the case $f_\mathrm{NL}=1$.

The generalization of the $f_\mathrm{NL}$ estimator to include polarization
in the case without binning was worked out in \cite{Yadav:2007rk}.
In that case the inner product (\ref{innerprodexact})
should be replaced by
\be
\langle B^A, B^B \rangle^\mathrm{no\ binning} =
\sum_{\ell_1 \leq \ell_2 \leq \ell_3}
\!\sum_{\tiny\begin{array}{c@{}c@{}c} p_1, & p_2, & p_3, \\ p_4, & p_5, & p_6\\ 
\end{array}} \!\!\!\!\!\!
B_{\ell_1 \ell_2 \ell_3}^{p_1 p_2 p_3, A}
(V^{-1})_{\ell_1 \ell_2 \ell_3}^{p_1 p_2 p_3 p_4 p_5 p_6}
 B_{\ell_1 \ell_2 \ell_3}^{p_4 p_5 p_6, B},
\label{innerprodexact_polar}
\ee
which involves the inverse of the covariance matrix given in 
(\ref{covarexact_polar}).
Computing this inverse simply implies inverting the three 
$2 \times 2$ matrices $\tilde{C}_\ell$ given in (\ref{Cmatrix_polar}).

Deriving an equivalent expression for the binned estimator is 
straightforward, as long as one keeps in mind that one should first bin
the elements of the covariance matrix (since that corresponds to the
covariance matrix of the binned bispectrum) and only afterwards compute 
the inverse. Trying to bin directly the elements of the inverse covariance
matrix (or one divided by these elements) is incorrect and leads to
wrong results (in particular for bins where $C_\ell^{TE}$ crosses zero).
So in the end the generalization of the binned bispectrum
estimator to include polarization is given by the prescription that the
inner product (\ref{innerprod}) should be replaced by
\be
\langle B^A, B^B \rangle^\mathrm{binned} =
\sum_{i_1 \leq i_2 \leq i_3}
\!
\sum_{\tiny\begin{array}{c@{}c@{}c} p_1, & p_2, & p_3, \\ p_4, & p_5, & p_6\\ 
\end{array}} \!\!\!\!\!\!
B_{i_1 i_2 i_3}^{p_1 p_2 p_3, A} 
(V^{-1})_{i_1 i_2 i_3}^{p_1 p_2 p_3 p_4 p_5 p_6}
B_{i_1 i_2 i_3}^{p_4 p_5 p_6, B},
\label{innerprod_polar}
\ee
involving the inverse of the binned covariance matrix given in 
(\ref{covarbinned_polar}).
However, since the multiplication with $N_\triangle^{\ell_1 \ell_2 \ell_3}$ in
combination with the binning couples the three $\tilde{C}_\ell$ matrices 
in (\ref{covarbinned_polar}) together,
the covariance matrix can only be inverted as a full $8\times 8$ matrix
that is no longer separable in $\ell$. Fortunately this non-separability
is irrelevant for the binned bispectrum estimator.

We can quantify how much the estimator variance increases due to binning, 
compared with an ideal estimator without binning:
\begin{equation}
R \equiv \frac{\Var ( \hat{f}_\mathrm{NL}^\mathrm{ideal})}
{\Var ( \hat{f}_\mathrm{NL}^\mathrm{binned})}
=
\frac{\langle B^\mathrm{th}, B^\mathrm{th} \rangle^\mathrm{binned}}
{\langle B^\mathrm{th}, B^\mathrm{th} \rangle^\mathrm{no\ binning}} .
\label{binning_error}
\end{equation}
$R$ is a number between 0 and 1. The closer $R$ is to 1, the better the binned
approximation for the template under consideration.
To show that $0 \leq R \leq 1$ we need to rewrite (\ref{binning_error}) in terms
of a single inner product definition. It can be checked straightforwardly that 
the binned inner product of the theoretical bispectrum can be rewritten as the 
exact inner product (no binning) of the bispectrum template defined below:
\begin{equation}
\langle B^\mathrm{th}, B^\mathrm{th} \rangle^\mathrm{binned}
= \langle B^\mathrm{bin}, B^\mathrm{bin} \rangle^\mathrm{no\ binning},
\end{equation}
where
\begin{equation}
B_{\ell_1 \ell_2 \ell_3}^{p_1 p_2 p_3, \mathrm{bin}} \equiv
\frac{1}{\Xi_{i_1 i_2 i_3}} \frac{g_{i_1 i_2 i_3}}{g_{\ell_1 \ell_2 \ell_3}}
\!\!\!\! \sum_{\tiny\begin{array}{c@{}c@{}c} p_4, & p_5, & p_6, \\ p_7, & p_8, & p_9\\ 
\end{array}} \!\!\!\!\!\!
V_{\ell_1 \ell_2 \ell_3}^{p_1 p_2 p_3 p_4 p_5 p_6} (V^{-1})_{i_1 i_2 i_3}^{p_4 p_5 p_6 p_7 p_8 p_9} 
B_{i_1 i_2 i_3}^{p_7 p_8 p_9, \mathrm{th}}
\end{equation}
with $(i_1,i_2,i_3)$ the bin triplet that contains the $\ell$-triplet
$(\ell_1,\ell_2,\ell_3)$.\footnote{This result follows from the identity 
(for any function $u_{\ell_1 \ell_2 \ell_3}$)
\begin{align}
\sum_{\ell_1 \leq \ell_2 \leq \ell_3}
\frac{1}{g_{\ell_1 \ell_2 \ell_3}} \, u_{\ell_1 \ell_2 \ell_3}
& = \frac{1}{6} \sum_{\ell_1,\ell_2,\ell_3} u_{\ell_1 \ell_2 \ell_3}
= \frac{1}{6} \sum_{i_1,i_2,i_3} 
\sum_{\ell_1\in\Delta_1} \sum_{\ell_2\in\Delta_2} \sum_{\ell_3\in\Delta_3} u_{\ell_1 \ell_2 \ell_3}
\nonumber\\ & = \sum_{i_1 \leq i_2 \leq i_3} \frac{1}{g_{i_1 i_2 i_3}}
\sum_{\ell_1\in\Delta_1} \sum_{\ell_2\in\Delta_2} \sum_{\ell_3\in\Delta_3} u_{\ell_1 \ell_2 \ell_3}.
\end{align}}
In addition it is simple to show that
\begin{equation}
\langle B^\mathrm{bin}, B^\mathrm{bin} \rangle^\mathrm{no\ binning}
= \langle B^\mathrm{bin}, B^\mathrm{th} \rangle^\mathrm{no\ binning}.
\end{equation}
Now we can rewrite $R$ as
\begin{equation}
R = \frac{\langle B^\mathrm{bin}, B^\mathrm{bin} \rangle^\mathrm{no\ binning}}
{\langle B^\mathrm{th}, B^\mathrm{th} \rangle^\mathrm{no\ binning}}
= \frac{\langle B^\mathrm{bin}, B^\mathrm{th} \rangle^\mathrm{no\ binning}}
{\langle B^\mathrm{th}, B^\mathrm{th} \rangle^\mathrm{no\ binning}}
= \frac{(\langle B^\mathrm{bin}, B^\mathrm{th} \rangle^\mathrm{no\ binning})^2}
{\langle B^\mathrm{th}, B^\mathrm{th} \rangle^\mathrm{no\ binning}
\langle B^\mathrm{bin}, B^\mathrm{bin} \rangle^\mathrm{no\ binning}}.
\end{equation}
From the first expression, given that $\langle x,x \rangle \geq 0$ for an
inner product, we see that $R \geq 0$. And the last expression implies
that $R \leq 1$ using the Cauchy-Schwarz inequality.

\subsection{Joint estimation}
\label{sec_joint}

If more than one of the above bispectrum shapes are expected to be
present in the data, then a joint estimation of the different $f_\mathrm{NL}$
parameters is required. For this the Fisher matrix 
\begin{equation}
F_{IJ} = \langle B^I, B^J \rangle,
\label{Fisher}
\end{equation}
where $I,J $ label the theoretical shapes (for example local and equilateral),
is a crucial quantity. 
The optimal estimation of the
$f_\mathrm{NL}$ parameter for shape $I$ is given by
\begin{equation}
\hat{f}_\mathrm{NL}^I = \sum_J (F^{-1})_{IJ} \langle B^J, B^\mathrm{obs} \rangle.
\label{fNL_joint}
\end{equation}
The estimate of the variance of $\hat{f}_\mathrm{NL}^I$
is $(F^{-1})_{II}$. If, on the other hand, the $\hat{f}_\mathrm{NL}^I$ parameters
would have been estimated independently using (\ref{fNL_estimator}) (as if
there is only one bispectrum shape present, but it is unknown which), then
their variance is given by $1/F_{II}$.

Another useful quantity to define is the symmetric correlation matrix 
\begin{equation}
C_{IJ} \equiv \frac{F_{IJ}}{\sqrt{F_{II} F_{JJ}}}
\label{corrmatrix}
\end{equation}
giving the correlation coefficients between any two bispectrum shapes.
By construction $-1 \leq C_{IJ} \leq +1$, with $C_{IJ}=-1,0,+1$ meaning that
the two shapes are fully anti-correlated, uncorrelated, or fully correlated, 
respectively.
Note that one could also define a correlation matrix using the inverse of
the Fisher matrix instead of the Fisher matrix itself in (\ref{corrmatrix}).
That would give us the correlation of the {\em $f_\mathrm{NL}$ parameters},
while (\ref{corrmatrix}) represents the correlation of the {\em templates}.
As an example we show the correlation coefficients between the templates of 
Section~\ref{sec_templates} in Table~\ref{tab_corr_coeff}.

\begin{table}
\begin{center}
\begin{tabular}{l|cccccc}
\hline
& Local & Equil & Ortho & LensISW & UnclustPS & CIB \\
\hline
Local & 1 & 0.21 & -0.44 & 0.28 & 0.002 & 0.006\\
Equilateral && 1 & -0.05 & 0.003 & 0.008 & 0.03\\
Orthogonal &&& 1 & -0.15 & -0.003 & -0.001\\
Lensing-ISW &&&& 1 & -0.005 & -0.03\\
Unclustered point sources &&&&& 1 & 0.93\\
CIB point sources &&&&&& 1\\
\hline
\end{tabular}
\end{center}
\caption{Correlation coefficients between the theoretical templates
of Section~\ref{sec_templates}, as defined in (\ref{corrmatrix}). 
The numbers are computed using the characteristics of the Planck experiment
and are for temperature. We see a large
correlation between local and orthogonal and between local and lensing-ISW.
Equilateral and orthogonal are mostly uncorrelated, and the correlation
between the point source templates and the primordial ones is negligible.}
\label{tab_corr_coeff}
\end{table}

Suppose that we had only two shapes with non-zero correlation, but the
amplitude of the second was fixed by theory (as is the case for example for
the lensing-ISW template that has no unknown amplitude parameter).
If the theory was fully trusted, it would be a shame to do a joint estimation,
with the associated increase in variance. In that case the influence of
the second shape on the first is more properly treated as a known bias that
can be subtracted without increasing the variance. The size of the bias can
be found from (\ref{fNL_joint}), by using the second equation ($I=2$) to 
eliminate $\langle B^{(2)}, B^\mathrm{obs} \rangle$ from the first equation 
($I=1$). After expressing
the elements of the inverse Fisher matrix in terms of the elements of the
Fisher matrix, the resulting equation for the first $f_\mathrm{NL}$ parameter
simplifies to:
\begin{equation}
\hat{f}_\mathrm{NL}^{(1)} = \frac{1}{F_{11}} \langle B^{(1)}, B^\mathrm{obs} \rangle
- \frac{F_{12}}{F_{11}} f_\mathrm{NL}^{(2)},
\end{equation}
the second term being the bias correction.
Here $f_\mathrm{NL}^{(2)}$ is the known $f_\mathrm{NL}$ parameter of the second 
shape, most likely equal to one if the known amplitude was included in the 
template (as is the case for example for the lensing-ISW template). The variance
of $\hat{f}_\mathrm{NL}^{(1)}$ is not influenced by the bias correction and remains
equal to $1/F_{11}$, the same as for a single shape. This result
can easily be generalized to more than two shapes.

\section{Masking and filling in}
\label{masking}

The region near the galactic plane and around extragalactic point
sources where reliable subtraction of contaminants is not possible
must be masked to prevent contamination of the primordial
bispectrum. Masking introduces a number of problems for estimating the
bispectrum because the process of filtering maps is nonlocal. If we
naively analyze a masked map in which the masked pixels are set to
zero --- or better yet, set equal to the average value of the unmasked
part of the map --- by filtering it, say with a high pass filter, we
would observe a deficit of small scale power around the edges of the
mask. A filter in frequency space moves around the small scale power
in real space. The power is smeared, so that if there is no small
scale power in the masked region, power from the unmasked region
escapes into the masked region without there being a compensating flux
returning from the masked region. Another edge effect tending to
increase the small scale power around the border of the unmasked
region results if there is a jump discontinuity. Such a discontinuity
contains spurious small scale power that bleeds into the unmasked
region after filtering. It is therefore important to introduce
artificially the right amount of small scale power into the masked
region and to avoid spurious jumps in the maps so that the two fluxes
cancel after filtering.\footnote{Large scale modes are much less
affected by the mask. Since these mode extend out over large parts of
the sky, they can be reconstructed reasonably accurately even when
some parts of the sky are missing. Furthermore edge effects are also
less important for a mask with larger holes. Consequently for a high
resolution experiment like Planck the use of inpainting algorithms has turned
out to be absolutely crucial, while for the lower resolution WMAP
experiment, which moreover had larger error bars, less care was
required.} This process of filling in the masked regions is also known
as `inpainting'.

Before showing quantitatively how masking affects the determination of
$f_\mathrm{NL}$, we first have to determine what the effect on the
error bars would be if we had none of these problems, but only less
data due to the reduced fraction of the sky. When the bispectrum is
determined according to (\ref{Bobsbinned}), it should be multiplied
with a factor $1/f_\mathrm{sky}$ to correct for the partial sky
coverage~\cite{Komatsu:2003iq}, where $f_\mathrm{sky}$ is the
fraction of the sky that is left unmasked. In practice this is done
automatically when the integral is replaced by a sum over the pixels:
the product of maps is summed over all unmasked pixels, divided by the
number of unmasked pixels, and multiplied by $4\pi$. In addition, the
partial sky coverage increases the variance of the estimator, the
theoretical estimate of which becomes $1/(f_\mathrm{sky} \langle
B^\mathrm{th,exp},B^\mathrm{th,exp}\rangle)$. If the mask is not too
large, this simple prescription for the variance works quite well.

\begin{table}
\begin{center}
\begin{tabular}{lcccccc}
\hline
& \multicolumn{3}{c}{No linear correction} &
\multicolumn{3}{c}{With linear correction} \\
& Local & Equil & Ortho & Local & Equil & Ortho \\
\hline
\multicolumn{4}{l}{No mask, isotropic noise} &&& \\
\multicolumn{1}{@{\hspace{1.3cm}}c@{\hspace{0.8cm}}}{$TTT$} & -0.1 $\pm$ 4.1 & 2 $\pm$ 58 & 5 $\pm$ 24
& -0.1 $\pm$ 4.1 & 3 $\pm$ 57 & 4 $\pm$ 25 \\
\multicolumn{1}{@{\hspace{0.5cm}}c@{}}{$EEE$} & 0.4 $\pm$ 24 & -11 $\pm$ 170 & 6 $\pm$ 92
& 0.4 $\pm$ 24 & -11 $\pm$ 171 & 7 $\pm$ 94 \\
\multicolumn{4}{l}{No mask, anisotropic noise} &&& \\
\multicolumn{1}{@{\hspace{0.5cm}}c@{}}{$TTT$} & 5.7 $\pm$ 84 & 2 $\pm$ 58 & 2 $\pm$ 35
& -0.2 $\pm$ 4.2 & 3 $\pm$ 57 & 4 $\pm$ 24 \\
\multicolumn{1}{@{\hspace{0.5cm}}c@{}}{$EEE$} & -23 $\pm$ 736 & -22 $\pm$ 193 & 15 $\pm$ 197
& 0.4 $\pm$ 24 & -20 $\pm$ 195 & 7 $\pm$ 94 \\
\multicolumn{4}{l}{Galactic mask, isotropic noise} &&& \\
\multicolumn{4}{l}{-- No filling in} &&& \\
\multicolumn{1}{@{\hspace{0.5cm}}c@{}}{$TTT$} & -0.2 $\pm$ 5.5 & 11 $\pm$ 78 & -1 $\pm$ 58
& 0.3 $\pm$ 5.1 & 6 $\pm$ 70 & 6 $\pm$ 32 \\
\multicolumn{1}{@{\hspace{0.5cm}}c@{}}{$EEE$} & 5 $\pm$ 32 & -5 $\pm$ 199 & 1 $\pm$ 108
& 2 $\pm$ 28 & -9 $\pm$ 202 & 3 $\pm$ 109 \\
\multicolumn{4}{l}{-- Diffusive filling in}&&& \\
\multicolumn{1}{@{\hspace{0.5cm}}c@{}}{$TTT$} & 0.8 $\pm$ 6.2 & 6 $\pm$ 70 & 4 $\pm$ 28
& 0.3 $\pm$ 4.6 & 7 $\pm$ 69 & 4 $\pm$ 29 \\
\multicolumn{1}{@{\hspace{0.5cm}}c@{}}{$EEE$} & 5 $\pm$ 31 & -8 $\pm$ 196 & 1 $\pm$ 109
& 2 $\pm$ 28 & -8 $\pm$ 198 & 2 $\pm$ 110 \\
\multicolumn{4}{l}{Point source mask, isotropic noise} &&& \\
\multicolumn{4}{l}{-- No filling in} &&& \\
\multicolumn{1}{@{\hspace{0.5cm}}c@{}}{$TTT$} & -0.7 $\pm$ 9.2 & 3 $\pm$ 73 & 6 $\pm$ 51
& -0.4 $\pm$ 8.4 & 3 $\pm$ 65 & 7 $\pm$ 36 \\
\multicolumn{1}{@{\hspace{0.5cm}}c@{}}{$EEE$} & 1 $\pm$ 27 & -7 $\pm$ 170 & 10 $\pm$ 92
& 0.1 $\pm$ 23 & -7 $\pm$ 170 & 9 $\pm$ 89 \\
\multicolumn{4}{l}{-- Diffusive filling in} &&& \\
\multicolumn{1}{@{\hspace{0.5cm}}c@{}}{$TTT$} & 0.2 $\pm$ 6.3 & 2 $\pm$ 59 & 5 $\pm$ 25
& -0.3 $\pm$ 4.3 & 3 $\pm$ 58 & 4 $\pm$ 24 \\
\multicolumn{1}{@{\hspace{0.5cm}}c@{}}{$EEE$} & -0.1 $\pm$ 26 & -0.1 $\pm$ 172 & 13 $\pm$ 98
& -0.5 $\pm$ 24 & -3 $\pm$ 173 & 12 $\pm$ 97 \\
\multicolumn{4}{l}{Gal + ps mask, anisotropic noise} &&& \\
\multicolumn{4}{l}{-- No filling in} &&& \\
\multicolumn{1}{@{\hspace{0.5cm}}c@{}}{$TTT$} & 0.3 $\pm$ 77 & 10 $\pm$ 93 & 3 $\pm$ 87
& -0.7 $\pm$ 9.4 & 5 $\pm$ 76 & 10 $\pm$ 39 \\
\multicolumn{1}{@{\hspace{0.5cm}}c@{}}{$EEE$} & -27 $\pm$ 719 & -11 $\pm$ 214 & 17 $\pm$ 247
& 2 $\pm$ 30 & -14 $\pm$ 207 & 4 $\pm$ 101 \\
\multicolumn{4}{l}{-- Diff. filling in of ps mask only} &&& \\
\multicolumn{1}{@{\hspace{0.5cm}}c@{}}{$TTT$} & 1.6 $\pm$ 85 & 10 $\pm$ 78 & -2 $\pm$ 70
& 0.02 $\pm$ 5.4 & 5 $\pm$ 71 & 7 $\pm$ 32 \\
\multicolumn{1}{@{\hspace{0.5cm}}c@{}}{$EEE$} & -27 $\pm$ 752 & -5 $\pm$ 213 & 16 $\pm$ 243
& 2 $\pm$ 31 & -13 $\pm$ 210 & 2 $\pm$ 109 \\
\multicolumn{4}{l}{-- Diff. filling in of both masks} &&& \\
\multicolumn{1}{@{\hspace{0.5cm}}c@{}}{$TTT$} & 2.7 $\pm$ 87 & 6 $\pm$ 72 & 3 $\pm$ 44
& -0.04 $\pm$ 5.0 & 6 $\pm$ 69 & 4 $\pm$ 29 \\
\multicolumn{1}{@{\hspace{0.5cm}}c@{}}{$EEE$} & -26 $\pm$ 756 & -9 $\pm$ 210 & 16 $\pm$ 242
& 2 $\pm$ 31 & -13 $\pm$ 208 & 1 $\pm$ 110 \\
\hline
\end{tabular}
\end{center}
\caption{Importance of filling in and of the linear
correction term for determining $f_\mathrm{NL}$ in the presence of
a mask and anisotropic noise. The results are based on a set of 100 Gaussian
CMB simulations at Healpix resolution $n_\mathrm{side}=2048$ with power 
spectrum according to the Planck 2013 release values. The simulations
include smoothing by a 5 arcmin FWHM Gaussian beam and noise
based on a white noise power spectrum with amplitude $1.5\times 10^{-17}$
for temperature and $6\times 10^{-17}$ for E polarization (in units made 
dimensionless by dividing by the CMB mean temperature $T_0=2.7255$~K).
Where relevant the noise has been made anisotropic by modulating it using
the hit count map of the Planck 143~GHz channel of the 2013 release.
The maps are analyzed with the binned bispectrum estimator using 54 bins 
and $\ell_\mathrm{max}=2500$, and 100 maps were used for the linear correction 
term. The masks used are shown in Fig.~\ref{fig_masks}.}
\label{table_filling_in}
\end{table}

\begin{figure}
\includegraphics[width=0.5\columnwidth]{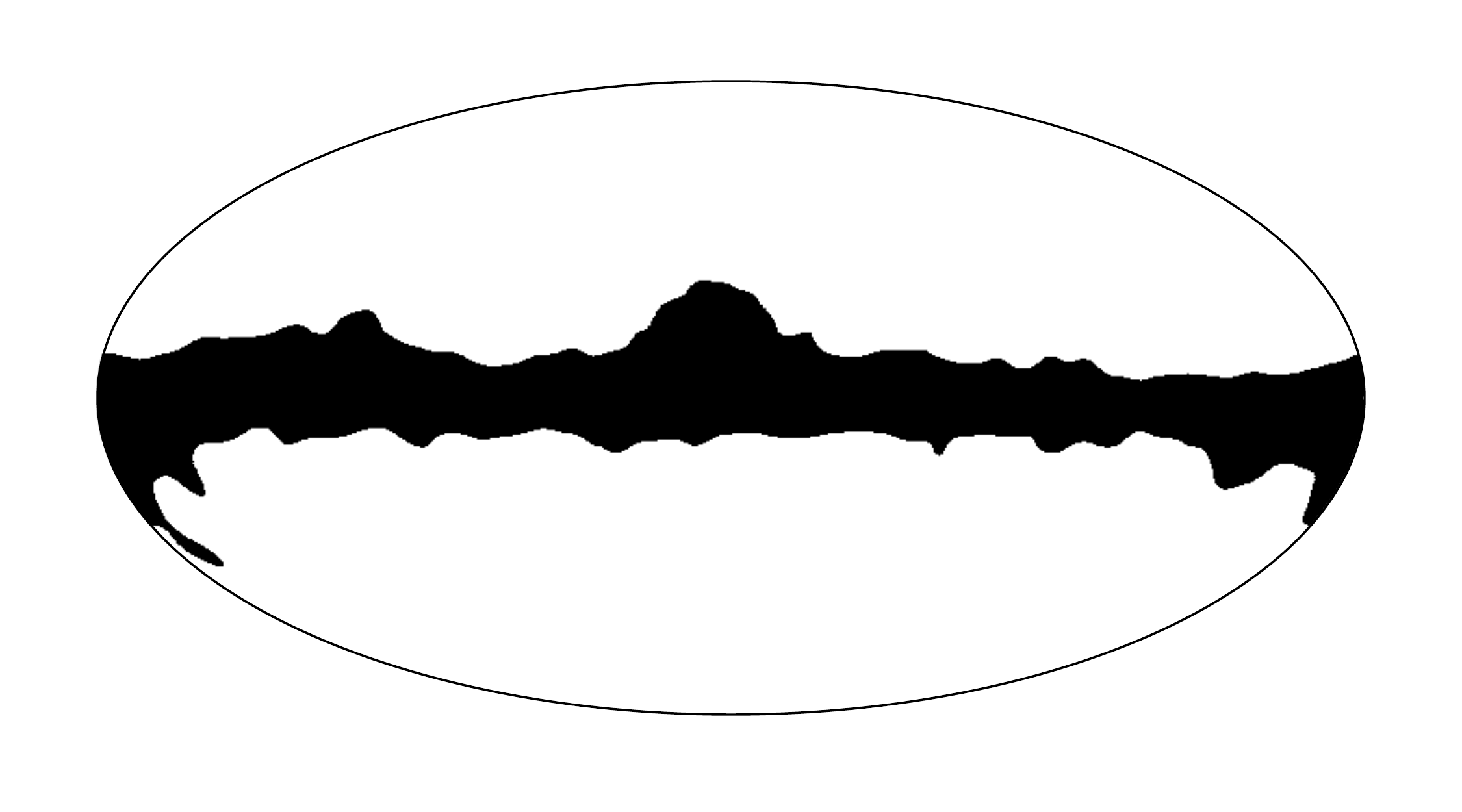}
\includegraphics[width=0.5\columnwidth]{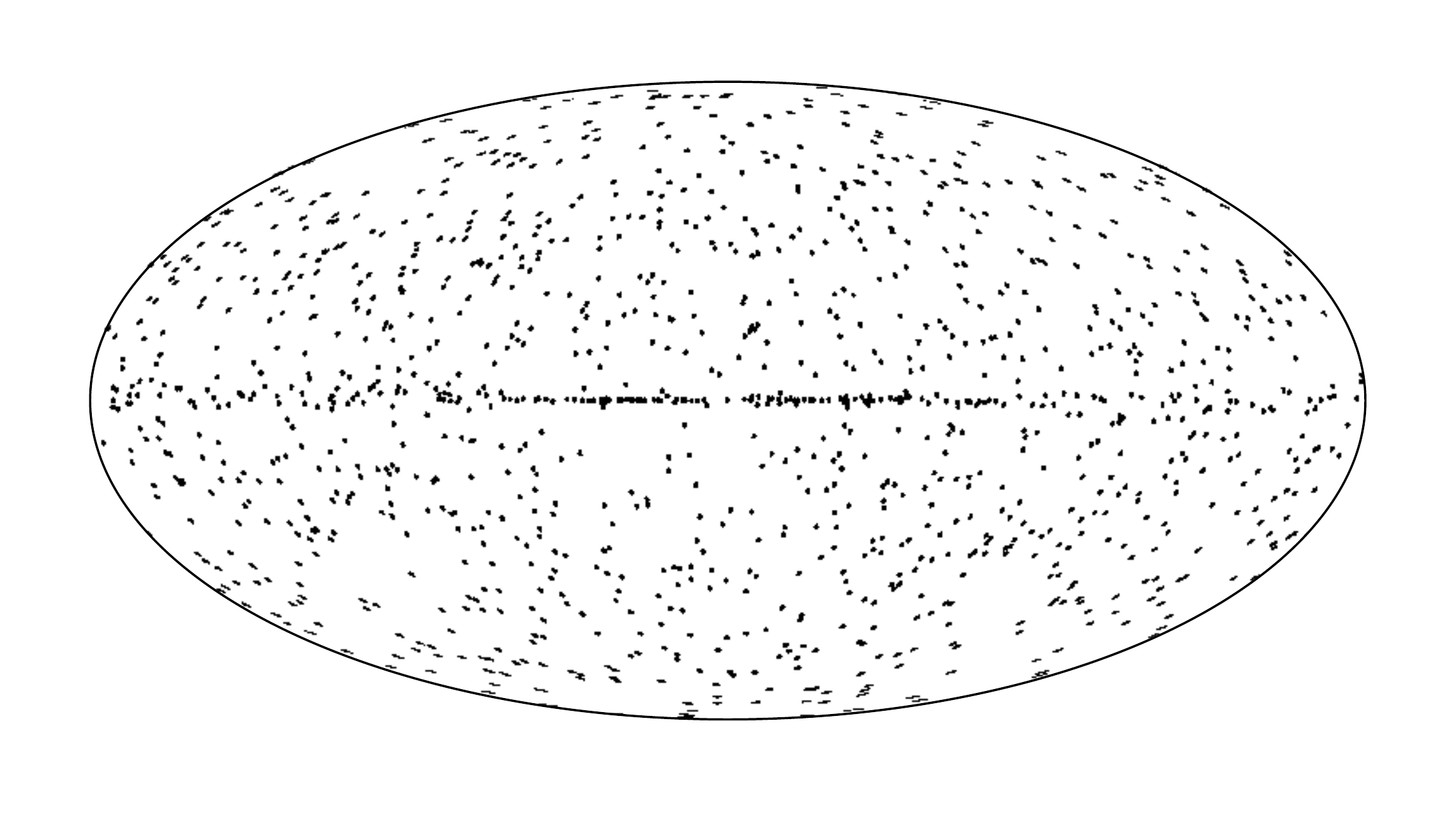}
\caption{The galactic (left) and point source mask (right) used in 
Table~\ref{table_filling_in}. The galactic mask is the $f_\mathrm{sky}=0.80$ 
galactic mask included in the Planck 2013 release, and the point source mask 
is the one based on the Planck LFI 30~GHz channel from the 2013 release 
(with a $4\sigma$ threshold level), which has $f_\mathrm{sky}=0.96$. 
The combined mask has $f_\mathrm{sky}=0.77$.}
\label{fig_masks}
\end{figure}

To illustrate quantitatively the problems encountered in determining
$f_\mathrm{NL}$ with a mask, we applied a series of tests to 
simulated CMB maps as described in Table~\ref{table_filling_in}. 
The masks used are shown in Fig.~\ref{fig_masks} while 
the details of the simulations are described in the caption of the Table. 
We find that when missing data in the masked regions are naively replaced
with the average
of the unmasked part of the map (the ``no filling in'' lines in the Table), 
the estimates of $f_\mathrm{NL}$ are unbiased but have much larger variance 
than expected, at least in temperature. The expected increase
in the standard deviation is only a factor $1/\sqrt{f_\mathrm{sky}}$ 
(i.e., $1.12$ for the galactic mask and $1.02$ for the point sources)
and in particular for the point source mask we observe
wider error bars in temperature for all three shapes.
Including the linear correction term (\ref{Bisp_lincorr}) in 
the estimator reduces this effect to some extent, but in temperature
this is clearly not enough.
The effect of the point source mask on the
local shape is exacerbated 
when the holes are smaller. For example, replacing the 
2013 Planck LFI 30~GHz point source mask with the 2013 Planck 
HFI 100~GHz 
channel mask (with a $5\sigma$ threshold level), which has a much smaller beam
and hence smaller holes ($f_\mathrm{sky}=0.99$), increases the 
``no filling in, no linear correction'' error bars for the local shape from 9.2 
to 29.5 (while the error bars for equilateral and orthogonal become smaller).
These results demonstrate the need for a suitable  filling in of the 
missing data in the masked regions of the temperature map, in particular 
for the point source mask.

\begin{figure}
\includegraphics[width=0.5\columnwidth]{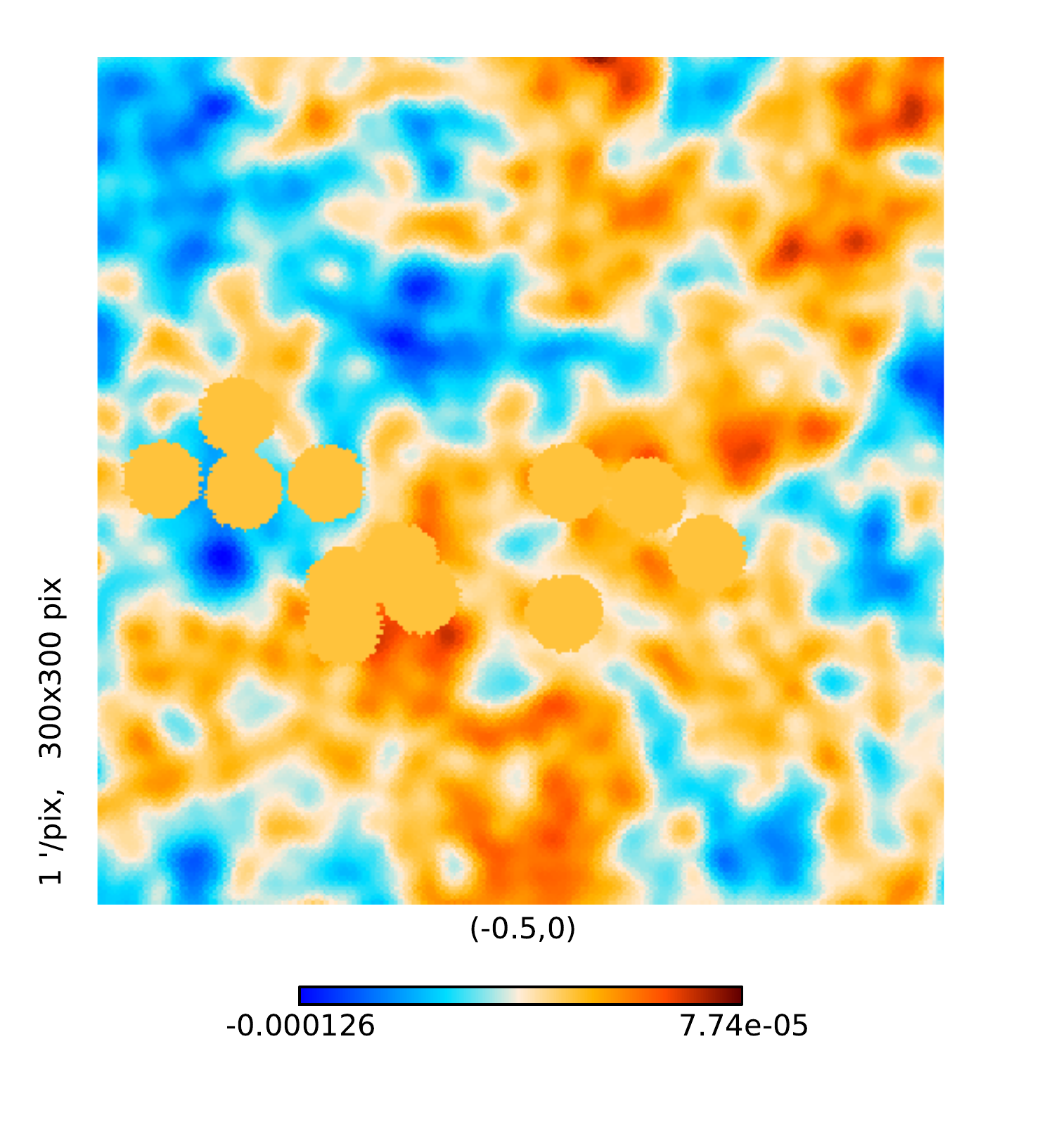}
\includegraphics[width=0.5\columnwidth]{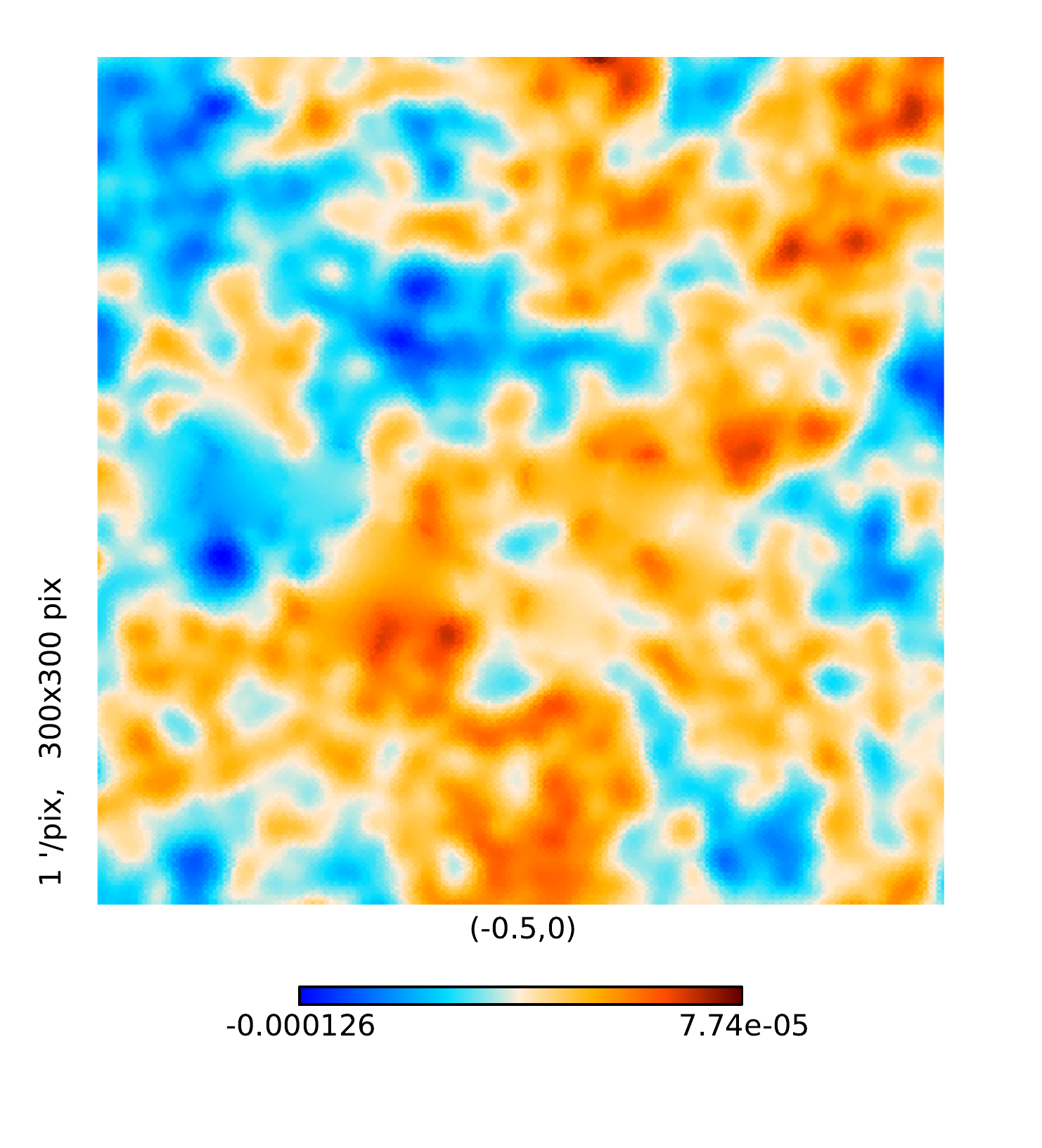}
\includegraphics[width=0.5\columnwidth]{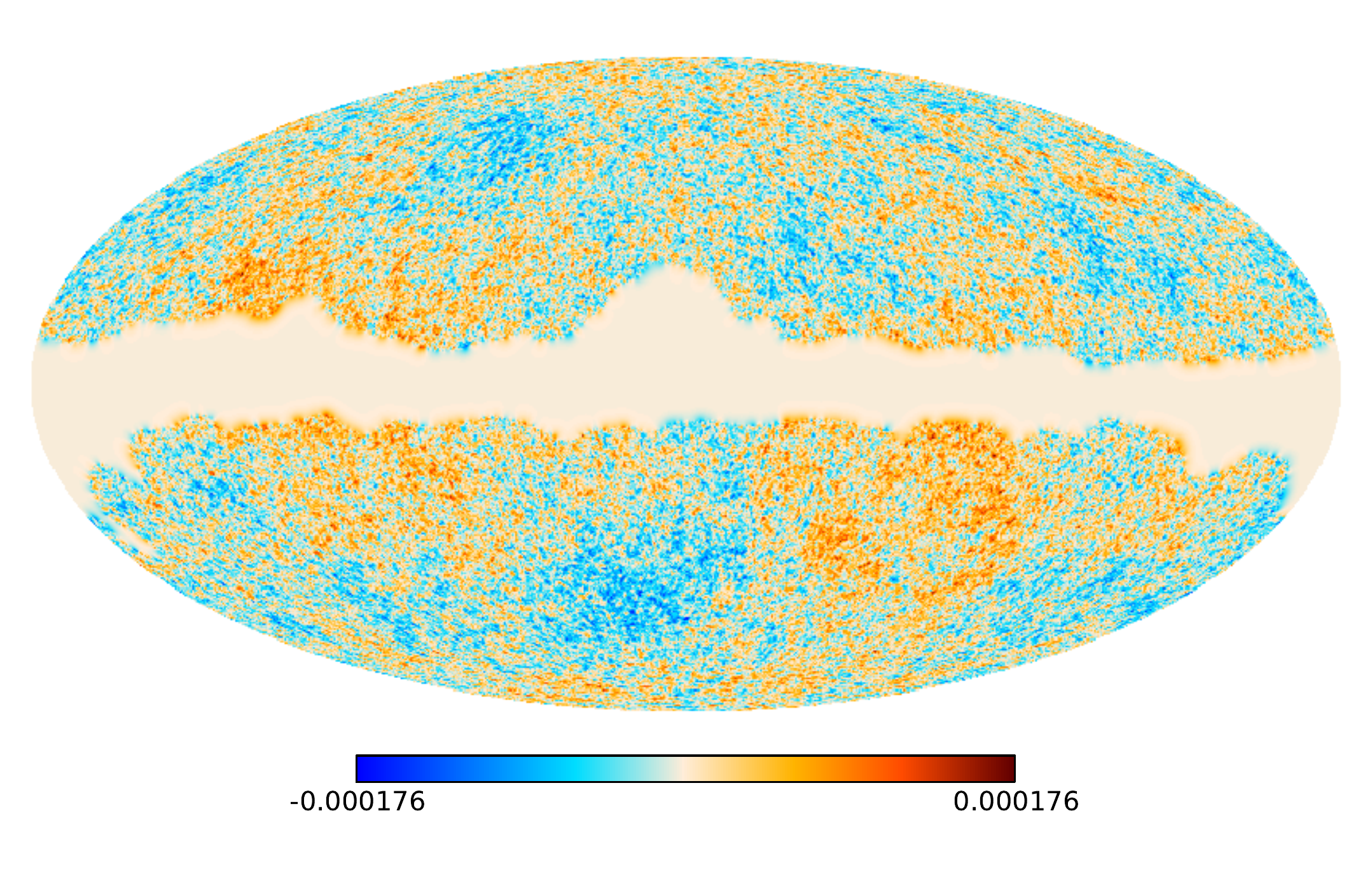}
\includegraphics[width=0.5\columnwidth]{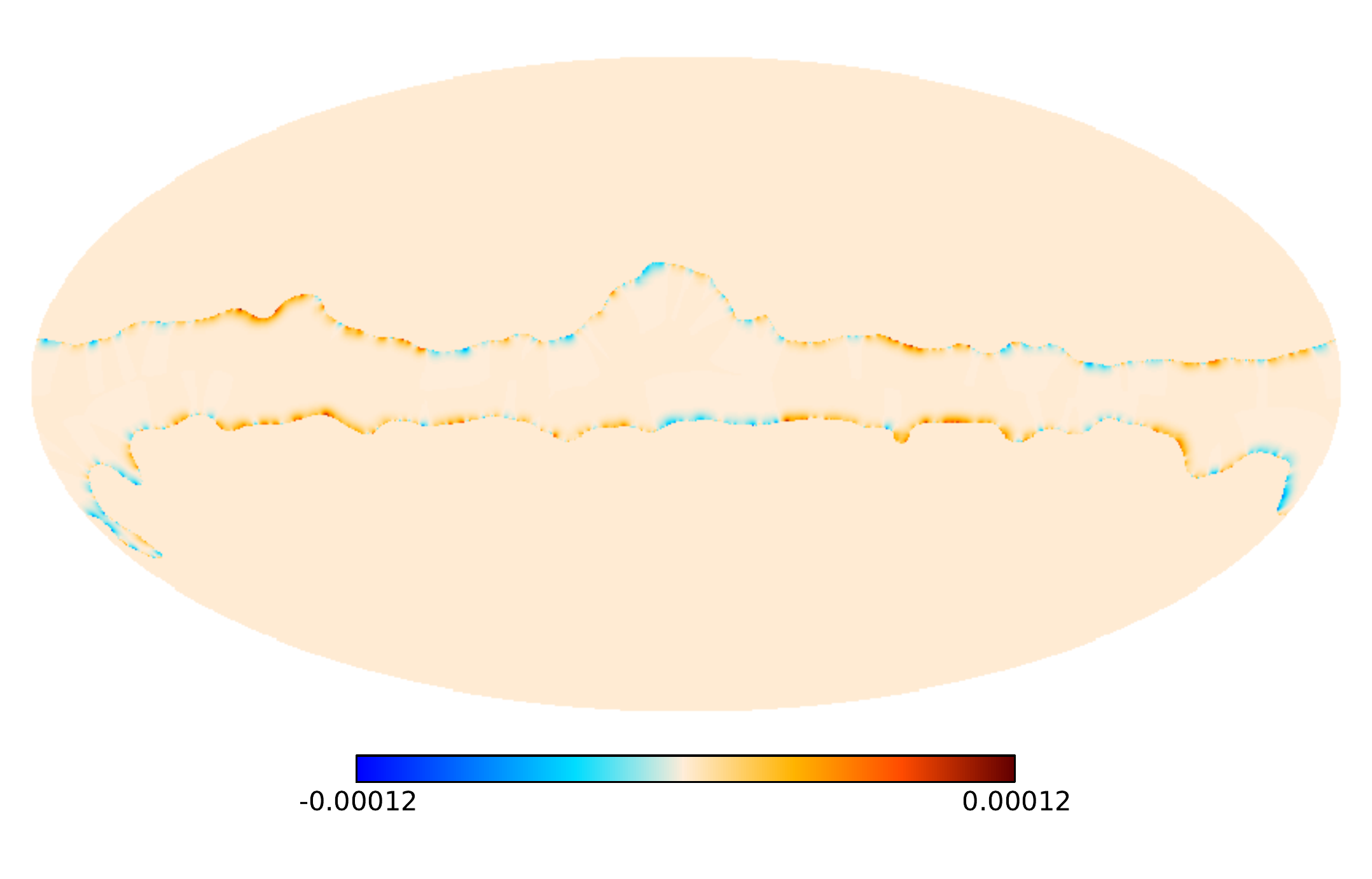}
\caption{The top two panels show a zoom of a Gaussian CMB map with holes 
from the 2013 Planck HFI 100~GHz point source mask, before (left) and after 
diffusive filling in (right) with 2000 iterations. The bottom left panel 
shows what filling in the galactic mask looks like, with the bottom right 
panel showing the difference between the maps with and without filling in.
The units are dimensionless ($\Delta T/T_0$).}
\label{fig_inpainted_maps}
\end{figure}

\begin{figure}
\includegraphics[width=0.5\columnwidth]{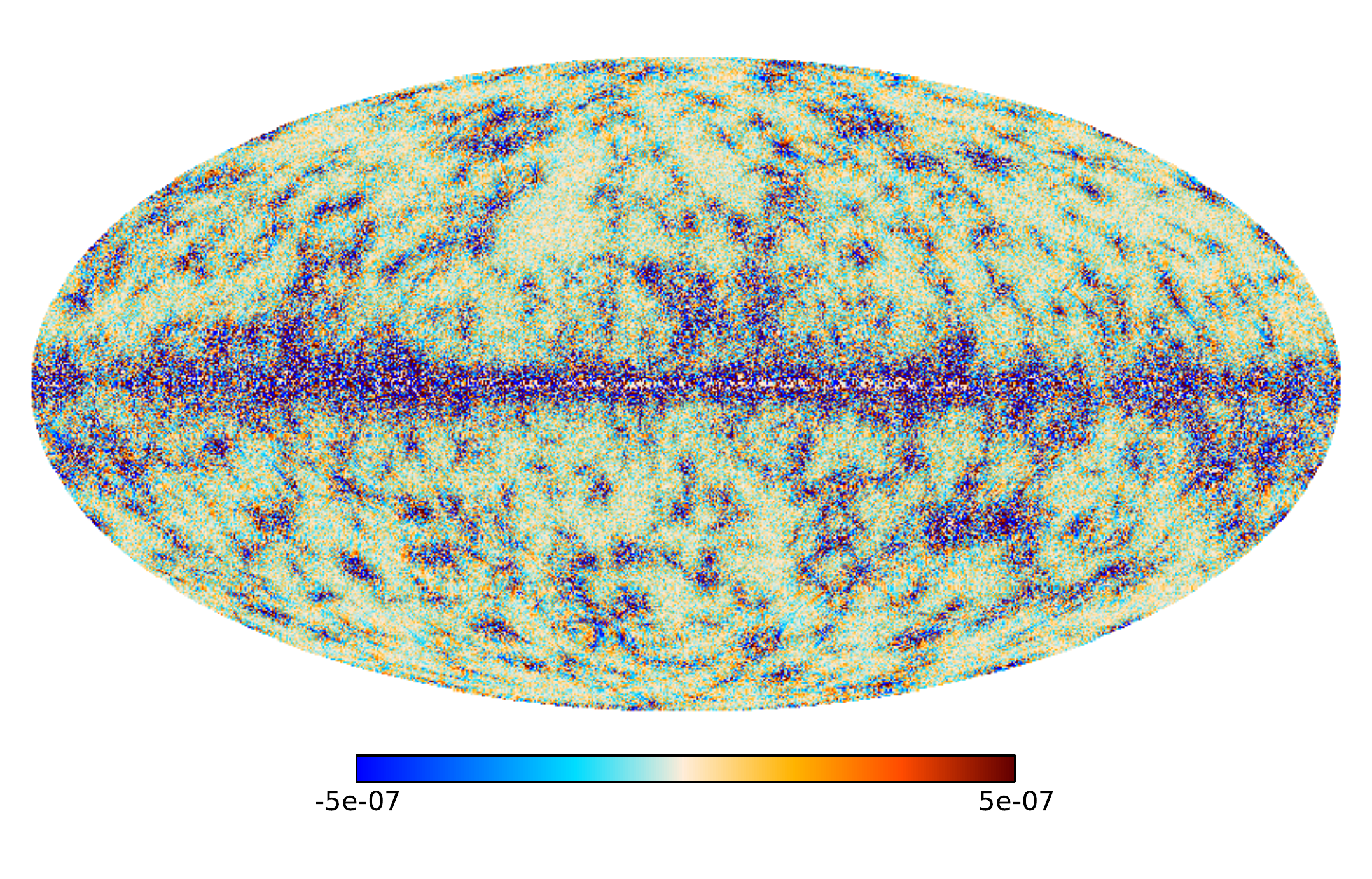}
\includegraphics[width=0.5\columnwidth]{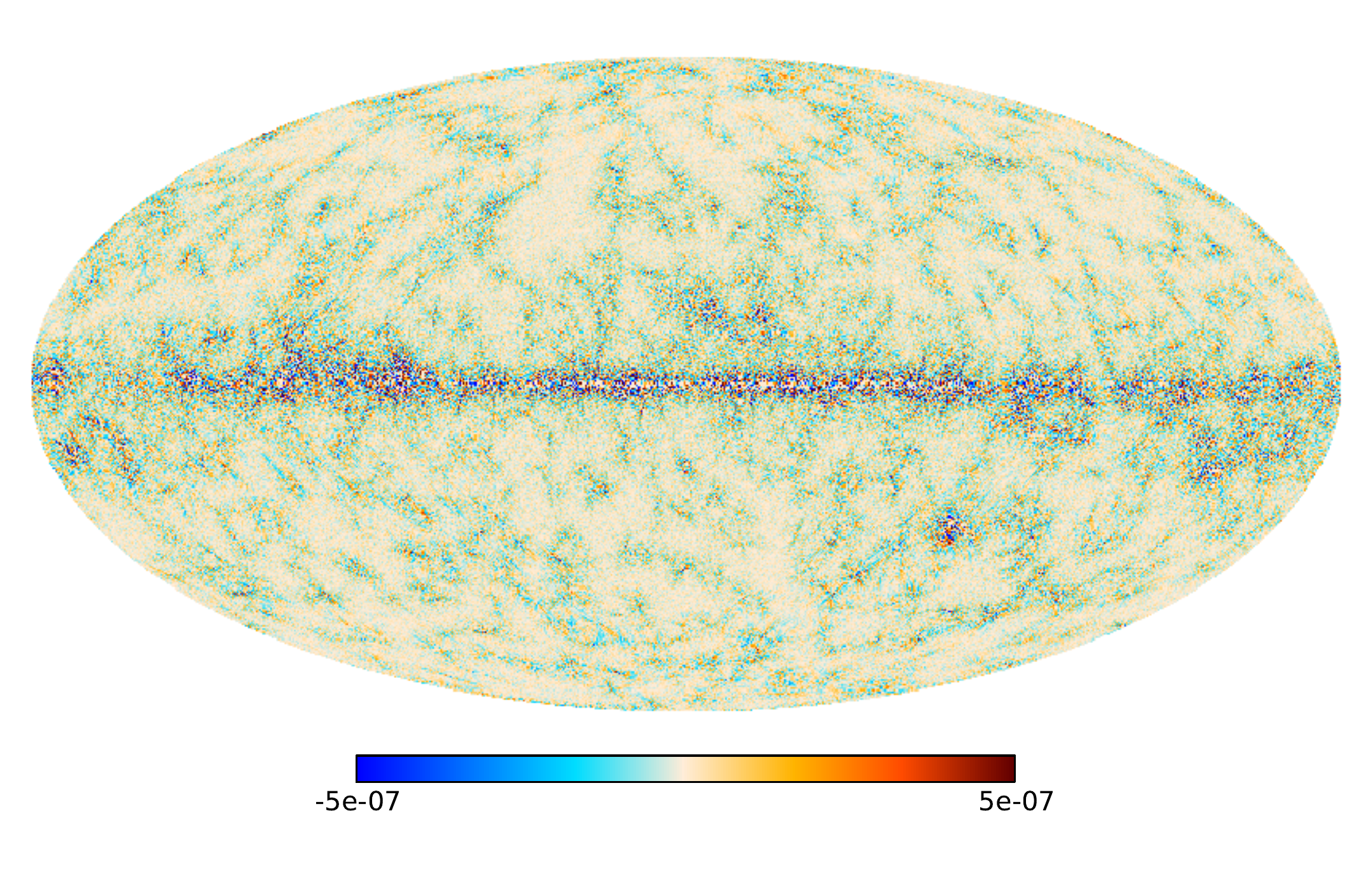}
\caption{Here we take one of the Gaussian CMB maps and filter out all
modes except $\ell\in [1500,1549]$ in three different ways. First we take
the unmasked map, filter it, and mask the result (1). Secondly, we take the
map, mask it with the 2013 Planck HFI 100~GHz point source mask, then 
filter it and mask again (2). Finally, the third map (3) is the same as the 
second one, except that we have performed diffusive filling in on the masked
map before filtering. On the left is shown the difference $(2)-(1)$ (no
filling in), while on the right we have $(3)-(1)$ (with filling in).
The units are dimensionless ($\Delta T/T_0$).}
\label{fig_inpainting}
\end{figure}

The simplest filling in method is diffusive filling in, which despite
its simplicity worked extremely well and was subsequently adopted for the 
other Planck bispectrum estimators (KSW and modal) as well. It became
the common method in both the 2013 and 2015 Planck releases.
After filling the masked regions with the average of the unmasked part of the
map as above, we fill each masked pixel with the average value of its
neighboring pixels and this procedure is iterated.
We found that 2000 iterations
sufficed for the Planck maps. One can implement the iterative procedure in
two different ways: compute the average of the neighbors on the current
iteration (Gauss-Seidel method, where some of the neighbor pixels will already
have been updated and others not) or on the previous iteration kept in a buffer
(Jacobi method, where all neighbor pixels will be on the previous iteration).
While the Gauss-Seidel implementation is anisotropic, we found 
that this has no impact on the results, while on the contrary the faster 
convergence of that implementation is an advantage.
This scheme solves a discretized version of
Laplace's equation for the pixels where there is no data
with the boundary of the unmasked region providing Dirichlet boundary
data. [See~\cite{Bucher:2011nf} for a discussion of how this scheme
is related to constrained random Gaussian realizations for filling in
the missing data.] While this sort of `harmonic averaging' is simple
to implement and dulls the sharp edges, it appears at first glance not to
remedy the problem of missing small-scale power described above, as the 
resulting maps have clearly visible bald spots, see 
Fig.~\ref{fig_inpainted_maps}. However, unlike
apodization which only dulls the edges, the diffusive filling in scheme 
does create small scale structure inside close to the boundary of the mask. 
Given that during harmonic transforms it is the wavelength of the modes
that determines how far they propagate, this is exactly what we need:
the short wavelengths can only propagate small distances and hence need
only be reconstructed close to the edges. Fig.~\ref{fig_inpainting}
illustrates both how the contamination due to the presence of the point
source mask is localized around the holes and how filling in remedies this 
problem. We use the Planck HFI 100~GHz point source mask (with the smaller 
holes) for this Figure, since the effect is larger and more apparent there.

After masking, filling in, and filtering the maps, we mask them once
again before integrating over products of maps.  The masked region is
never directly used in the calculation of the bispectrum, but the
filling in is crucial to avoid the influence of the masked region
spreading out over the sky when filtering the maps, as explained
above. In addition the average of the filtered maps outside the mask 
is subtracted to remove any monopole. If this is not done small-scale power
(whose origin is from the two-point function) will combine with this
monopole to masquerade as (local) bispectral power, and this `aliasing' 
can be a large effect.

Other more sophisticated filling in techniques include nonlinear methods based
on sparsity (see~\cite{Abrial2007, Abrial:2008mz, Perotto:2009tv}) or
constrained Gaussian realizations~\cite{Bucher:2011nf}. Alternatively,
and even better for bispectrum determination, one can perform a full
inverse covariance weighting (Wiener filtering) of the maps 
(see e.g.,~\cite{Smith:2009jr,Elsner:2012fe}).
However as the results in Table~\ref{table_filling_in} show, these methods
do not appear necessary, as a combination of diffusive filling in and
the linear correction term leads to results that are effectively optimal for 
the temperature maps (meaning they cannot be distinguished from the optimal
results within the error bars). For $E$ polarization the situation is even 
simpler, at least at the Planck resolution and sensitivity. Not even diffusive 
filling in is required. Just applying the linear correction term appears 
sufficient. However as a precaution we also applied diffusive filling 
in to the $Q$ and $U$ maps for the Planck analysis.

\begin{figure}
\includegraphics[width=0.5\columnwidth]{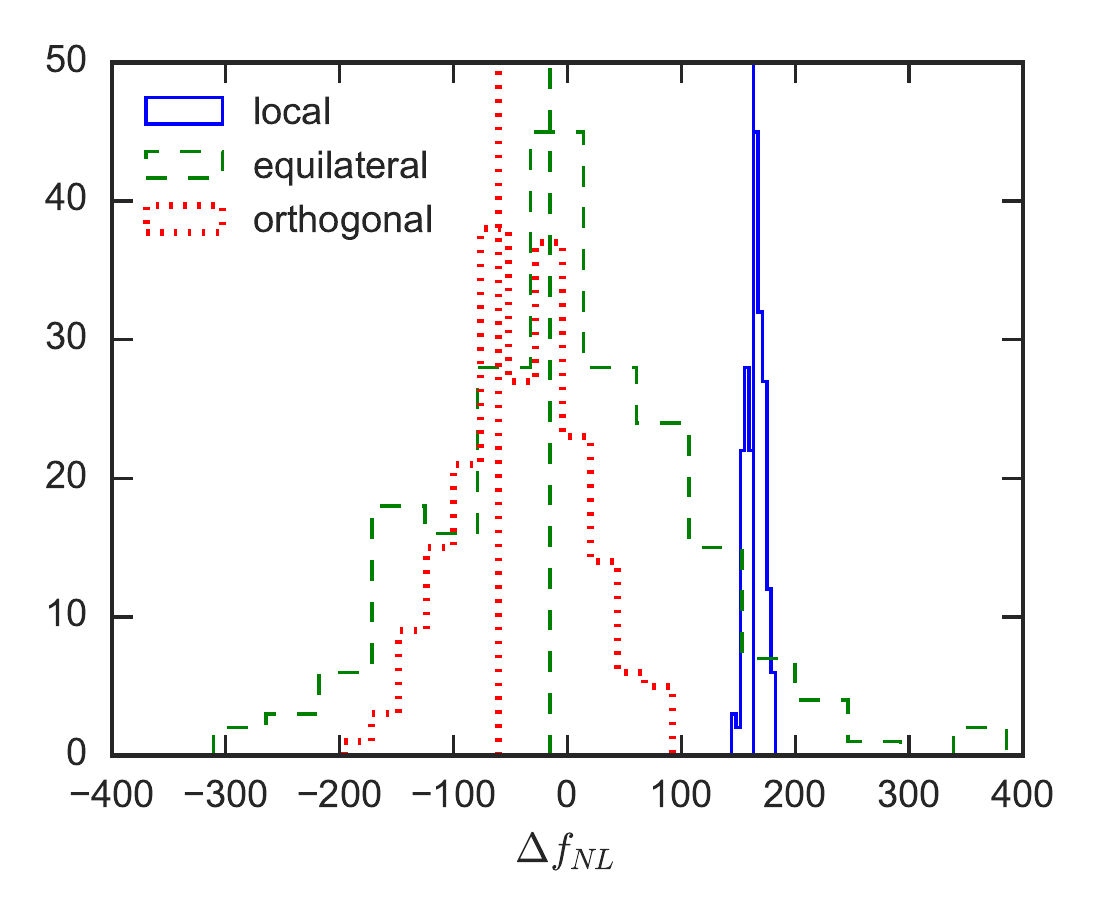}
\includegraphics[width=0.5\columnwidth]{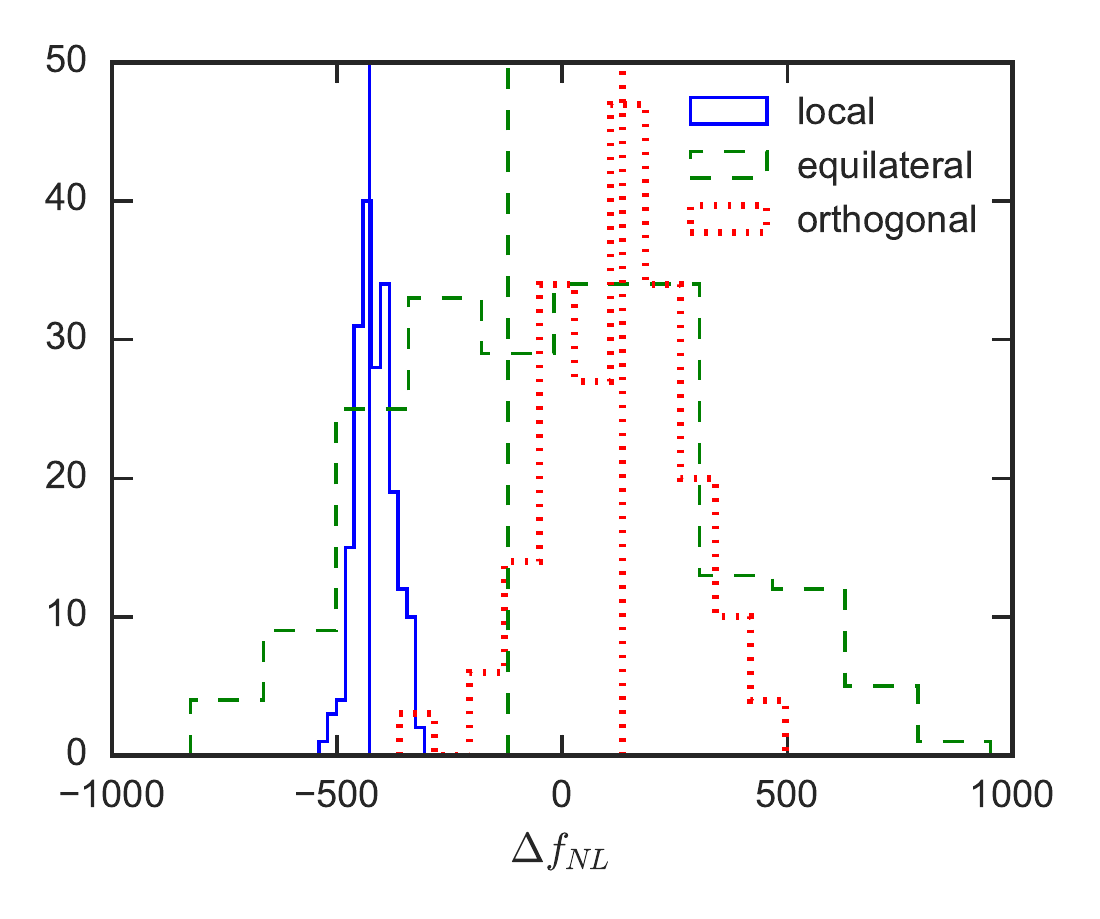}
\caption{Histogram of the contribution of 199 individual Gaussian maps
to the linear correction part of $f_\mathrm{NL}$ for one of the maps from the 
``no mask, anisotropic noise'' case of Table~\ref{table_filling_in},
for both temperature (left) and $E$ polarization (right).
Results are shown for the local (blue), equilateral (green), and orthogonal 
(red) shapes. The vertical lines correspond to the cubic (uncorrected)
part of $f_\mathrm{NL}$ for that map.}
\label{fig_hist_lincorr}
\end{figure}

\begin{figure}
\includegraphics[width=0.5\columnwidth]{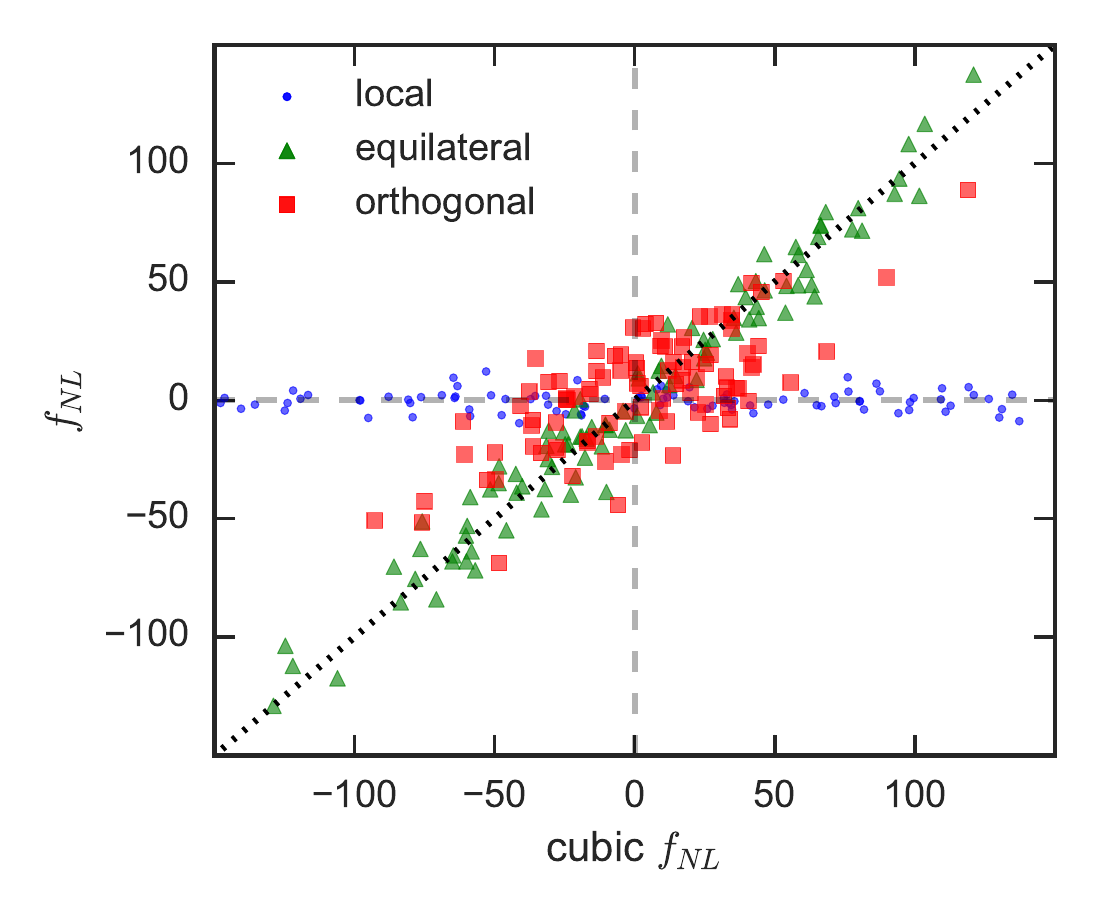}
\includegraphics[width=0.5\columnwidth]{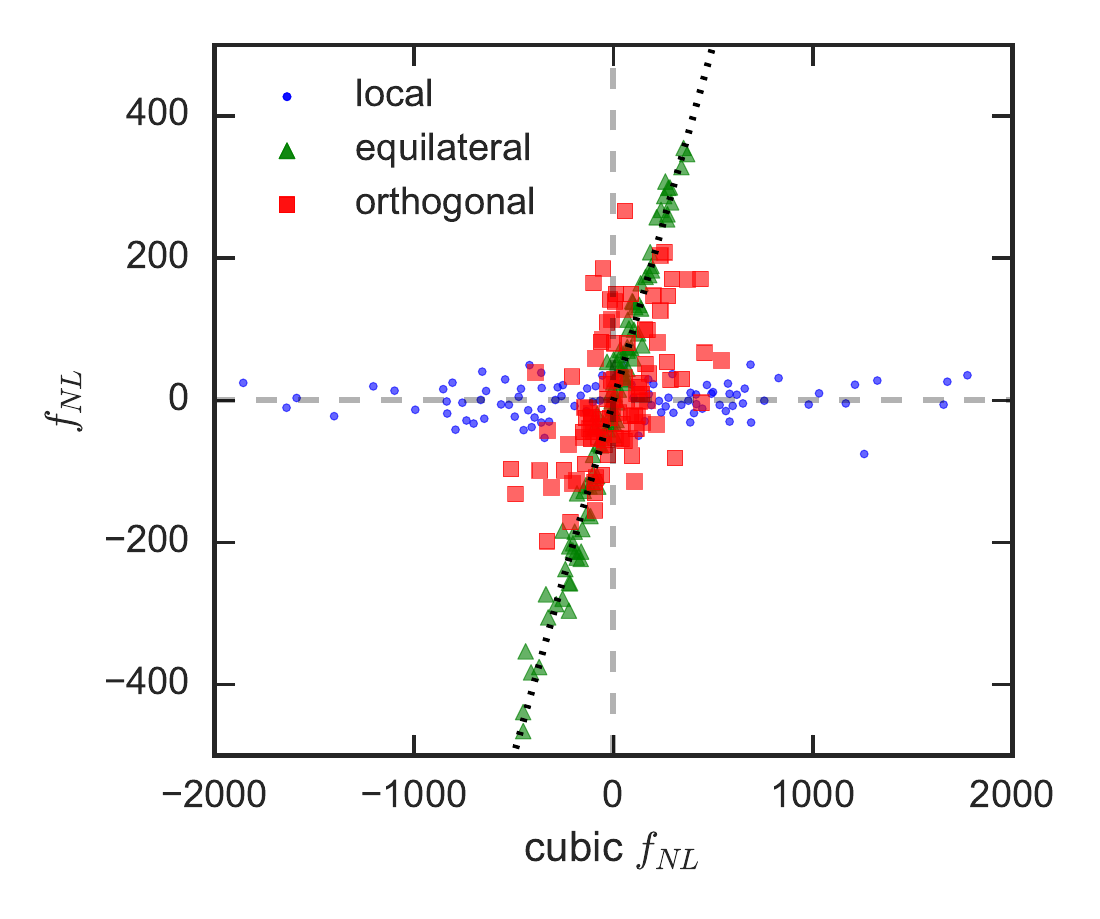}

\caption{Scatter plot of the values of $f_\mathrm{NL}$ without linear correction
(horizontal axis) against the values including linear correction (vertical
axis), for the 100 maps of the ``no mask, anisotropic noise'' case of 
Table~\ref{table_filling_in}. Results are
shown for temperature (left) and $E$ polarization (right) and for the
local (blue), equilateral (green), and orthogonal (red) shapes.
We clearly see the huge effect of the linear correction term for the
local shape (where a wide range of uncorrected values all map onto a small
range of corrected values, without any apparent correlation), some effect 
for orthogonal, and no effect for equilateral
(where the corrected and uncorrected values are very similar and highly
correlated).}
\label{fig_scatter}
\end{figure}

Table~\ref{table_filling_in} also highlights the importance of the linear
correction term when there is anisotropic noise. While there is hardly any
impact for the equilateral shape and no bias for any shape, for the local 
shape the error bars simply explode when we add anisotropic noise to the 
map, both for temperature and for the $E$ polarization mode. (For an explanation
see Section~\ref{sec_lincorr}.) However including the linear correction 
term described in Section~\ref{sec_lincorr} suffices to recover the same 
error bars as in the ideal case. 
As can be seen from (\ref{Bisp_lincorr}), the linear correction to the
bispectrum of a given map, and hence to the $f_\mathrm{NL}$ parameters 
via (\ref{fNL_estimator}), involves the average over a large number of 
Gaussian maps. In Fig.~\ref{fig_hist_lincorr} we show the histogram of
the individual contributions of 199 Gaussian maps to the linear correction
part of $f_\mathrm{NL}$ for one of the maps from the 
``no mask, anisotropic noise'' case of Table~\ref{table_filling_in}.
The corresponding mean values are for $T$: local $165 \pm 0.5$, equilateral
$-4 \pm 8$, orthogonal $-40 \pm 4$, and for $E$:
local $-415 \pm 3$, equilateral $-20 \pm 24$, orthogonal $119 \pm 11$.
As expected we see a hugely significant linear correction for local, a very
significant correction for orthogonal (due to the large correlation with
local), and no significant correction for equilateral. The error bars
on the linear correction term for a single map are much smaller (in this 
case of 199 maps about a factor 7) than the error bars on the values of the 
different $f_\mathrm{NL}$ parameters determined from 100 maps in
Table~\ref{table_filling_in}, indicating that 
we have used enough maps to determine the linear correction.
Another way of representing the importance of the linear
correction term in the presence of anisotropic noise is shown in 
the scatter plot of Fig.~\ref{fig_scatter}.

\section{Implementation of the estimator}
\label{implementationsec}

A significant advantage of the binned bispectrum estimator is that it 
divides the bispectral analysis and determination of $f_\mathrm{NL}$ into 
three separate parts, the first two of which are completely independent. 
The first, slow, part is the
computation of the raw binned bispectrum of the map under consideration, 
including its linear correction. The second, much faster, part involves the
computation and subsequent binning of the theoretical bispectrum templates
one wants to test and of the expected bispectrum covariance.
Finally, the third, extremely fast, part (that runs in less than about a 
minute) is where the different analyses (for example for different templates) 
are carried out using the raw binned bispectrum from part 1 and the
quantities from part 2 as an input. In the case of $f_\mathrm{NL}$ determination,
this last part corresponds to the evaluation of the sum over the bins and
polarization indices in the inner product (\ref{innerprod_polar})
used in (\ref{fNL_estimator}).

This approach has several advantages. Firstly, the full
(binned) bispectrum is a natural output of the code 
and can be studied on its own without a particular template in mind. 
Such an analysis will be the subject of the last Sections of the paper. 
Secondly, there is no need for the bispectrum template to be separable, 
since nowhere in the method does the need arise to split up the template 
into a separable form.
Thirdly, once the bispectrum of a map has been computed, modifications to 
the theoretical analysis (like for example testing additional templates) is 
fast, since there is no need to rerun the observational part (which consumes 
by far the most time). This is in contrast with competing estimators such as 
the KSW estimator, where the theoretical and observational
steps are mixed together (a separation is instead made in terms of 
$\ell_1,\ell_2,\ell_3$), so that the full code 
has to be rerun for any new template.
Fourthly, with the binned bispectrum estimator the dependence
of $f_\mathrm{NL}$ on $\ell$ is obtained almost for free, simply by leaving 
out bins from the sum when computing the final inner product. In particular
this has been used to study the dependence on $\ell_\mathrm{min}$ and 
$\ell_\mathrm{max}$ in the Planck analysis.
Finally, the binned bispectrum estimator compares
favorably to the other estimators in terms of speed: it is very fast
on a single map.

The only disadvantage of this method is that the templates
that can be studied accurately have to be reasonably smooth, or if not then
any rapid changes should be limited to a small part of $\ell$-space, in order 
for the template to be well approximated by a binned template with a not 
too large number of bins.\footnote{There are indications that the binned
bispectrum estimator might even perform well for oscillating templates that
do not satisfy these criteria. For the so-called constant feature 
model~\cite{Ade:2015ava} with a primordial bispectrum proportional to
$\sin(\omega(k_1+k_2+k_3)+\varphi)/(k_1 k_2 k_3)^2$, taking $\varphi=0$
and $\omega=100$, we find an overlap of 94\% for $T$-only with the
standard Planck binning (i.e.\ not optimized for this template). This will be 
investigated in more detail in a future publication.} 
For most primordial and foreground templates studied
so far, this is not a problem. Moreover, even for
templates that do not satisfy this criterion, the binned bispectrum estimator
could still perform quite well. For example, among the templates discussed in
Sections~\ref{sec_templates} and~\ref{sec_isocurv}, only the lensing-ISW 
template cannot be easily
binned. For a typical Planck binning the overlap is of the order of 60--70\%
(as opposed to 95\% or higher for all the other templates considered). 
Nevertheless the binned bispectrum estimator gives unbiased results even 
for this template, with error bars that are only slightly widened.

The code has been written mainly in Python, using some routines 
written in C. 
It is run on the computers of the Centre de Calcul de l'Institut 
National de Physique Nucl\'eaire et de Physique des Particules (CC-IN2P3)
in Lyon, France\footnote{\tt http://cc.in2p3.fr} and any explicit remarks about
computing time refer to that system.

\subsection{Theoretical part}

The theoretical part of the code consists of two steps: first determining 
the unbinned theoretical bispectrum and power spectrum, and second, computing 
from these spectra the binned bispectrum templates $B^\mathrm{th}_{i_1 i_2 i_3}$
and the inverse of the binned covariance matrix 
$(V^{-1})_{i_1 i_2 i_3}^{p_1 p_2 p_3 p_4 p_5 p_6}$, see (\ref{fNL_estimator}) and 
(\ref{innerprod_polar}). This also requires experimental inputs in 
the form of the beam transfer function $b_\ell$ and the noise power spectrum 
$N_\ell$.

The first step is in some sense not really part of the estimator code. 
We have a code to compute all the bispectra discussed in 
Sections~\ref{sec_templates} and~\ref{sec_isocurv}, but in principle an 
explicitly computed theoretical bispectrum from any source could be used here. 
In our code we use the radiation transfer functions $\Delta_\ell^{p\, I}$
(with $p$ the polarization index and $I$ the isocurvature index) computed
with CAMB (slightly modified to write them to file, since these are not a 
normal output of CAMB) to compute the primordial templates (\ref{Bth}).
For separable templates, this is a fast calculation, since the triple integral
over $k_1,k_2,k_3$ becomes a product of single integrals. For non-separable
templates a brute force calculation is much slower, but while one
might look for smarter ways to compute such bispectrum templates, it should
not be forgotten that (for a given cosmology) for use in the binned bispectrum
estimator, a template has to be computed only once. Hence even
a slow calculation might be acceptable.
While this code can also compute the power spectra from the radiation transfer
functions according to (\ref{Clth}), in practice we use the power 
spectra computed by CAMB. These power spectra are used in the covariance 
matrix and some foreground bispectrum templates. 
This code is similar to the code described in 
\cite{Bucher:2009nm}, except for the inclusion of additional primordial 
templates and the generalization to polarization and isocurvature. Since the 
calculation of the $N_\triangle^{\ell_1 \ell_2 \ell_3}$ in real time is fast enough, 
these are no longer precomputed and stored.
The primordial bispectra are precomputed only on a grid (with $\Delta\ell$ 
increasing to about 10 at high $\ell$). This is denser than the binning, and 
thus accurate enough for the smooth local, equilateral, and orthogonal 
templates.

The second step involves the binning of the bispectrum templates and the
covariance matrix. The calculation of the covariance
matrix from the power spectra as well as the calculation of the foreground
templates is done directly in this step. As was seen in 
Section~\ref{sec_templates}, the foreground templates are simpler to
compute than the primordial templates, since there are no integrals, so there
is no need to precompute them, the required values can be computed in real 
time while binning. As for the precomputed primordial templates, since these 
have been precomputed only on a grid, other values are computed by 
three-dimensional linear interpolation.
While we developed a tetrahedral integration scheme to speed up the
calculation of all binned quantities, as described in \cite{Bucher:2009nm},
recently we have moved away from using it. Given that the theoretical 
computation is much faster than the observational computation, there is no
point in making additional approximations to speed it up. Performing an
exact calculation of the binned quantities (where the quantities are
explicitly computed for each value of $\ell_1,\ell_2,\ell_3$ and then
summed over the bin) is fast enough. We can thus also
directly compute the overlap between the binned and the exact template using 
(\ref{binning_error}).

The final output of this step consists of two files: one containing the binned
theoretical bispectrum for all requested shapes, polarization and 
isocurvature components; and another containing 
the inverse of the binned covariance matrix for all
polarization components. In addition the exact Fisher matrix (\ref{Fisher})
(without binning) is produced to allow for the estimation of the accuracy 
of the binning approximation using (\ref{binning_error}).

\subsection{Choice of binning}

The choice of binning is an important
part of the implementation. In theory the idea is very simple: one chooses
the binning that makes the overlap parameter $R$ defined in 
(\ref{binning_error}) as close to one as possible.
In practice this is not so simple, since both the number of bins
and all the bin boundaries are free parameters. Fortunately $R$ does not
depend strongly on the exact binning choice. Moreover, one does not need 
$R=1$ to obtain results statistically indistinguishable from the
exactly optimal results. For example, even with $R=0.95$, which is about the
lowest overlap for any of the templates considered in the Planck analysis 
(except for lensing-ISW), the increase in the standard
deviation is only 2.6\%. This should be compared to the 5\% uncertainty
in the standard deviation due to its determination from 200 maps.
Note that the code allows the use of separate binnings for the $T$-only,
the $E$-only, and the full $T+E$ analyses, although for reasons related
to time a single binning was used for the Planck analysis.

We developed three optimization tools: one that checks which bin boundary can be
removed with the smallest decrease of $R$ (reducing the number of bins by
one), one that checks where a bin boundary can be added with the largest
increase in $R$ (increasing the number of bins by one; the bin boundary
is added in the exact center of an existing bin), and one that tries
moving all the bin boundaries by a given amount (relative to the size of
the bin) and tells for which bin this increases $R$ the most (leaving the number
of bins unchanged). For all of them one can indicate which shapes
and polarizations (meaning $T$ and/or $E$) should be taken into account.
These three tools are then used iteratively to optimize the
binning used as starting point, until no more significant improvements are 
obtained (as defined by a certain threshold in the change of $R$).
The starting point is arbitrary. 
For example a simple log-linear binning (with bin sizes increasing
logarithmically at low $\ell$, up to a certain value of $\ell$, after
which the binning becomes linear) or a binning that has already been
partially optimized in another way can be used. The latter could for example 
be done using the method described in \cite{Bucher:2009nm}, which can 
provide a good starting point. (That method produces suboptimal binnings 
and can benefit significantly from the procedure
described here.)
While this method can likely be optimized further, for the 
Planck analysis the binning obtained in this way produces 
effectively optimal results.

\subsection{Observational part}

The observational component of the code consists of two parts: one to compute
the cubic part of the bispectrum of the map according to (\ref{Bobsbinned}),
and the other to compute the linear correction according to 
(\ref{Bisp_lincorr}). First the map is fully prepared, which can be as simple
as reading an existing map and doing the masking and filling in, or involve the
creation of CMB and noise realizations. It is then saved in the form 
of $a_{\ell m}$'s for later use with the linear correction term, or for 
reproducibility in the case of generated random realizations.

The maps are then filtered according to (\ref{Tmapbinned}). This leads
to some practical issues that had to be resolved, since in principle
we need to hold twice ($T$ and $E$) 50--60 maps (one for each bin) of
Planck resolution ($n_\mathrm{side}=2048$) in memory for this
calculation.  However our computer system has a limit of
16~GB per processor, which makes this impossible. We managed to save
space in two ways. In the first place, while all the preprocessing of
the maps is done in double precision, the final filtered maps are only
kept in single precision, which saves a factor two in
memory. Tests have shown that this has no significant impact on the
final result for $f_\mathrm{NL}$.
Secondly, it is unnecessary to use $n_\mathrm{side}=2048$
precision for the maps that contain only low-$\ell$ bins. Hence the
filtered maps of bins up to about $\ell=800$ are produced at
$n_\mathrm{side}=1024$, which saves a factor of four in memory for those 
maps (the number of pixels in the map is $12 \, n_\mathrm{side}^2$), as well 
as speeding up the final computation where three maps have to be multiplied and
summed (see (\ref{Bobsbinned})). Using the nested Healpix\footnote{\tt 
http://healpix.sourceforge.net} 
format, it is easy to multiply maps of different $n_\mathrm{side}$ together.

We have developed two different ways of computing the linear correction term 
of a map. In the first method, which was used for the Planck analysis and
the analyses for this paper, each job treats one of the {\em Gaussian} maps 
(see (\ref{Bisp_lincorr})), which is preprocessed and filtered as above, 
and the filtered maps are held in memory. Then a filtered map 
of only the first bin of the {\em observed} map is created and all required sums
of products involving that map are computed. Next this process is repeated
for the second bin of the observed map, etc.\footnote{In an earlier version 
of the code these filtered maps of the observed map, which are also produced 
during the cubic calculation, were saved to disk at that time, and then read in
here. However, the required I/O turned out to make this actually slower than 
when these filtered maps are recreated on the fly, which also has the advantage
of using much less disk space.} The final result of this job
is a temporary file with a linear correction term computed with
just one Gaussian map. Once all jobs have finished (with the results for the
other Gaussian maps), the results are summed and averaged to obtain the
final linear correction term for the map. 
This whole process (preprocessing the map and computing the cubic and linear 
terms) for a single map at Planck resolution for all $T+E$ (including
mixed) components takes a few hours, which is quite fast compared to other
bispectrum estimators. (Computing the theoretical part is much faster and 
requires only a single job, so can easily be done on the side.) 
With this method one can simply add more Gaussian maps to the linear 
correction term at a later stage if required, and investigate its convergence 
as a function of the number of Gaussian maps. 
However, this first method of computing the linear correction term scales
very badly with the number of observed maps. Since the object
$\langle M_{i_1}^{p_1, G} M_{i_2}^{p_2, G} \rangle$ in (\ref{Bisp_lincorr}) is too 
large to compute directly
and save to file, if one has a set of similar maps (for example to
compute error bars), the linear correction term has to be recomputed 
for each map in the same way as above, making this a very slow process.

For this reason we recently developed another way to compute the linear
correction term. This second method is based on the observation that while
the object $\langle M_{i_1}^{p_1, G} M_{i_2}^{p_2, G} \rangle$ (consisting of 6612
maps for a full $T+E$ calculation in the case of 57 bins) is too large to
handle, saving it in the form of $a_{\ell m}$'s is doable. Moreover, we make
use of the fact that when multiplying several masked maps together (all with
the same mask), it is enough if only one of the maps is masked. Hence if
the observed map in (\ref{Bisp_lincorr}) is properly masked, the Gaussian maps
can be left unmasked (since the Gaussian maps are based on simulations, they
are full-sky maps). This has the advantage that no filling-in needs to
be performed on these maps, which would otherwise be required before conversion
to $a_{\ell m}$'s, as explained in section~\ref{masking}. By limiting the
number of considered bins per job in such a way that both the filtered maps 
for those bins and all the product maps involving those bins can be kept in 
memory at the same time, one job can compute the full average for
the considered bins by treating one Gaussian map after the other. Only
at the end are the final maps converted to $a_{\ell m}$ format and written to 
disk. This precomputation for the linear correction term can be run with a 
modest number of jobs (about 100) in a reasonable amount of time (less than
a day for 200 maps). Once the precomputation has finished, the linear
correction for any map can be quickly computed using (\ref{Bisp_lincorr}).
Each job reads in a number of product maps (i.e.\ for certain values of
$i_1$ and $i_2$; the number being determined by memory considerations), 
and converts them back to pixel space. They are then multiplied with the
filtered observed maps as explained above for the first method. The main 
difference is that the results are now for the full average of all the 
Gaussian maps, instead of for a single one. Another (small) advantage of this
second method is that at this step we only need to multiply two maps together
and not three. Once all jobs are finished, the temporary files containing
results for different $i_1$-$i_2$ bins are combined to get the full linear
correction for the observed maps.
While this second method with precomputation is slower if one is only 
interested in a single map, its much better scaling with the number of 
maps makes it by far the preferred method when dealing with a set of maps,
for example to compute error bars.

The final result of this part are two files for each map, one containing the 
binned cubic-only bispectrum of the map and the second its linear correction,
both containing all requested polarization components. These can then be
combined with the results from the theoretical part to compute $f_\mathrm{NL}$
according to (\ref{fNL_estimator}), which takes less than a minute even when
producing convergence plots and dependence on $\ell$ as well, or
be studied directly without the assumption of a theoretical template, as
discussed in Section~\ref{smoothingsec}.

\section{Smoothed binned bispectrum}
\label{smoothingsec}

The previous Sections described how the binned bispectrum
of a map can be analysed parametrically by computing the
$f_\mathrm{NL}$ parameters corresponding to a selection of theoretically
motivated templates. But one advantage of the binned bispectrum
estimator is that the full (binned) three-dimensional bispectrum is a
direct output of the code, which can be studied 
non-parametrically, thus searching for any deviations from
Gaussianity even when no suitable template is available.  
Here we describe the smoothing procedure that must first be applied
to the binned bispectrum in order to enhance the 
signal-to-noise of any possible non-Gaussian features, which 
otherwise would remain hidden in the noise. 
The next Section describes the statistical analysis subsequently 
applied to this smoothed binned bispectrum to assess the statistical 
significance of any non-Gaussian features appearing as extreme values. 

We first normalize the binned data by dividing by the square root of the 
expected
bin variance, so that each bin triplet in the absence of a bispectral signal 
would have noise obeying a normalized Gaussian distribution. Thus for the 
bin triplets for which there is data, we define
\begin{align}
&{\cal B}_{i_1 i_2 i_3}^{TTT} =
\frac{B_{i_1 i_2 i_3}^{TTT} }{\sqrt{V_{i_1 i_2 i_3}^{TTTTTT}}},
\qquad
{\cal B}_{i_1 i_2 i_3}^{EEE} =
\frac{B_{i_1 i_2 i_3}^{EEE} }{\sqrt{V_{i_1 i_2 i_3}^{EEEEEE}}},
\nonumber\\
&{\cal B}_{i_1 i_2 i_3}^{T2E} =
\frac{B_{i_1 i_2 i_3}^{T2E} }{\sqrt{\mathrm{Var}(B^{T2E})_{i_1 i_2 i_3}}},
\qquad
{\cal B}_{i_1 i_2 i_3}^{TE2} =
\frac{B_{i_1 i_2 i_3}^{TE2} }{\sqrt{\mathrm{Var}(B^{TE2})_{i_1 i_2 i_3}}}.
\end{align}
For the mixed $T$ and $E$ components we analyzed only the combinations
$B^{T2E} \equiv TTE+TET+ETT$ and $B^{TE2} \equiv TEE+ETE+EET$, with 
corresponding variance Var($B^{T2E}$) = 
Var($TTE$) + Var($TET$) + Var($ETT$) + 2 Cov($TTE,TET$) + 2 Cov($TTE,ETT$)
+ 2 Cov($TET,ETT$), and similarly for Var($B^{TE2}$). 
This projection entails a loss of information but allows the same analysis to
be used as for $TTT$, as described below.
Results for the unsymmetrized mixed bispectra will be discussed in a future
publication.

As discussed in Section~\ref{sec_binned_bispectrum}, only bin triplets 
containing $\ell$'s that satisfy both the parity condition and the
triangle inequality contain data. However, among the bin triplets containing 
data, we noticed that some triplets systematically produced outliers. It 
turned out that
these bin triplets contained very few valid $\ell$-triplets [for example,
the hypothetical bin triplet $([50,100],[50,100],[200,300])$ would contain only 
one valid $\ell$-triplet (100,100,200), since the triangle inequality imposes
that $l_3 \leq l_1+l_2$]. While the theoretical variance calculation is exact, 
the computation of the observed bispectrum using Healpix spherical harmonic 
transforms contains some numerical inaccuracies, so that the 
bispectrum in points outside the triangle inequality is not zero but
contains leakage.\footnote{This results because the pixelization breaks the 
spherical symmetry as must be the case with any pixelization of the sphere.}
For bin triplets like the above example with many $\ell$-triplets violating 
the triangle inequality, a significant mismatch between
the theoretical and the actual standard deviation of the bispectrum in that bin
is observed. 
The obvious solution is to remove such bin triplets from the data. Moreover,
the statistical analysis described
in the next Section assumes that bin triplets contain many valid $\ell$-triplets
in order for Gaussian statistics to apply to the noise from cosmic variance,
which constitutes another reason to exclude such triplets. 
After some experimentation, we adopted the
selection criterion that the ratio of valid $\ell$-triplets to the ones 
satisfying only the parity condition (but not the triangle inequality) in a
bin triplet should be at least 1\%, finding this a good threshold for 
rejecting systematic outliers. The results are insensitive
to the precise threshold used. For the Planck binning with 57 bins (which
is used for the results in this Section and the next), this 
criterion excluded 293 out of 13020 bin triplets.

If we were looking for a sharp bispectral feature of a linewidth
narrow compared to the binwidth, there would be no
motivation to smooth. We would simply examine the statistical
significance of the extreme values of the renormalized binned
bispectrum described above taking into account the look elsewhere
effect. However for broad features, as are likely to arise from
galactic foregrounds, smoothing increases statistical significance 
by averaging over and thus diminishing the noise.
One approach would be to use binning with a range of bin widths, but
this approach has the disadvantage that the statistical significance
for detecting a feature depends on how it is situated relative to
the neighbouring bin boundaries. Instead we rather smooth using a
Gaussian kernel and renormalize so that in the absence of a signal the
single pixel distribution function is again a unit Gaussian. For a
Gaussian kernel $K_{\sigma_\mathrm{bin}}$ of width $\sigma_\mathrm{bin}$, 
we have
\begin{equation}
\mathcal{B}^{p_1 p_2 p_3, \text{smoothed}}_{i_1i_2i_3} = 
\sum_{i'_1} \sum_{i'_2} \sum_{i'_3} 
K_{\sigma_{\text{bin}}}(i_1-i'_1, i_2-i'_2, i_3-i'_3)  
\mathcal{B}^{p_1 p_2 p_3} _{i'_1i'_2i'_3} 
\end{equation}
where the Gaussian smoothing kernel 
\begin{equation}
K_{\sigma_{\text{bin}}}(
\Delta i_1, 
\Delta i_2, 
\Delta i_3) 
=
\left( 2\pi {\sigma _\mathrm{bin}}^2\right) ^{-3/2} ~
\exp \left[ 
-\frac{1}{2}\frac{
{\Delta i_1}^2
+{\Delta i_2}^2 
+{\Delta i_3}^2 
}{
{\sigma_\mathrm{bin} }^2}
\right]
\end{equation}
is used. Numerically the kernel is applied in the Fourier domain.

Without boundaries this smoothing and renormalization procedure would
be straightforward.  However near the boundary the Gaussian 
smoothing kernel would extend
into the region where there is no data.  To minimize 
boundary effects, we first extend the fundamental domain
(where $i_1 \leq i_2 \leq i_3$) to the five identical domains
obtained by permuting $(i_1, i_2, i_3)$ and pad with zero data 
beyond the boundaries of this extended domain as well
as for triplets inside the domain for which there is no data.  
The smoothing causes power to leak out into 
the zero padded regions, and to correct for this leakage, we construct a mask 
consisting of ones in the domain of definition and zeros outside. After 
smoothing the signal-to-noise bispectrum $\mathcal{B}$, we renormalize by 
dividing by the mask that has undergone the same smoothing procedure. 
For the bin triplet statistic to be a Gaussian of unit variance, we generate 
1000 Monte Carlo realizations going through the same procedure and compute 
the variance, with which we divide our smoothed renormalized bispectrum.

\begin{figure}
\centering
\includegraphics[trim = 3mm 0mm 12mm 0mm,clip, width=0.31\columnwidth]{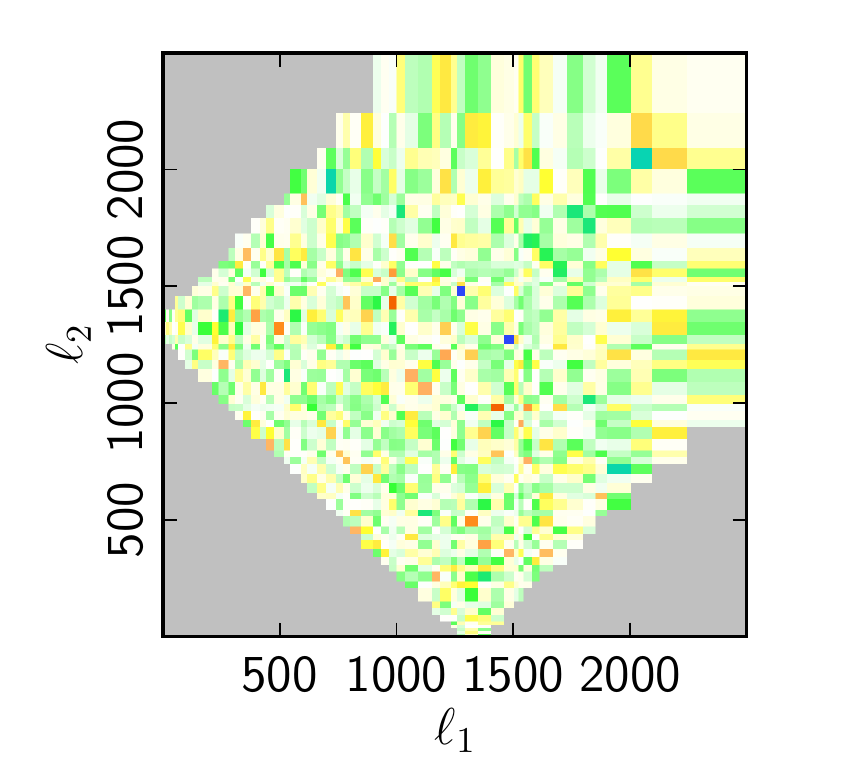}
\includegraphics[trim = 3mm 0mm 12mm 0mm,clip, width=0.31\columnwidth]{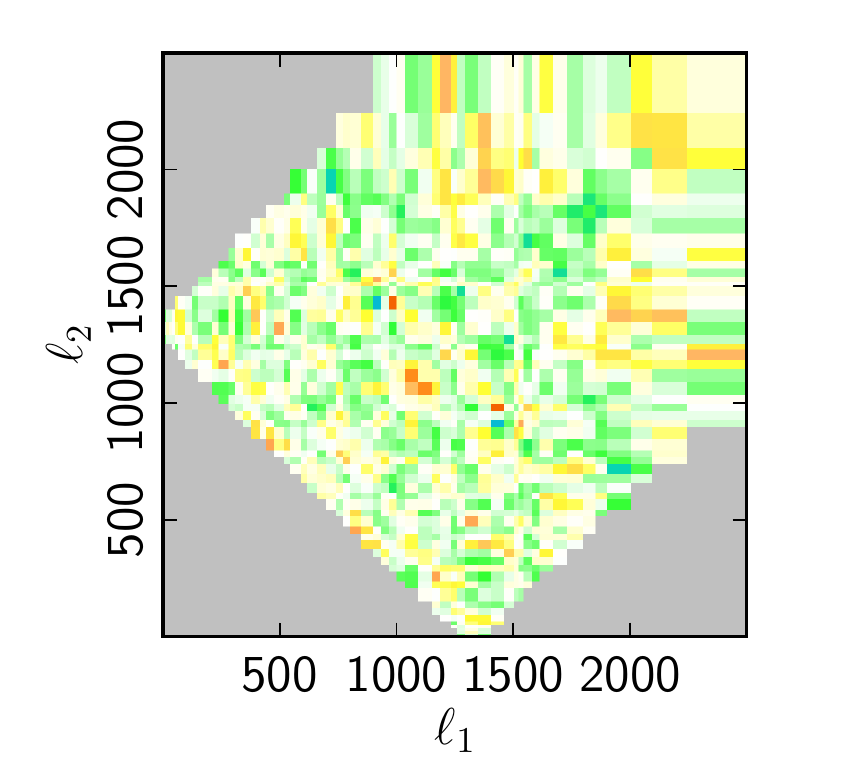}
\includegraphics[trim = 3mm 0mm 12mm 0mm,clip, width=0.31\columnwidth]{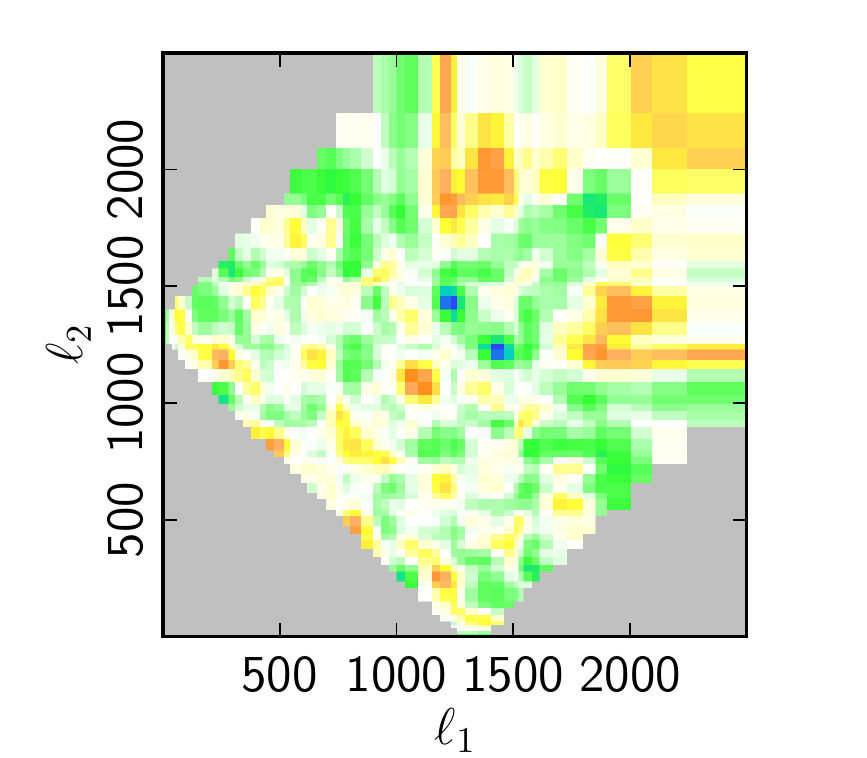}
\includegraphics[trim = 3mm 0mm 12mm 0mm,clip, width=0.31\columnwidth]{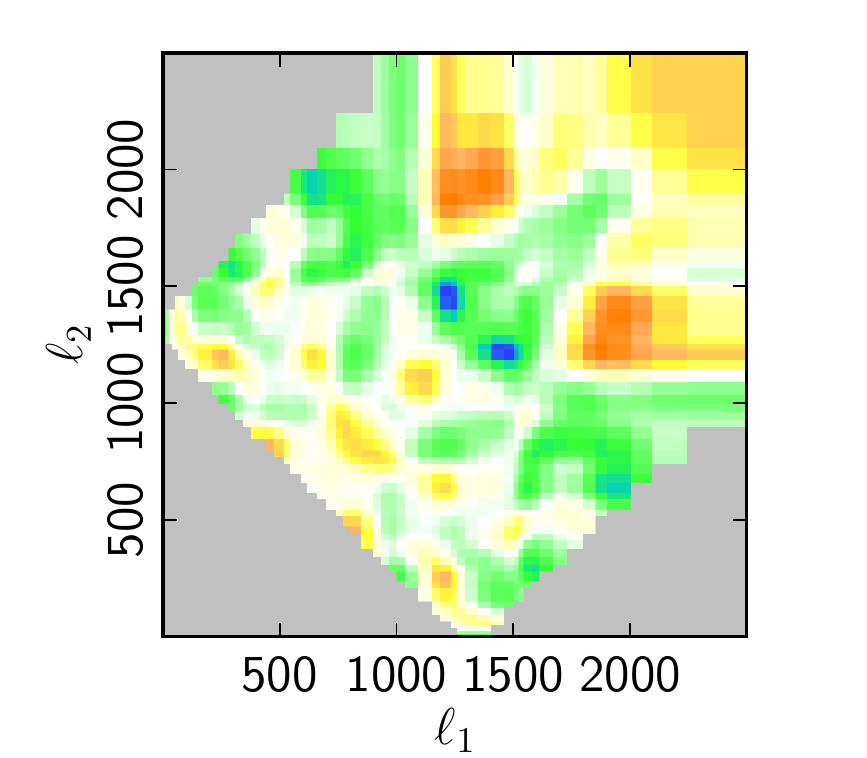}
\includegraphics[trim = 3mm 0mm 12mm 0mm,clip, width=0.31\columnwidth]{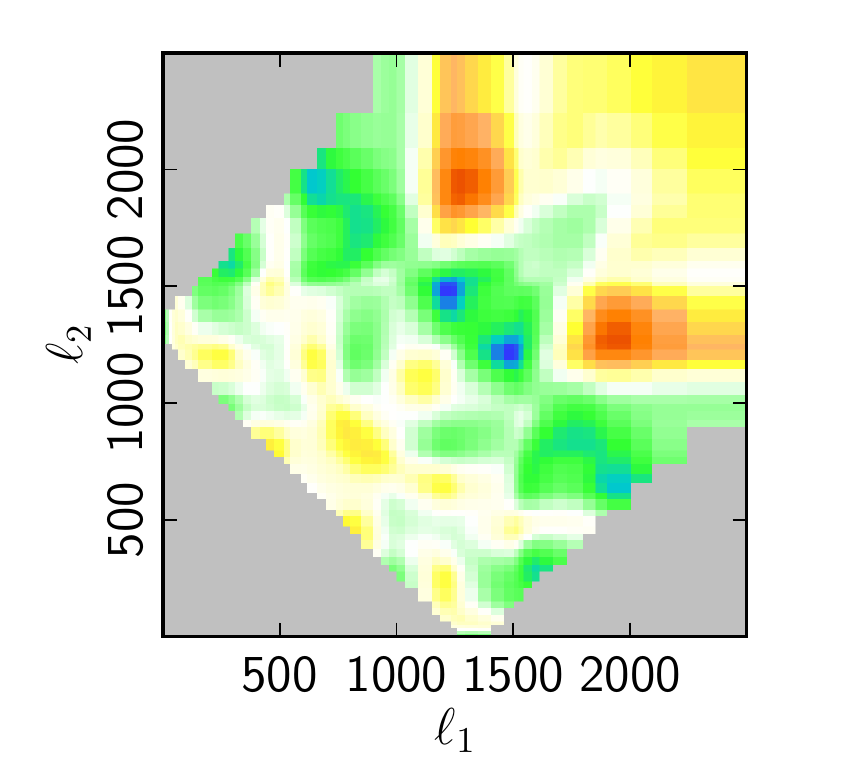}
\includegraphics[trim = 3mm 0mm 12mm 0mm,clip, width=0.31\columnwidth]{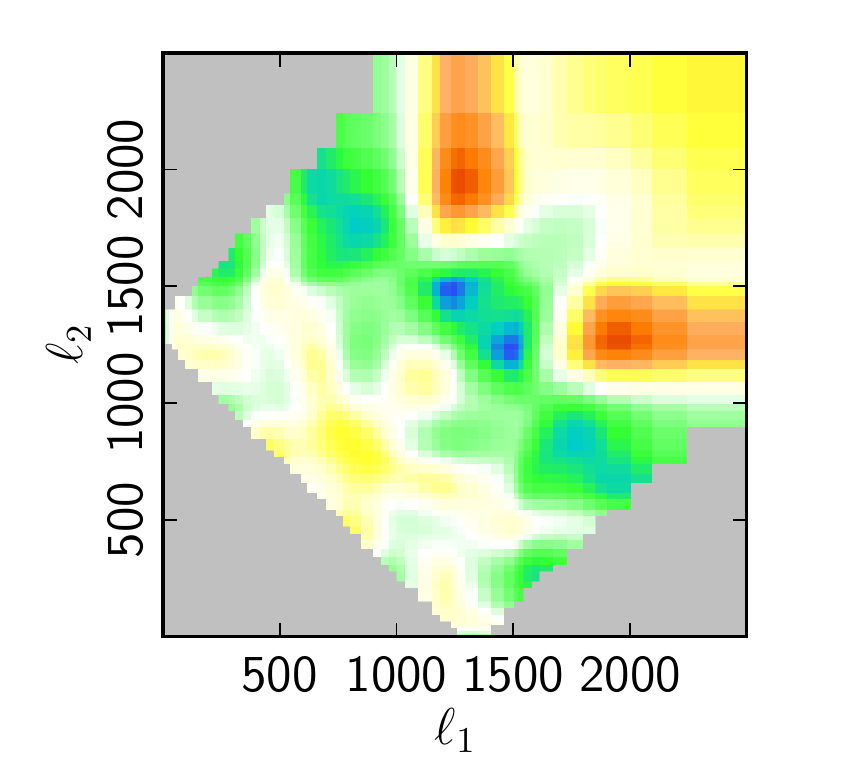}
\includegraphics[trim = 3mm 0mm 12mm 0mm,clip, width=0.31\columnwidth]{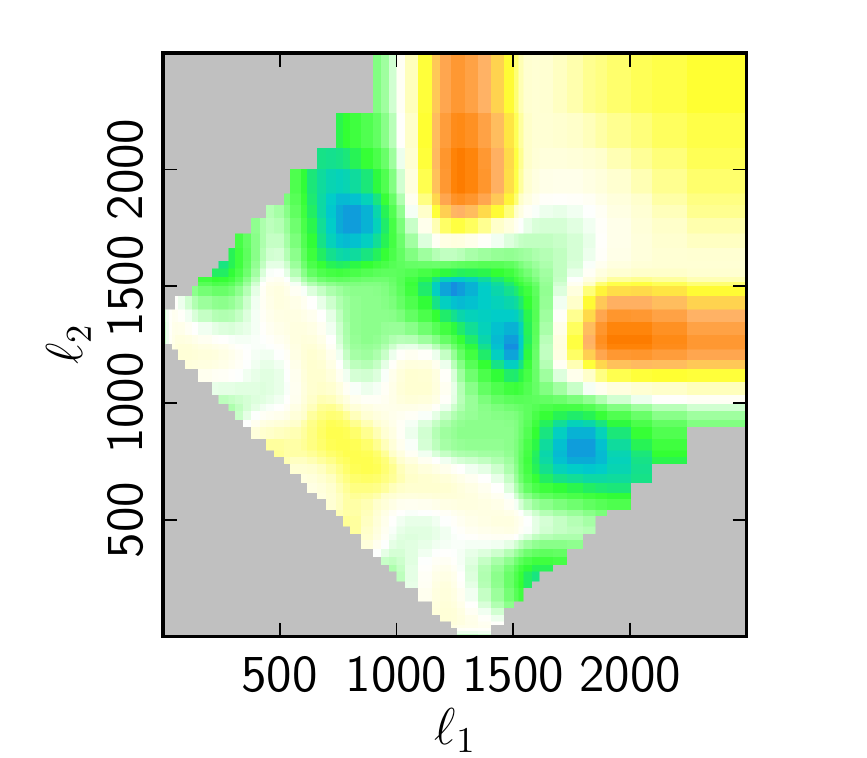}
\includegraphics[trim = 0mm 0mm 25mm 10mm,clip,width=0.75\columnwidth]{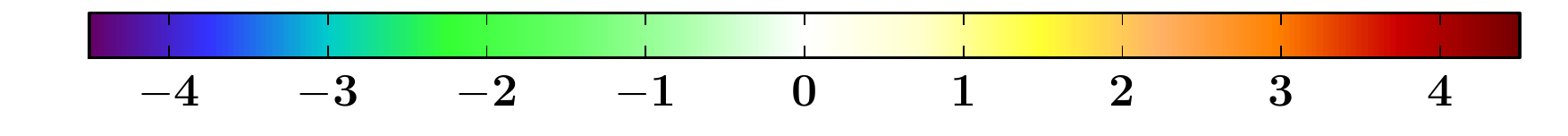}
\caption{Effect of different smoothing lengths on the bispectrum of one of 
the Gaussian maps from Table~\ref{table_filling_in} with 
galactic and point source masks and anisotropic noise. From left to right 
and top to bottom are shown: no smoothing, smoothing using $\sigma_{\mathrm{bin}} 
= $ 0.5, 1, 1.5, 2, 2.5 and 3. 
The slices correspond to $\ell_3 \in [1291, 1345]$. }
\label{figure_smolen}
\end{figure}

The result using different smoothing lengths is
illustrated in Fig.~\ref{figure_smolen} as two-dimensional
slices showing $\mathcal{B}$ as a function of $\ell_1$ and $\ell_2$ for a
fixed bin in $\ell_3$. 
As a further illustration, to show what certain known non-Gaussian
features look like in this representation, we present in 
Appendix~\ref{App:AppendixA} slices of the smoothed  
theoretical template bispectra from Sections~\ref{sec_templates} 
and~\ref{sec_isocurv}.
With the color scale used in Fig.~\ref{figure_smolen}, both dark red and dark 
blue represent 
extreme values with small $p$ values if Gaussianity is assumed, and thus 
suggest the presence of statistically significant bispectral non-Gaussianity.
A correct analysis of the significance would also take into account the look 
elsewhere effect --- that is, that the small probability to exceed, calculated 
for a fixed bin, is too small because it does not reflect that an improbable 
value could have occured in any of a number of bins. 
The analysis of this issue is complicated by the correlations between the 
bins that result from the smoothing, an issue analyzed in the next Section.

\section{Statistical analysis}
\label{statanalsec}

This Section addresses how to assess the statistical significance of
possible bispectral non-Gaussianity in the binned bispectrum that has
been smoothed according to the procedure described in the previous
Section. Even in the case of a Gaussian sky, where before the
smoothing over bins has been applied the single-bin distribution
function is a normalized Gaussian for which the values in distinct
bins are statistically independent, after this smoothing procedure
fluctuations in neighboring bins become correlated, complicating the
analysis. We consider here how to deal with this complication.

In the absence of smoothing, we face the following statistical
problem. We have a binned bispectrum that has been rescaled so that we
have $N$ bins and the bispectrum value in each bin $x_i,$ where
$i=1,\ldots , N$, has a probability distribution function well
approximated by a normalized Gaussian distribution. Moreover, values in
different bins are almost statistically independent. The quadratic
correlation vanishes, but some of the higher order joint correlations
do not precisely vanish, a feature that we shall neglect here. The
corrections to Gaussianity and to statistical independence are
suppressed when $\ell $ is large and when there are many $\ell$-triplets 
containing data in a bin.  Thus we have the distribution function
\begin{eqnarray}
p(x_1,\ldots ,x_N)=(2\pi )^{-N/2}~\exp \left[ -\frac{1}{2}\sum _{i=1}^N {x_i}^2
\right] ,
\label{pdfMV}
\end{eqnarray}
and since we are interested in extreme values, we define two new derived 
statistics
\begin{eqnarray}
X_\mathrm{min}=\min (x_1,\ldots , x_N); \qquad
X_\mathrm{max}=\max (x_1,\ldots , x_N),
\end{eqnarray}
and accordingly define the $p$-values
\begin{eqnarray}
p_\mathrm{min}(X)=P(X_\mathrm{min}<X); \qquad 
p_\mathrm{max}(X)=P(X_\mathrm{max}>X)
\end{eqnarray}
where $X_\mathrm{min}$ and $X_\mathrm{max}$ are the derived random variables 
defined above. If either of these $p$-values are extremely small, then we 
have evidence of bispectral non-Gaussianity directly in the unsmoothed 
binned bispectrum, and this $p$-value can be converted into a $\sigma $ 
for the normal distribution using the inverse error function as is customary.

For this simple unsmoothed case it is not hard to give the probability 
distribution function for the extreme value statistics $X_\mathrm{min}$ and 
$X_\mathrm{max}$. 
Given the (complementary) cumulative distribution function for the normal 
distribution (integrating from right to left)
\begin{eqnarray}
\Phi (x)=\frac{1}{\sqrt{2\pi }}\int _x^{+\infty }dt~\exp \left[ -\frac{1}{2}t^2
\right] ,
\label{ccdf}
\end{eqnarray}
the analogous distribution for the maximum extreme value for $N$ variates is 
given by 
\begin{eqnarray}
\Phi_\mathrm{max}(X_\mathrm{max}; N)=1-\Bigl( 1- \Phi (X_\mathrm{max})\Bigr) ^N
\label{evCDF}
\end{eqnarray}
and we may straightforwardly obtain an analogous expression for the case
of the minimum value. (Below we shall only give results for the case of
the maximum.) For $X\gg 1$ we obtain an approximation to 
$\Phi_\mathrm{max}(X_\mathrm{max}; N)$ by inserting the following expression
\cite{Abramowitz}
\begin{eqnarray}
\ln \Bigl[ \Phi (X)\Bigr] \approx -\left[ 
\frac{X^2}{2}+ \ln (X)+ \frac{1}{2}\ln (2\pi ) 
\right] 
\label{evCDFapprox}
\end{eqnarray}
into (\ref{evCDF}). 
%Expression~(\ref{evCDFapprox}) is found by performing
%the change of variables $t=u+x$ in~(\ref{ccdf}) and realizing that, since $x$
%is very large, the main contribution to the integral will come from values
%of $u$ very close to zero. Hence $\exp(-u^2/2)$ can be approximated by 1 and
%the integral computed explicitly.

When we consider extreme values of multivariate Gaussian distributions
with correlations, there is, as far as we know, no way of obtaining an
analytic result for the extreme value distribution for 
$\Phi_\mathrm{max}(X_\mathrm{max})$. After the smoothing described in the previous
Section is applied, the probability distribution defined in
(\ref{pdfMV}) must be replaced with
\begin{eqnarray}
p({\bf x})=(2\pi )^{-N/2}~\exp \left[ -\frac{1}{2} {\bf x}^T {\bf C}~{\bf x} 
\right] 
\end{eqnarray}
where the correlation matrix ${\bf C}$ has all ones on the diagonal,
but also a lot of positive off-diagonal elements as the result of the
smoothing process, rather than all zeros away from the diagonal. It is
these off-diagonal elements that prevent us from solving analytically for the
extreme value statistic probability distribution function.

Instead we postulate an Ansatz to approximate the cumulative
distribution function (CDF) of the extreme value statistic $\Phi_\mathrm{max}$, 
which has one adjustable parameter $N_\mathrm{eff}$, the effective
number of independent bins, which will be smaller than the actual
number of bins $N$ as the result of the smoothing. The Ansatz states
that the CDF given in (\ref{evCDF}) [and approximated using
(\ref{evCDFapprox})] holds where $N$ has been replaced with
$N_\mathrm{eff}$. For a given level of smoothing, we fit $N_\mathrm{eff}$ to the
tail of the CDF, which has been determined empirically by Monte Carlo
simulations. We then assess the quality of the approximation, in
particular in the tail of the distribution where $X$ is very large,
which is the range of values of particular interest here.  It should
be stressed that we do not need a good approximation to the entire
CDF. It suffices to have an approximation that works well
asymptotically, in the extreme tail of the distribution where
$p$-values cannot feasibly be obtained by Monte Carlo methods. Thus
the Ansatz serves as an asymptotic approximation for the tail of the
distribution.

\begin{figure}
\begin{center}
\includegraphics[width=7.5cm]{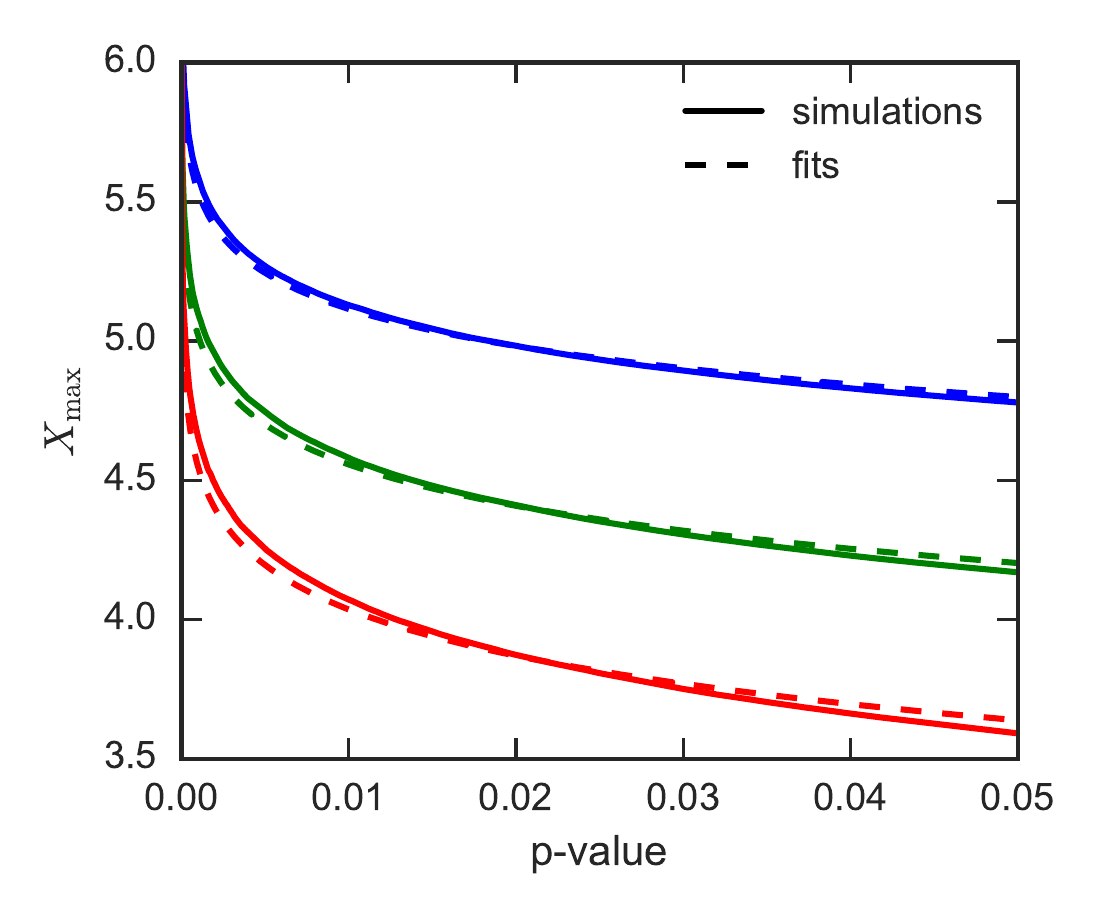}
\includegraphics[width=7.5cm]{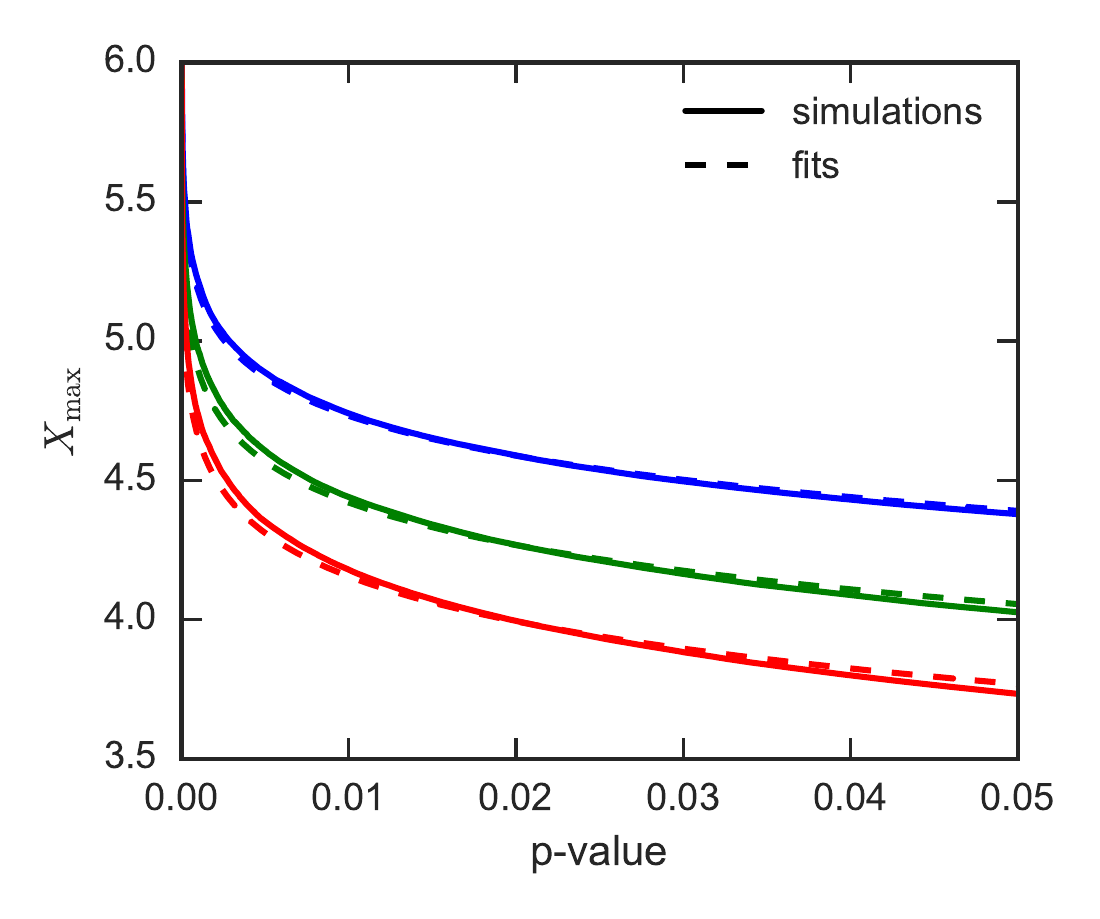}
\end{center}
\caption{Fitting the extreme values of the smoothed bispectrum.
The horizontal axis gives the $p$-value and the vertical axis
the maximum bin deviation in number of $\sigma$.
The solid curves show the CDFs for the smoothed $64^3$ cubes 
(figure on the left) for $\sigma_\mathrm{smoothing} =2, 5,$ and $10$ 
(respectively from top to bottom blue, green and red)
and for the more realistic situation (figure on the right) for 
$\sigma_\mathrm{smoothing} = 1, 2,$ and $3$ as described in the main text.
We show only the 5\% most extreme values, since we are
interested in the extreme tail of the distribution. The dashed curves show 
non-linear least square fits to these curves using the Ansatz where a 
distribution with a smaller number $N_\mathrm{eff}$ degrees of freedom is used 
to approximate the distribution with correlations between bins. Here (from 
top to bottom in the figure on the left) $N_\mathrm{eff} = 61900, 3709,$ and 
$351$ are used to represent the $64^3 = 2.6\times 10^5$ degrees of freedom
with various degrees of correlation as the result of the smoothing (see the main
text for the values of the figure on the right). The fit
provides a good approximation, especially at the small $p$-values that
are of the greatest interest for this application.
}
\label{extremeDist}
\end{figure}

To demonstrate the validity of our Ansatz
in a simplified context very similar to the case of interest,
we generate a three-dimensional periodic cubic lattice filled with $64^3$ 
independent realizations of a normal Gaussian random variable. This cube
is then smoothed using a Gaussian smoothing kernel with 
widths $\sigma _\mathrm{smoothing}=2, 5,$ and $10$. The smoothed cube is rescaled
so that the variable at each lattice point has unit variance.
For each smoothing width, the extreme value statistic (maximum) is taken 
for $10^6$ realizations and only the greatest $5\% $ of the extreme values 
are retained. Fig.~\ref{extremeDist} (left) shows the empirical CDF for the 
extreme values, which are compared to the functional form of the Ansatz for the 
best-fit values of $N_\mathrm{eff}$ according to
the approximation given in~(\ref{evCDFapprox}).

The above discussion demonstrates that an extreme value distribution
for $N_\mathrm{eff}$ independent Gaussian variates can be used to approximate
the distribution for $N_\mathrm{bin}$ variates with correlations due to
smoothing. However, the geometry of the allowed bins is complicated
and must be taken into account. Thus simulations must be carried out
to determine $N_\mathrm{eff}$ for a given smoothing scale in the more realistic
case of the actual bins used for the bispectrum analysis. We show that the 
Ansatz is still a good approximation also in this case in 
Fig.~\ref{extremeDist} (right). We generate random
numbers in the domain of definition of the binned bispectrum, and
smooth it as for the real data. Restricting to those bin triplets that
contain enough valid data (see the discussion in
Section~\ref{smoothingsec}), these simulations are a good
approximation to a Gaussian CMB map's bispectrum.

We now illustrate this method by applying it to a realistic situation. 
We add a point source map to a simulated Gaussian CMB map with anisotropic
noise, generated as described in the caption of
Table~\ref{table_filling_in}. The point source simulation was created
with the Planck Sky Model, at 143~GHz, with a beam with a FWHM of 5
arcmin, and contains faint infrared sources, as described
in~\cite{Delabrouille:2012ye}, and faint radio sources with the
improved parameters described in~\cite{Ade:2015via}. The galactic and
point source masks were applied as described in Section~\ref{masking}.

The binned bispectrum of this map was evaluated applying the linear
correction and the filling in procedure, and the $f_{\rm NL}$'s were
determined individually for each of the templates described in
Section~\ref{sec_templates}.  The unclustered point
source contribution was detected with high significance in this
contaminated map: $b_{\rm ps} = (64.8\pm 0.8)\times 10^{-29}$. This
signal is much stronger than the one detected in the {\em cleaned} Planck maps, 
but of the same order of magnitude as the forecast at
217~GHz (see \cite{Ade:2015ava}).  No statistically significant
detection of a nonzero $f_{\rm NL}$ was obtained for the other
templates, with the exception of the CIB template.  But the CIB
bispectrum has significant overlap with the unclustered point
source bispectrum (see Table~\ref{tab_corr_coeff}), so this result is
not surprising.  The nonzero result for $b_\mathrm{CIB}$ disappears in a joint 
analysis of the unclustered point source and CIB templates.
Finally we smooth the bispectrum with a few different values of the smoothing
length, namely $\sigma_{\rm smoothing}=1,2,3$. 

Apart from studying the contaminated bispectrum, we can also try to remove the 
estimated point source contribution from the measured binned bispectrum, 
simply by subtracting the corresponding smoothed template (\ref{Bunclust}) 
with the measured amplitude. We can then check if there are remaining 
non-Gaussian features in this cleaned bispectrum using the method described 
above.
Table~\ref{ataTable} gives the maximum bin values for the
bispectrum before and after the template cleaning is applied. The
minima are not given because the inclusion of the point sources tends
to gives a positive bispectral contribution. The $p$-values were calculated 
using simulations of $10^6$ Gaussian realizations and fitting the CDF for 
$N_\mathrm{eff}$ to the empirical distribution.  We found that the smoothing 
lengths $\sigma_\mathrm{smoothing} =1, 2,$ and $3$ correspond to 
$N_\mathrm{eff}=8664, 1948,$
and $588$ respectively.
We see that a highly statistically significant detection is found using the 
above procedure on the uncleaned bispectrum. We also observe that the 
template cleaning
procedure is successful; however, some detectable unsubtracted residual remains.
This residual, however, has little overlap with the known 
theoretically motivated primordial templates.

\begin{table}
\begin{center}
\begin{tabular}{c|llll}
\hline
Smoothing  & \multicolumn{2}{c}{Before} & \multicolumn{2}{c}{After}\\
length     & \multicolumn{2}{c}{template cleaning} 
& \multicolumn{2}{c}{template cleaning}\\
($\sigma_\mathrm{smoothing}$) &   $X_{max}$ & $p$-value  &  $X_{max}$ & $p$-value \\
\hline
1 & 33.4 & $e^{-553}$ & 5.1 & $1.5 \times 10^{-3}$ \\
2 & 51.2 & $e^{-1308}$ & 3.6 & 0.15 \\
3 & 59.2 & $e^{-1751}$ & 4.4 & $3.3 \times 10^{-3}$  \\
\hline
\end{tabular}
\end{center}
\caption{Maximum bin values and associated $p$-values for the bispectrum of
a simulated Gaussian CMB + point sources map before and after cleaning of the
point source contribution, for various values of the smoothing 
length. See main text for a more detailed explanation.}
\label{ataTable}
\end{table}

\section{Conclusions}
\label{conclusionsec}

This paper gives a detailed description of the binned bispectrum
estimator, which is one of the three estimators used for the official
2013 and 2015 Planck analyses. The estimator determines the full
three-dimensional bispectrum of a map, binned in harmonic space. This
binned bispectrum can then be combined with a library of theoretical
bispectrum templates to determine the so-called $f_\mathrm{NL}$
parameters. This aspect of the binned bispectrum estimator was first
described in~\cite{Bucher:2009nm}, but many new developements took
place since that paper was published and are detailed here. These
include the treatment of the masked sky using diffusive filling in, a
method which was subsequently adopted by the other bispectrum
estimators, and the generalization to include polarization.
The binned bispectrum estimator code is fast, has a convenient modular
structure, separability of the templates is not required, and the dependence 
of $f_\mathrm{NL}$ on $\ell$ is obtained automatically, thus providing 
additional information.

Moreover, the binned bispectrum of the map can also be used directly for
non-parametric (blind) non-Gaussianity searches. This is useful to investigate
the non-Gaussianity of those foregrounds for which there is no theoretically
motivated template, or to check for any non-Gaussian primordial signal in a
cleaned map beyond the standard templates. For this purpose the binned
bispectrum is first smoothed with a Gaussian kernel to increase the 
statistical significance of any broad features and make them stand out
above the noise. The smoothing complicates the statistical analysis of
the significance of non-Gaussian features by introducing correlations between
neighbouring bins. In this paper we described how to address this complication
using approximated extreme value statistics. We illustrated this procedure
on the bispectrum of a realistic map that contains a Gaussian CMB
and radio point sources, both before and after cleaning the bispectrum with
the point source template.

Further applications of the binned bispectrum estimator can be found in
the Planck analyses of 2013~\cite{Ade:2013ydc} and 2015~\cite{Ade:2015ava}.
In those papers we determined the $f_\mathrm{NL}$ values for various theoretical
templates, and showed the full smoothed binned bispectrum of the cleaned Planck
sky map. A statistical analysis of the Planck smoothed bispectrum according
to the method described here will be given in the next Planck release.
Applications of the binned bispectrum estimator to foreground maps in order
to determine their smoothed bispectra and characterize their non-Gaussianities
are planned for a future paper. These bispectra can then be used as templates
in a parametric estimation to investigate if contamination remains in cleaned
maps.

To conclude we add some comments regarding the possibility of extending our
binned estimator methodology to the trispectrum. The reduced trispectrum 
depends on five scalar variables, which we can represent in the following
way. First there are the lengths of the sides of the quadrilateral
$\ell_1$, $\ell_2$, $\ell_3$, and $\ell _4$. Since the quadrilateral must close,
we can write (in the flat-sky approximation for simplicity):
$\boldsymbol{\ell}_1+ \boldsymbol{\ell}_2+\boldsymbol{\ell}_3
+\boldsymbol{\ell}_4=\boldsymbol{0}$.
But unlike in the case of a triangle, for which the lengths of the sides
create a rigid polygon with all angles between the sides determined,
for the quadrilateral an `internal brace' is needed to render it rigid.
We must additionally specify one more parameter, which we could for example 
take to be $\ell_{12}=|\boldsymbol{\ell}_1+ \boldsymbol{\ell}_2|$,
and now the quadrilateral would be completely rigid, at least in the 
two-dimensional context that is of interest here. 
To generalize the technique using filtered maps developed in this paper
for the bispectrum to the trispectrum, we cannot simply multiply four filtered
maps and take the average. We must, with the parameterization used above 
for the trispectrum, after taking the products of the filtered $\ell_1$ and 
$\ell_2$ maps, apply the $\ell_{12}$ filter before multiplying by the 
filtered $\ell_3$ and $\ell_4$ maps, and then take the average in the final 
step. In other words, writing $F_1$ to indicate filtering with respect to 
the $\ell_1$ bin, etc., and $M$ for the CMB map:
\be
\mathrm{Trispectrum}_{i_1 i_2 i_3 i_4 i_{12}} = 
\int d\hat{\Omega} \, F_{12}(F_1(M)F_2(M)) F_3(M)F_4(M).
\ee
This extension will be explored in a future paper.

\vspace{0.5cm}

{\bf Acknowledgements:} We thank Ata Karakci for
providing the simulated radio point source maps generated using the
Planck Sky Model, Guillaume Patanchon for useful discussions, and
Franz Elsner for useful remarks regarding the lensing-ISW template.
The Planck non-Gaussianity working group, in particular the estimator subgroup
consisting of James Fergusson, Michele Liguori, Alessandro Renzi, and BvT, 
provided
%was 
a stimulating environment for estimator development.
% that benefitted all its participants.
We acknowledge the use of the CAMB ({\tt http://camb.info}) and 
Healpy (Python version of HEALPix~\cite{Gorski:2004by}) software packages.
We gratefully acknowledge the IN2P3 Computer 
Centre ({\tt http://cc.in2p3.fr}) for providing 
computing resources.
BR acknowledges funding from the French Minist\`ere de l'enseignement sup\'erieur 
et de la recherche and from the Research Council of Norway.

\appendix

\section{Two-dimensional sections of the smoothed theoretical bispectra} 
\label{App:AppendixA}

\begin{figure}
\centering
\includegraphics[trim = 3mm 0mm 12mm 0mm,clip,width=0.19\columnwidth]{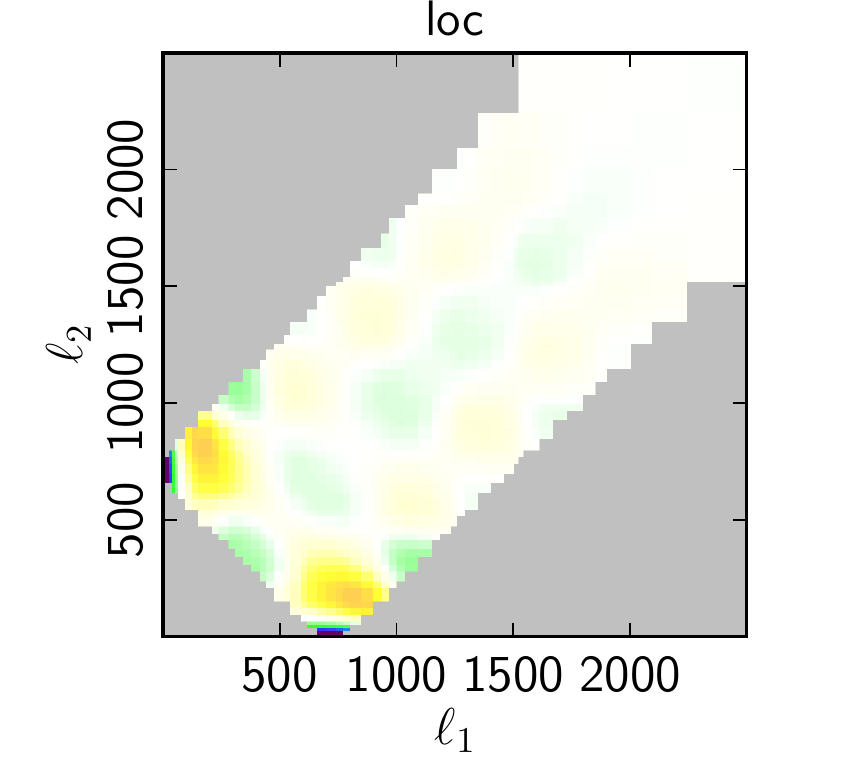}
\includegraphics[trim = 3mm 0mm 12mm 0mm,clip,width=0.19\columnwidth]{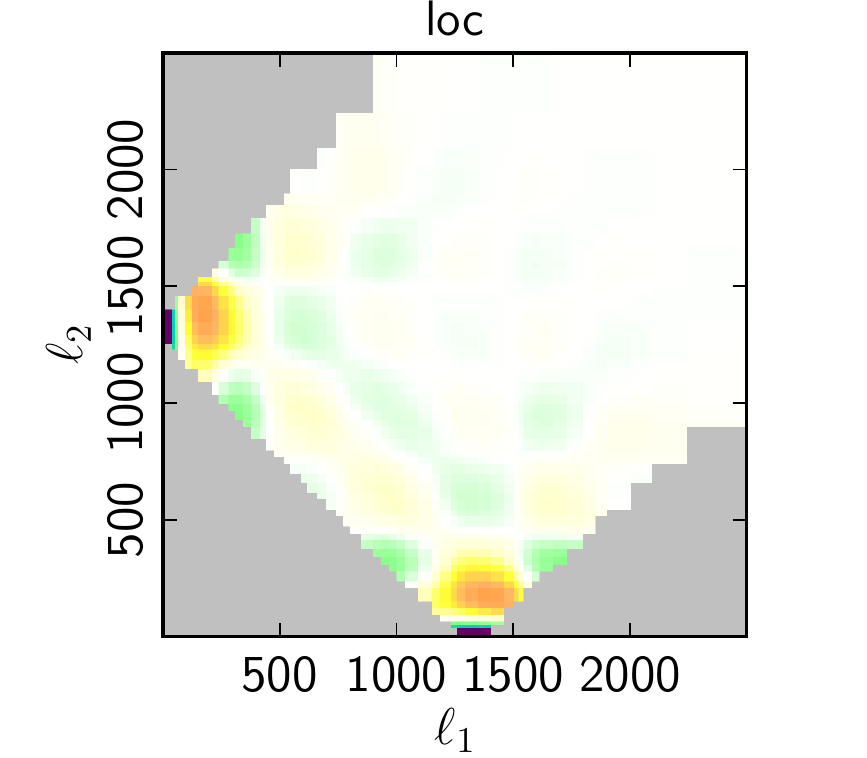}
\includegraphics[trim = 3mm 0mm 12mm 0mm,clip,width=0.19\columnwidth]{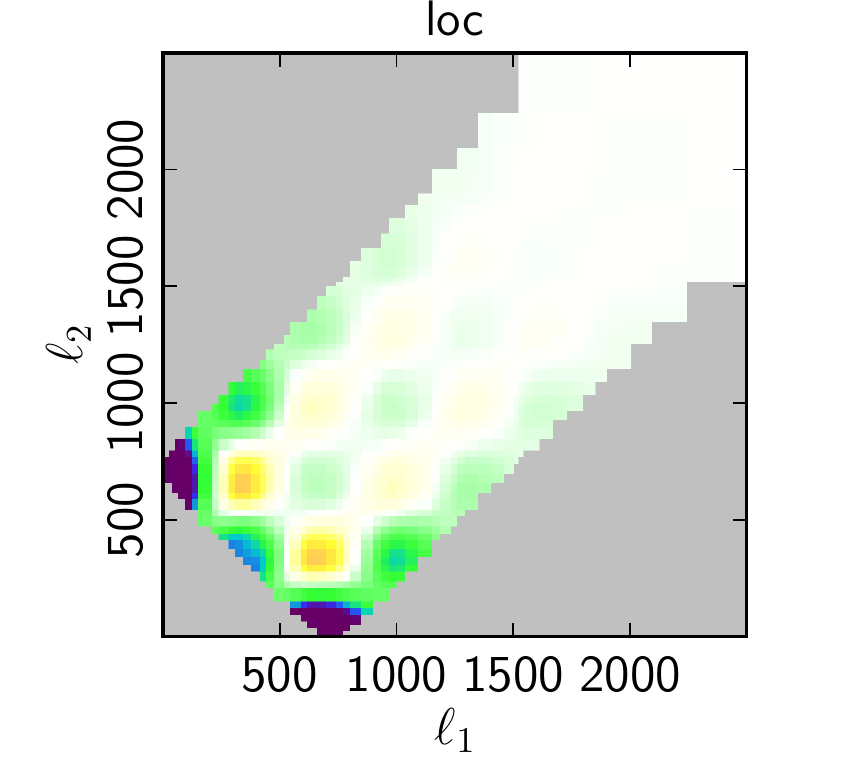}
\includegraphics[trim = 3mm 0mm 12mm 0mm,clip,width=0.19\columnwidth]{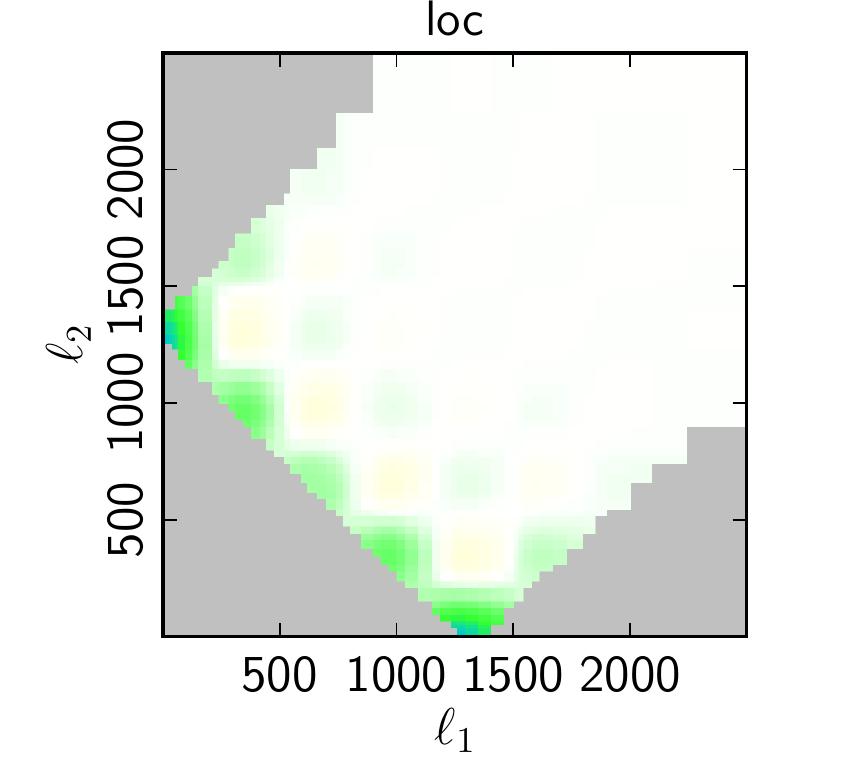}

\includegraphics[trim = 3mm 0mm 12mm 0mm,clip,width=0.19\columnwidth]{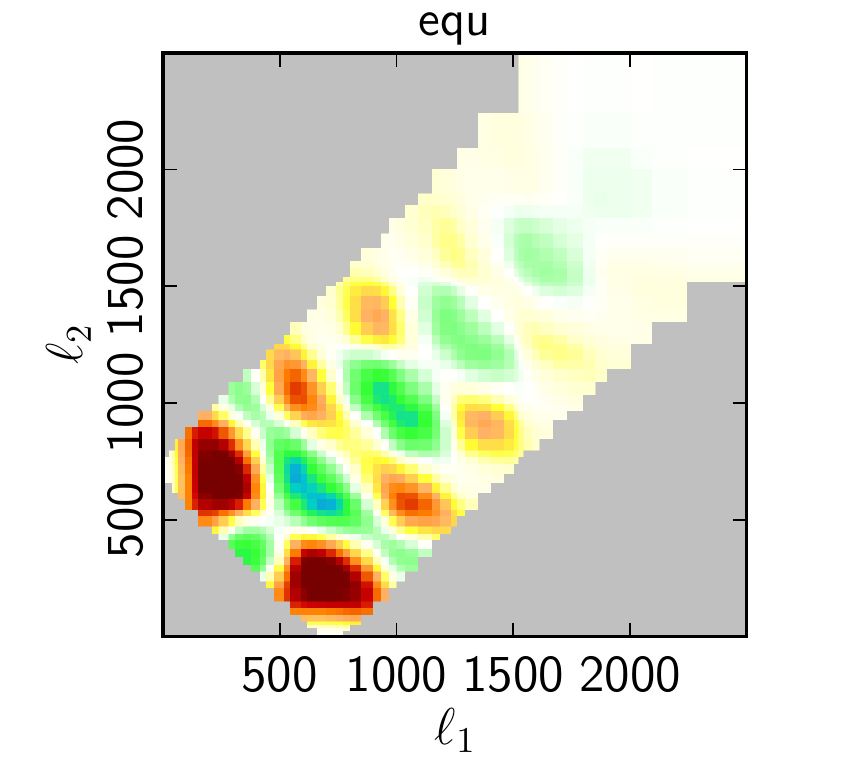}
\includegraphics[trim = 3mm 0mm 12mm 0mm,clip,width=0.19\columnwidth]{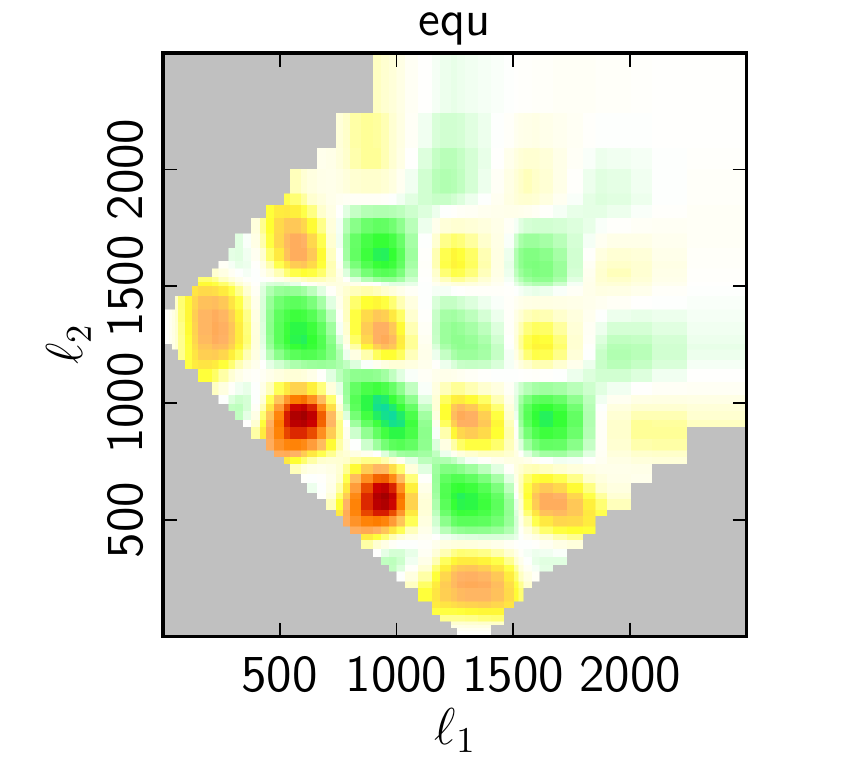}
\includegraphics[trim = 3mm 0mm 12mm 0mm,clip,width=0.19\columnwidth]{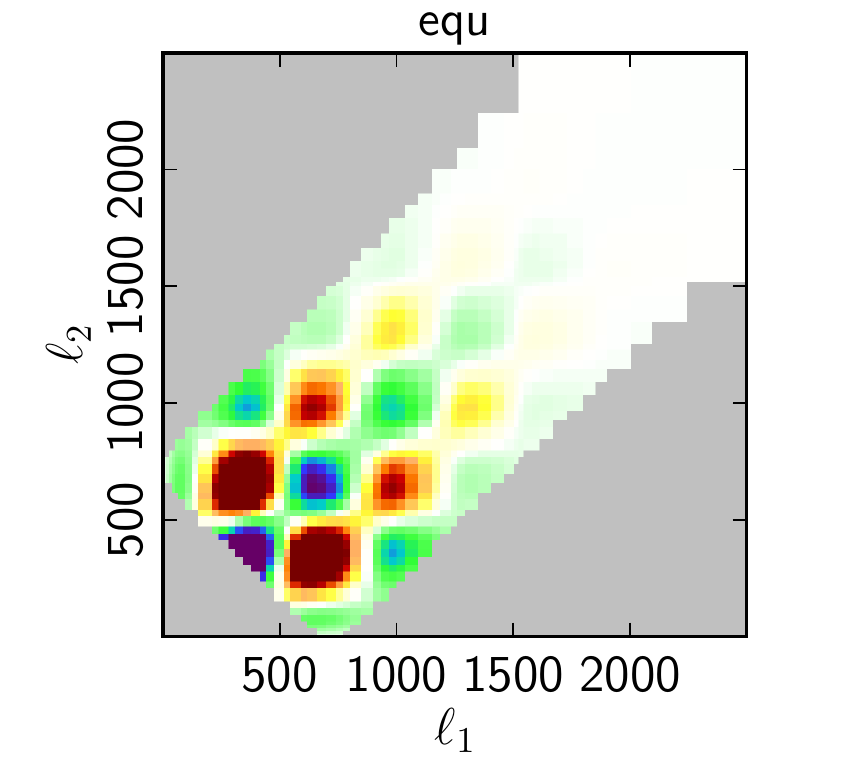}
\includegraphics[trim = 3mm 0mm 12mm 0mm,clip,width=0.19\columnwidth]{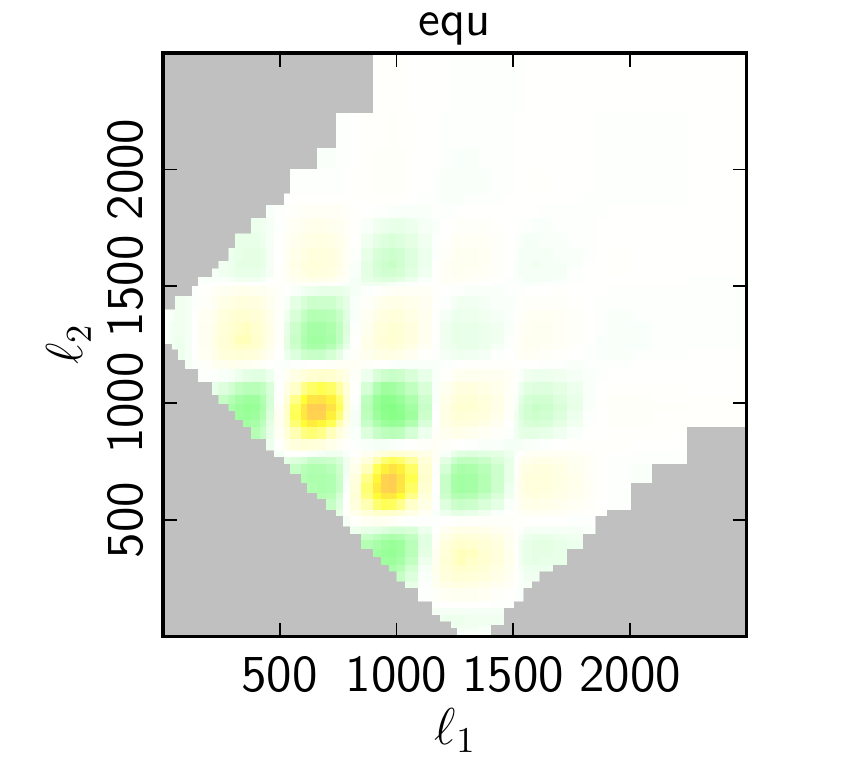}

\includegraphics[trim = 3mm 0mm 12mm 0mm,clip,width=0.19\columnwidth]{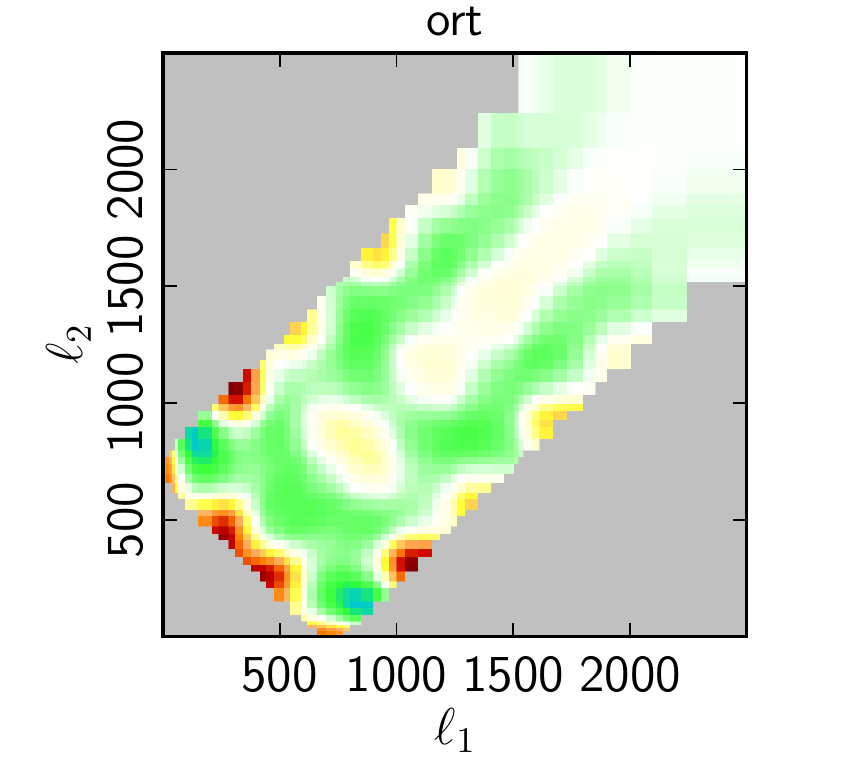}
\includegraphics[trim = 3mm 0mm 12mm 0mm,clip,width=0.19\columnwidth]{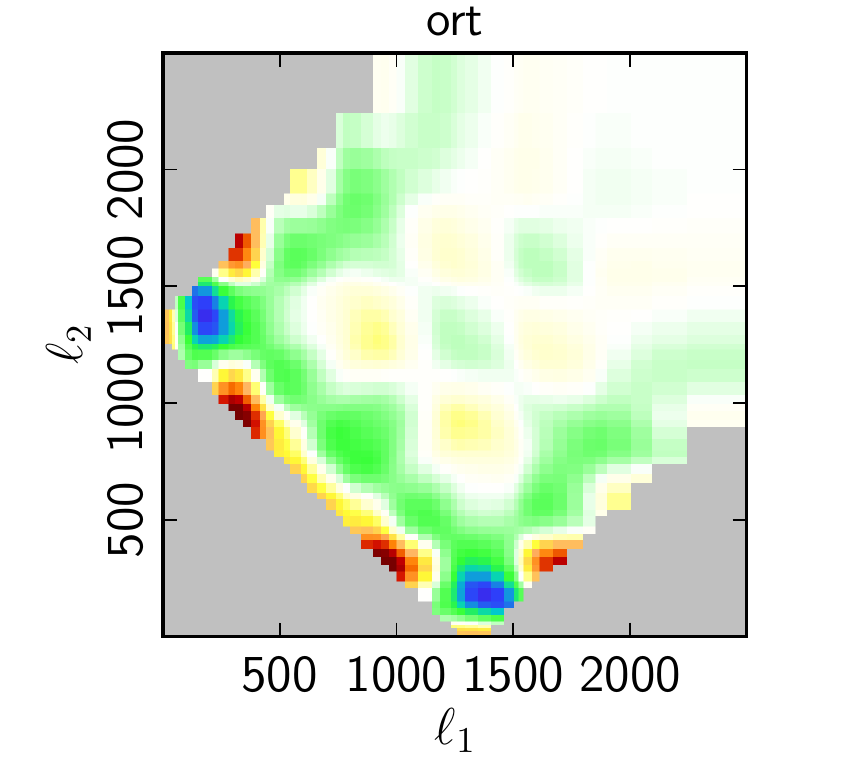}
\includegraphics[trim = 3mm 0mm 12mm 0mm,clip,width=0.19\columnwidth]{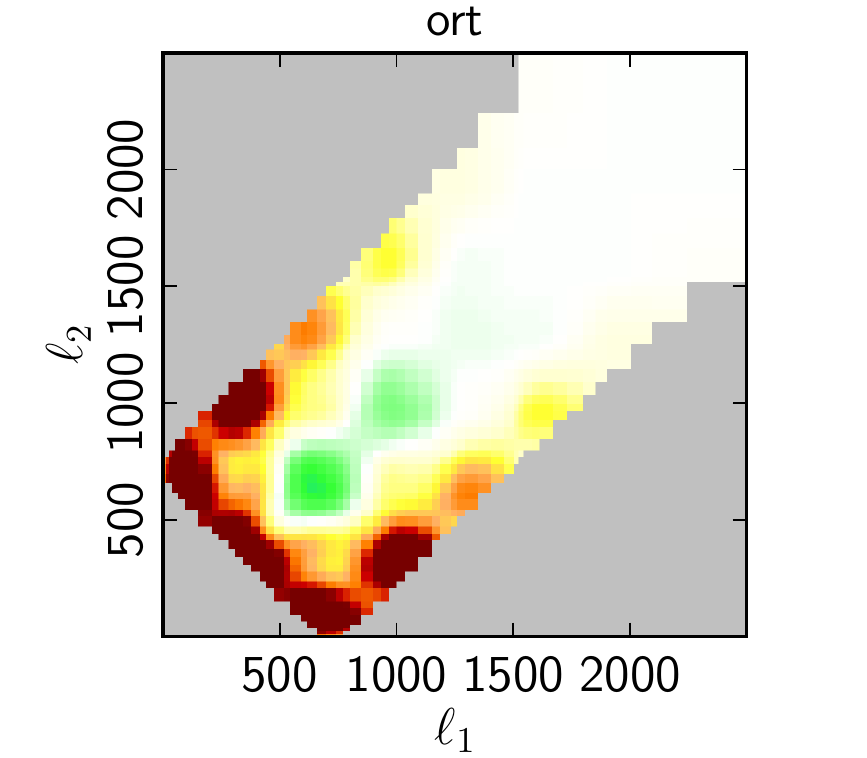}
\includegraphics[trim = 3mm 0mm 12mm 0mm,clip,width=0.19\columnwidth]{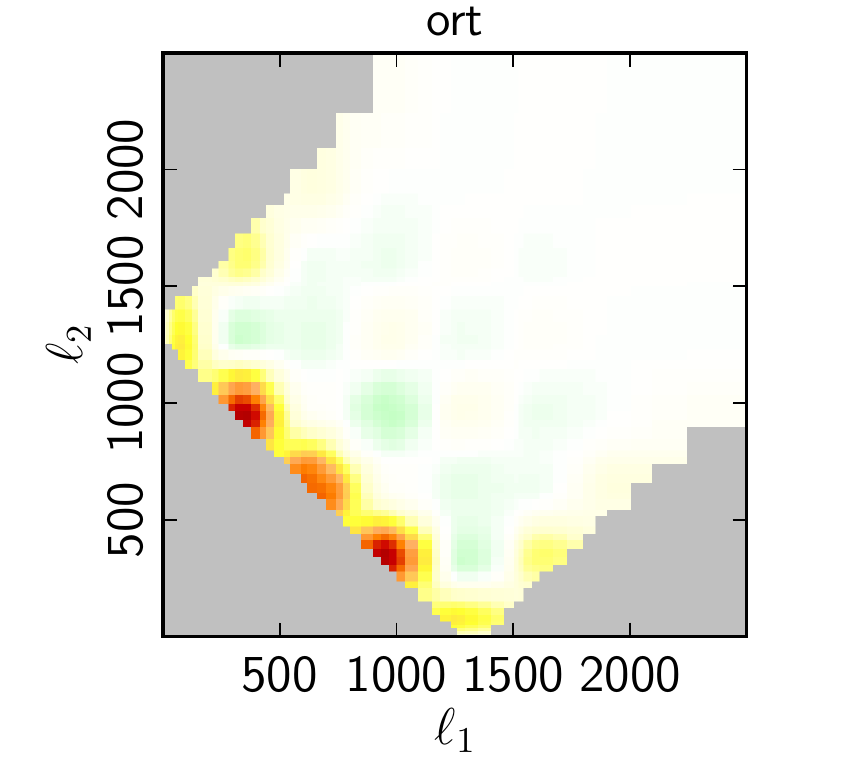}

\includegraphics[trim = 3mm 0mm 12mm 0mm,clip,width=0.19\columnwidth]{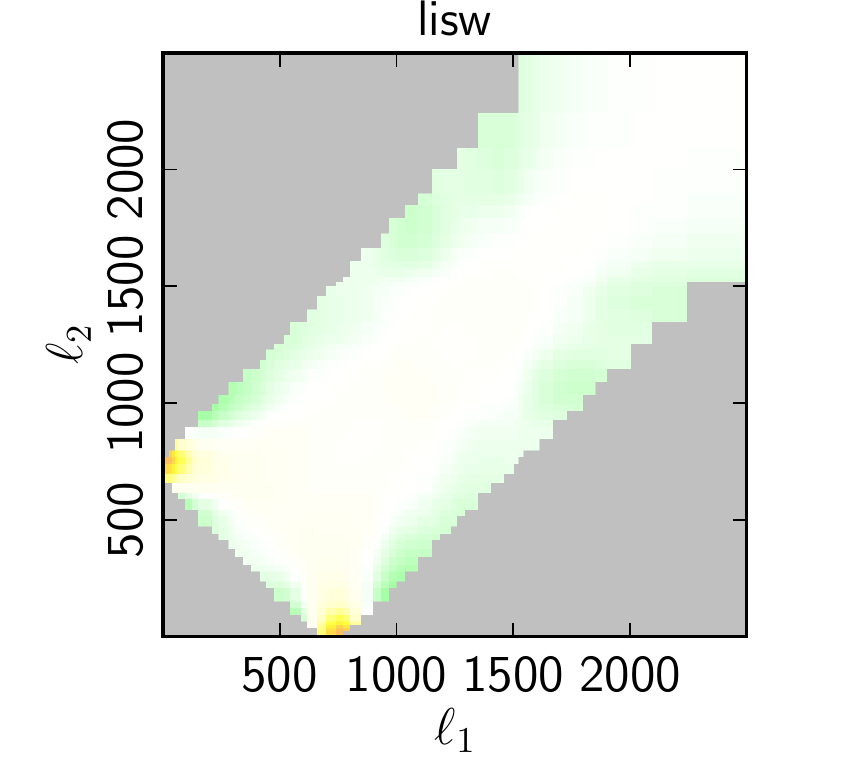}
\includegraphics[trim = 3mm 0mm 12mm 0mm,clip,width=0.19\columnwidth]{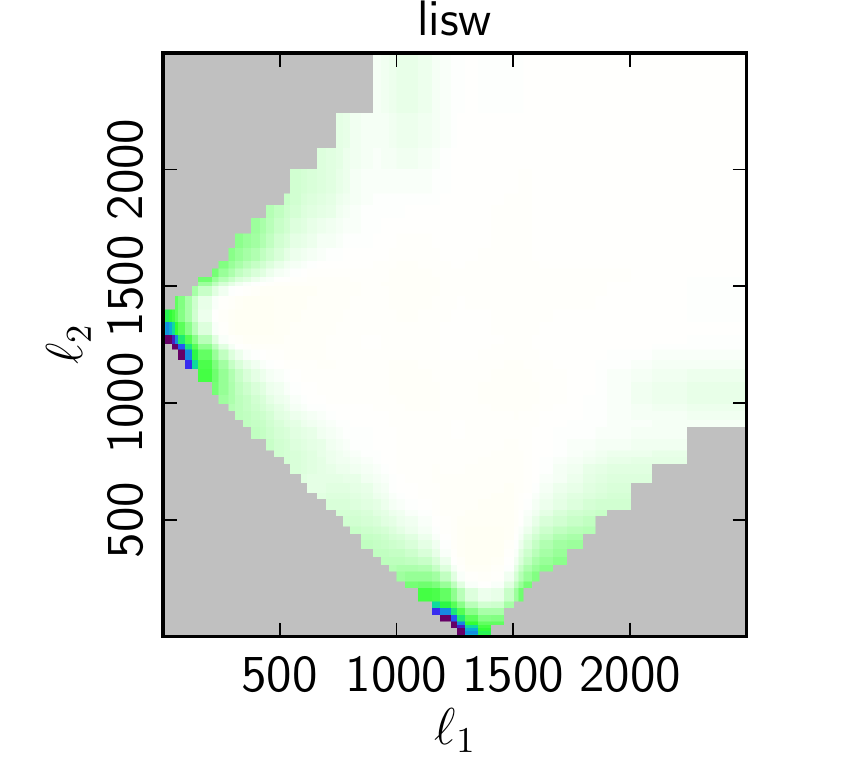}
\includegraphics[trim = 3mm 0mm 12mm 0mm,clip,width=0.19\columnwidth]{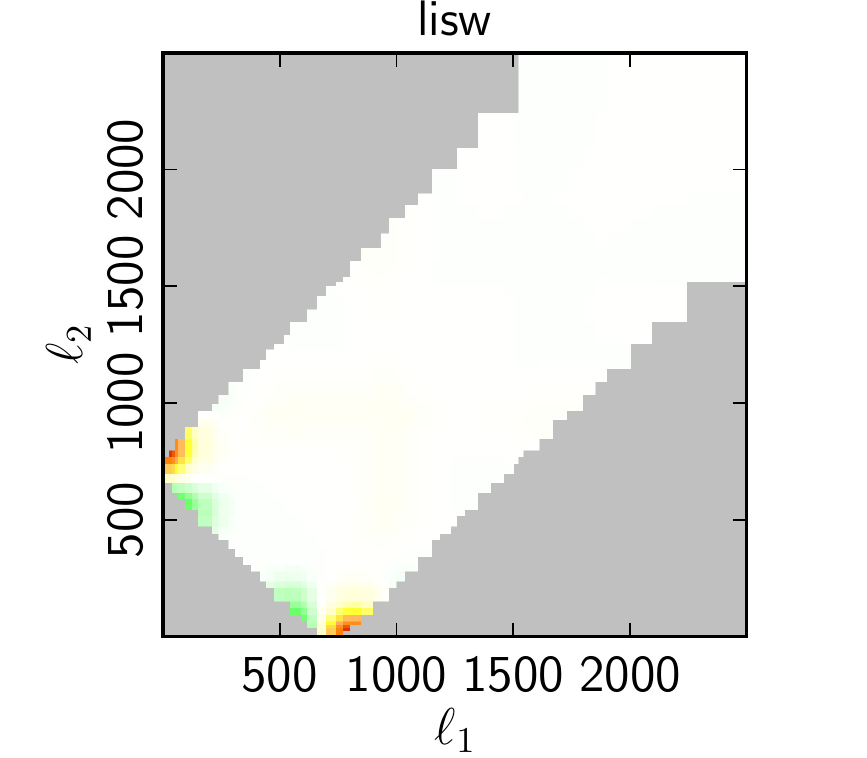}
\includegraphics[trim = 3mm 0mm 12mm 0mm,clip,width=0.19\columnwidth]{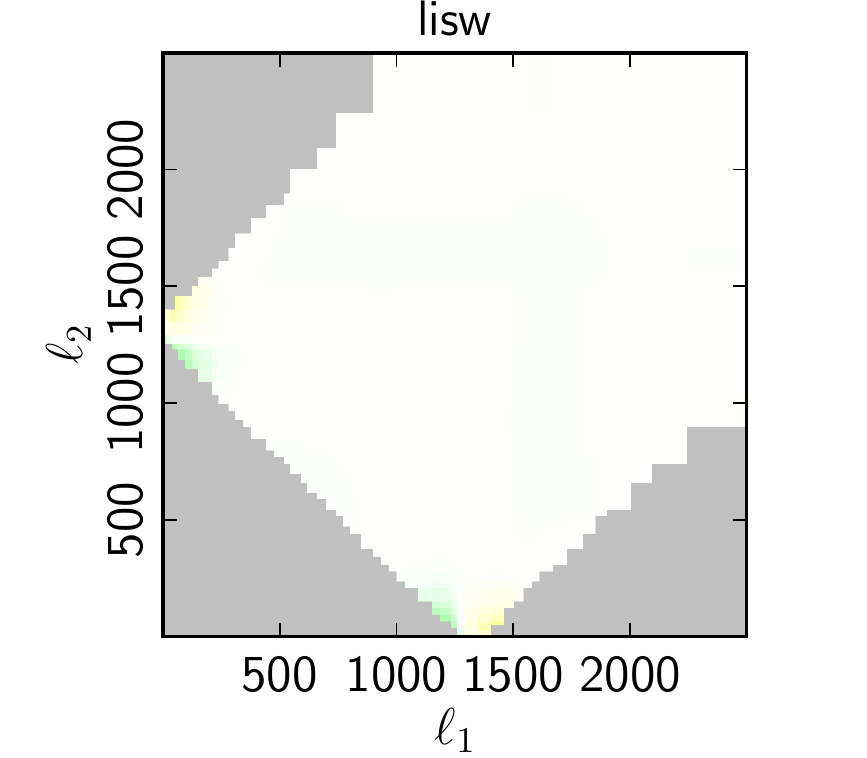}

\includegraphics[trim = 3mm 0mm 12mm 0mm,clip,width=0.19\columnwidth]{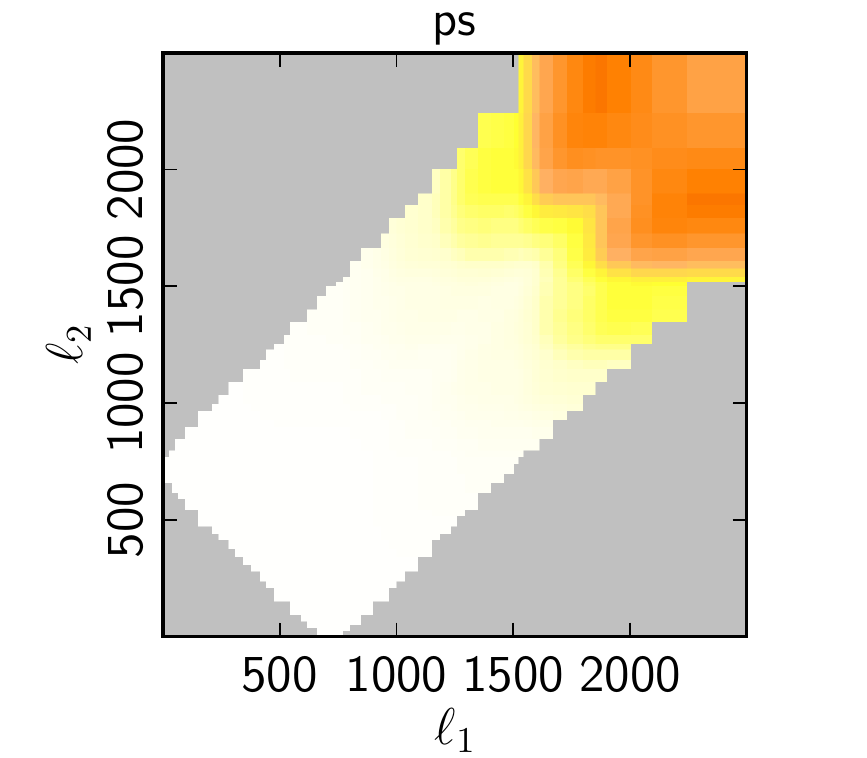}
\includegraphics[trim = 3mm 0mm 12mm 0mm,clip,width=0.19\columnwidth]{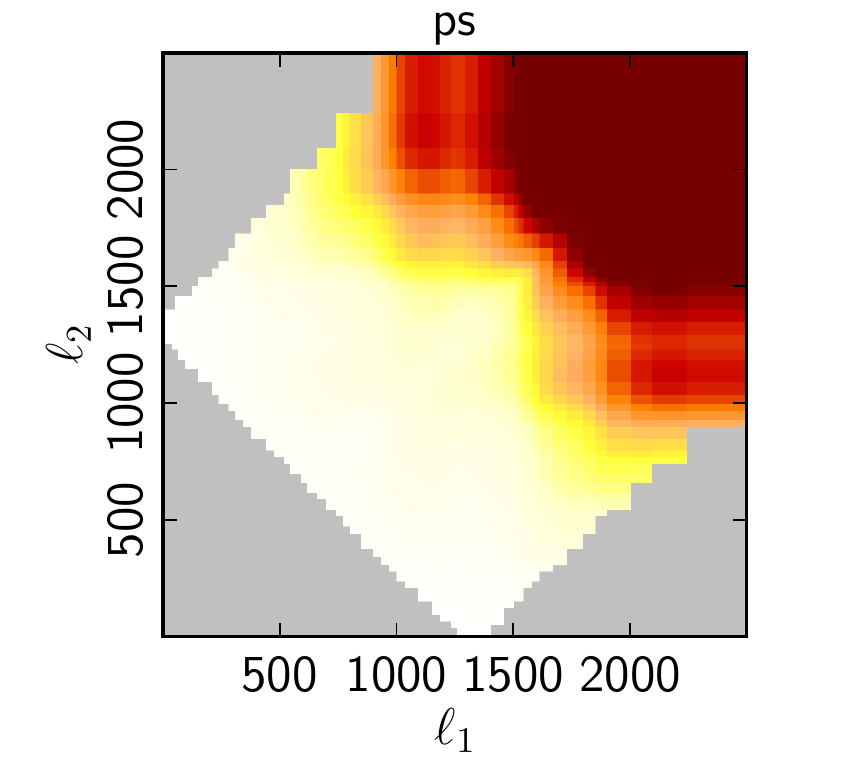}
\includegraphics[trim = 3mm 0mm 12mm 0mm,clip,width=0.19\columnwidth]{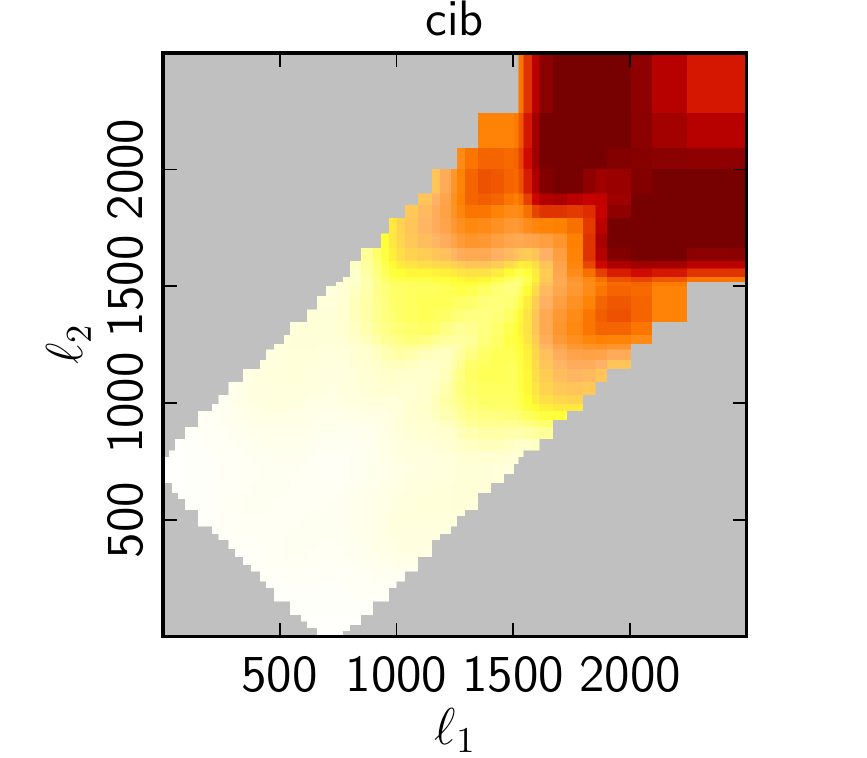}
\includegraphics[trim = 3mm 0mm 12mm 0mm,clip,width=0.19\columnwidth]{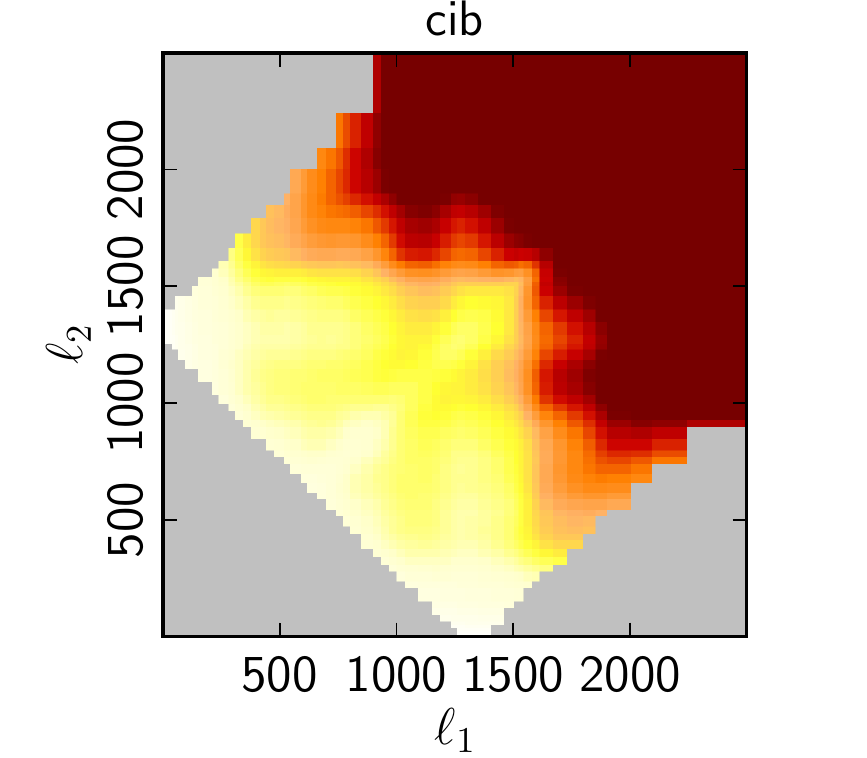}
\includegraphics[trim = 0mm 0mm 25mm 10mm,clip,width=0.75\columnwidth]{figures/colorbars_lin_signifrenorm_martincolor.pdf}

\caption{Smoothed theoretical templates for (from top to bottom) the local, 
equilateral, orthogonal,
lensing-ISW, unclustered point sources, and CIB (clustered) point sources 
shapes, as defined in Section~\ref{sec_templates}. The templates are normalized
by the expected standard deviation of the bispectrum (using the beam and noise
characteristics of Section~\ref{masking}), so that a dimensionless 
signal-to-noise bispectrum results. Moreover, they have all been multiplied
by such $f_\mathrm{NL}$ values as would give a $30\sigma$ detection.
The first two columns are for $TTT$, respectively for $\ell \in$ [700, 741] and 
$\ell \in$ [1291, 1345], while the two last columns show the same 
$\ell$ ranges, but for $EEE$. Except for the last line, which contains both
point source templates for $TTT$ only.}
\label{fig:theor_smoo}
\end{figure}

To illustrate the smoothed signal-to-noise bispectrum $\mathcal{B}$ that we 
introduced in Section~\ref{smoothingsec} and to show what certain known
types of non-Gaussianity look like, we present here slices of the 
theoretical template bispectra that we discussed in 
Sections~\ref{sec_templates} and~\ref{sec_isocurv}.
In Fig.~\ref{fig:theor_smoo} we show the normalized smoothed templates
(with smoothing length $\sigma_\mathrm{bin}=2$)
for six theoretical shapes: local, equilateral, orthogonal, lensing-ISW,
unclustered point sources, and CIB (clustered) point sources. All are shown 
both for $TTT$ and $EEE$, except the point source templates which are 
$TTT$-only, and for two different slices: $\ell_3 \in$ [700, 741] and 
$\ell_3 \in$ [1291, 1345]. 
The templates have been normalized with the expected standard deviation 
assuming the standard Planck cosmology and the same noise and beam 
properties as in Section~\ref{masking}.
Moreover, the templates have been multiplied by such values of $f_\mathrm{NL}$ 
as would lead to a $30\sigma$ detection given those noise and
beam properties. Obviously that means different values of $f_\mathrm{NL}$
for each template as they all have different error bars. Explicitly
these error bars $\sigma_{f_\mathrm{NL}}$ are for $TTT$: local 4.7, equilateral 61,
orthogonal 32, lensing-ISW 0.21, unclustered point sources $7.0\times 10^{-30}$,
CIB point sources $3.6\times 10^{-28}$; and for $EEE$: local 24, equilateral 178,
orthogonal 95, lensing-ISW 3.0.

In the three primordial templates, we see the bispectral acoustic
oscillations. For the local shape, as expected, the signal is peaked
in the squeezed limit (i.e.,\ for a small $\ell_1$ and a
large $\ell_2$ and $\ell_3$). The equilateral shape, on the other hand, is
peaked for configurations where all $\ell$'s are equal, although that is
difficult to see in these Figures since they show just one bin in $\ell_3$
and the value is also modulated by the acoustic oscillations.
The orthogonal shape explores configurations in the equilateral and 
flattened ($\ell_1+\ell_2 \approx \ell_3$ with $\ell_1 \approx \ell_2$)
domains, the latter being visible in the Figures, although again modulated
by the acoustic oscillations. We also see its anti-correlation with the 
local shape (see Table~\ref{tab_corr_coeff}).
The lensing-ISW signal is highly concentrated in the squeezed limit,
and hence has a large correlation with the local shape.
The point source templates are featureless and the statistical weight
is concentrated for configurations where all three $\ell$'s are large.

\begin{figure}
\centering
\includegraphics[trim =3mm 0mm 12mm 0mm,clip,width=0.19\columnwidth]{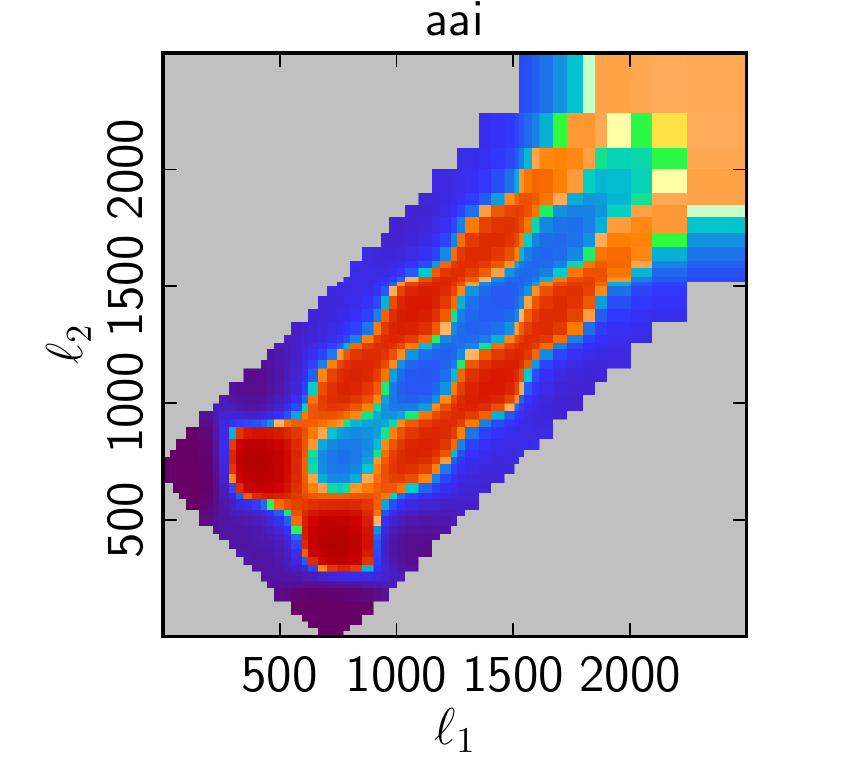}
\includegraphics[trim =3mm 0mm 12mm 0mm,clip,width=0.19\columnwidth]{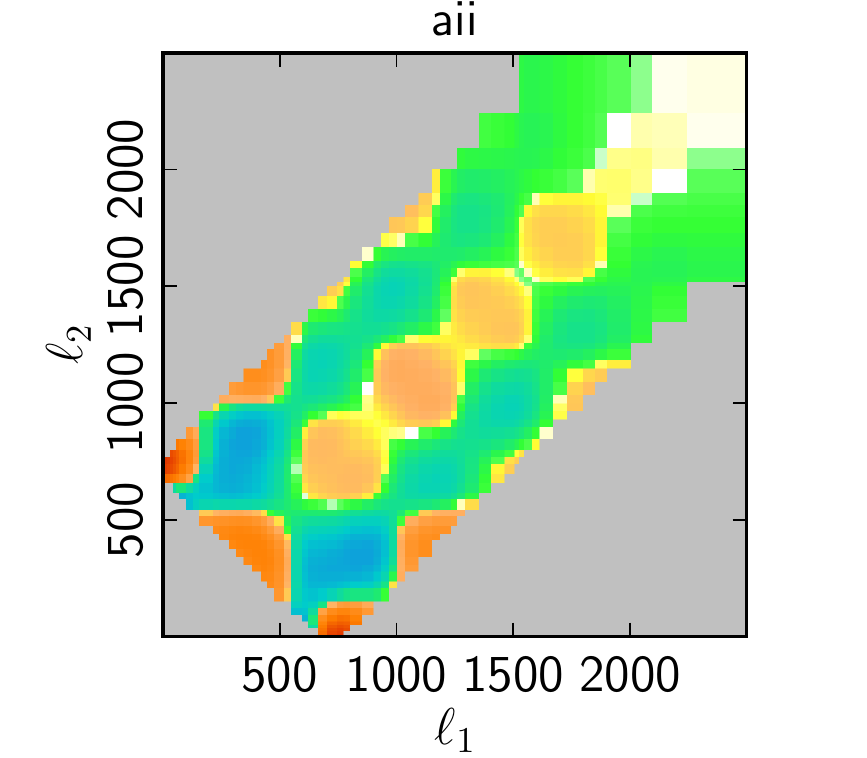}
\includegraphics[trim =3mm 0mm 12mm 0mm,clip,width=0.19\columnwidth]{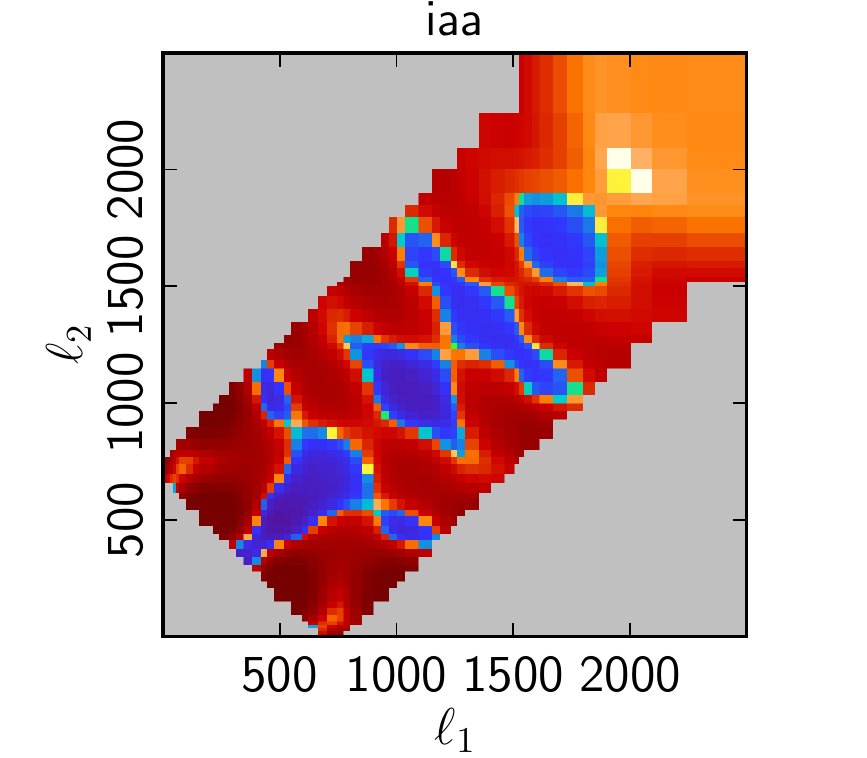}
\includegraphics[trim =3mm 0mm 12mm 0mm,clip,width=0.19\columnwidth]{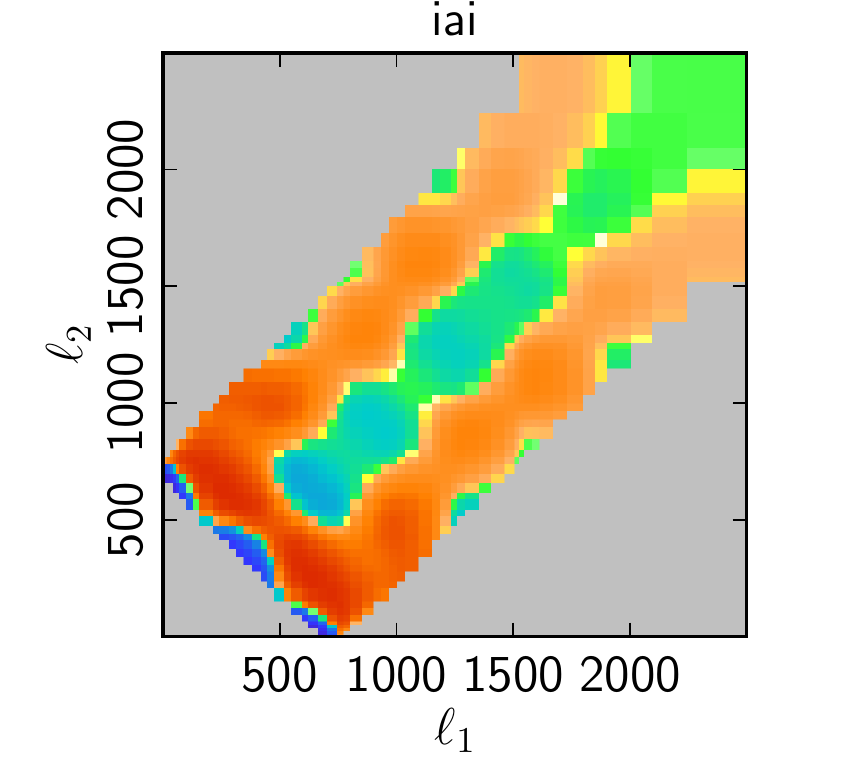}
\includegraphics[trim =3mm 0mm 12mm 0mm,clip,width=0.19\columnwidth]{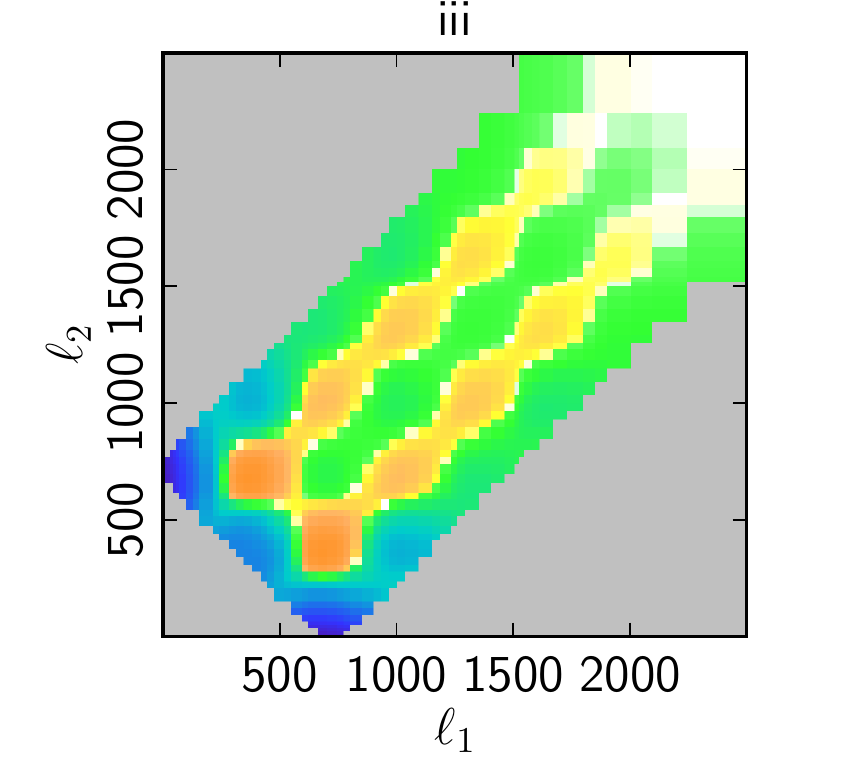}

\includegraphics[trim =3mm 0mm 12mm 0mm,clip,width=0.19\columnwidth]{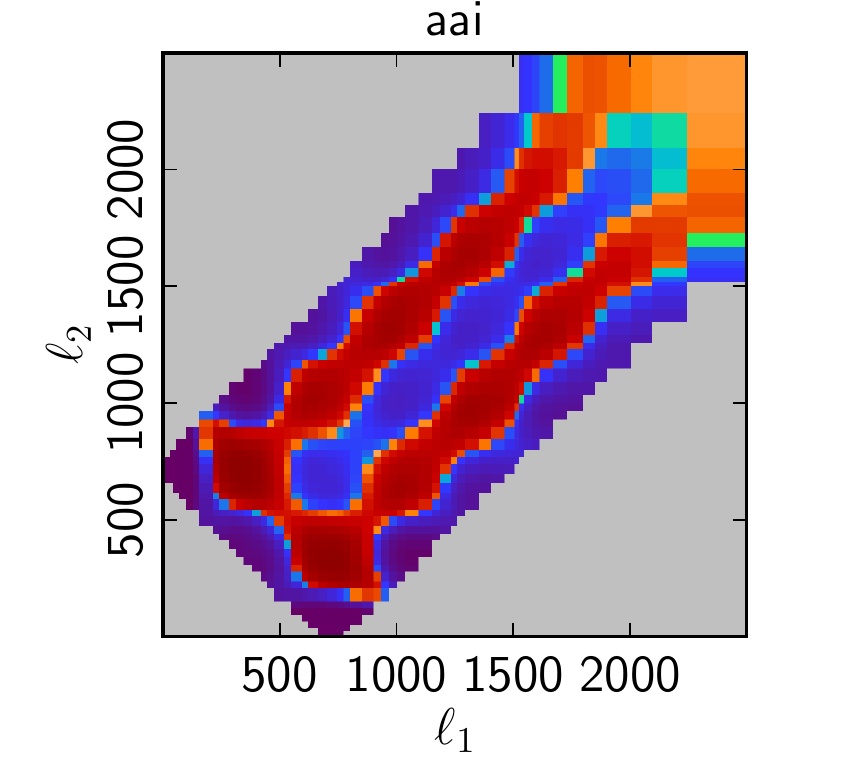}
\includegraphics[trim =3mm 0mm 12mm 0mm,clip,width=0.19\columnwidth]{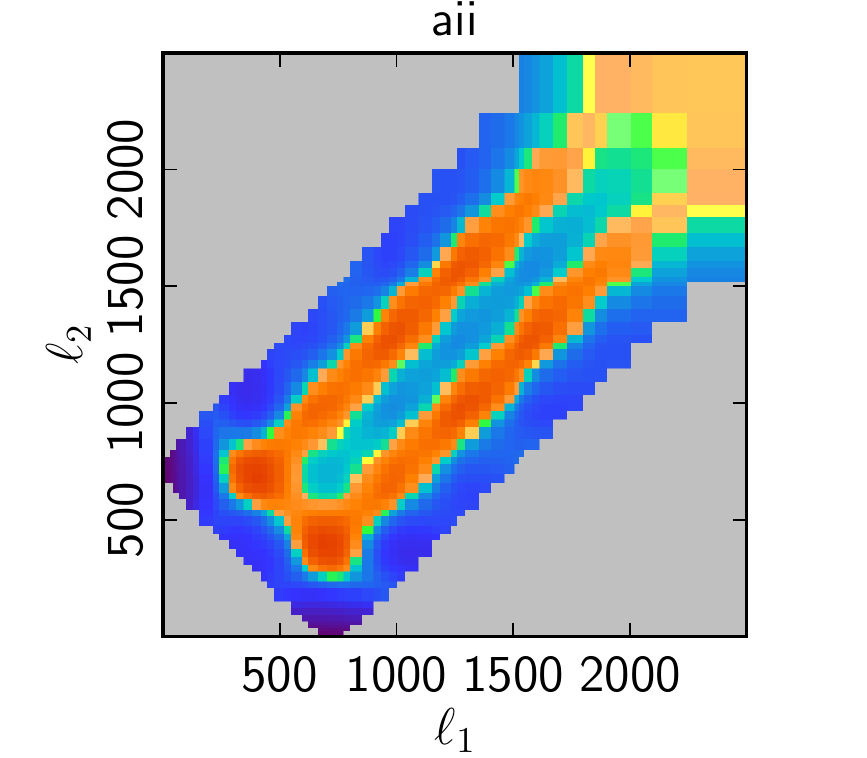}
\includegraphics[trim =3mm 0mm 12mm 0mm,clip,width=0.19\columnwidth]{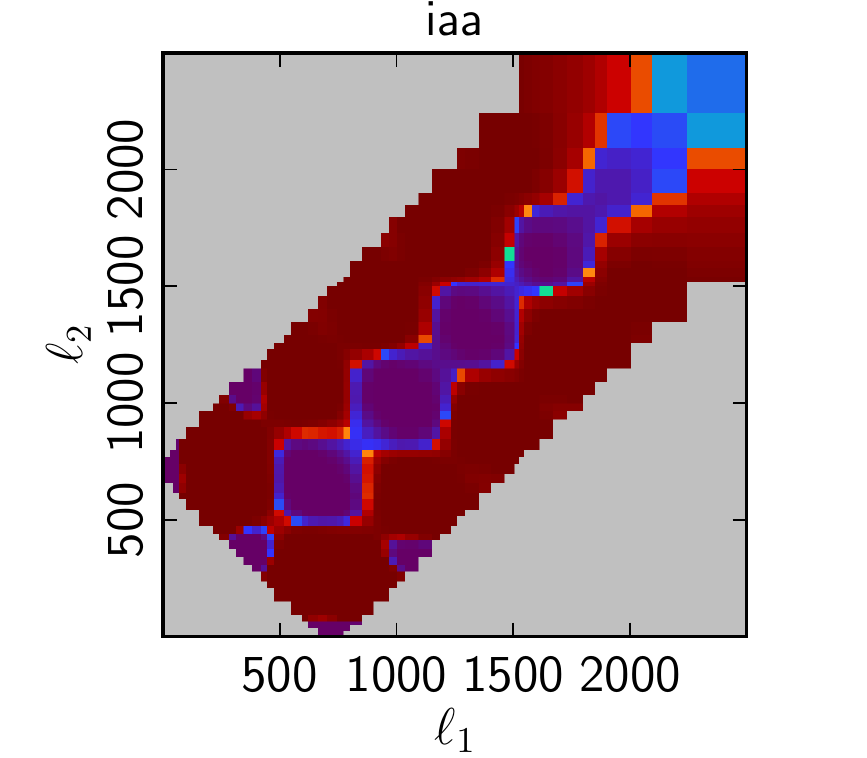}
\includegraphics[trim =3mm 0mm 12mm 0mm,clip,width=0.19\columnwidth]{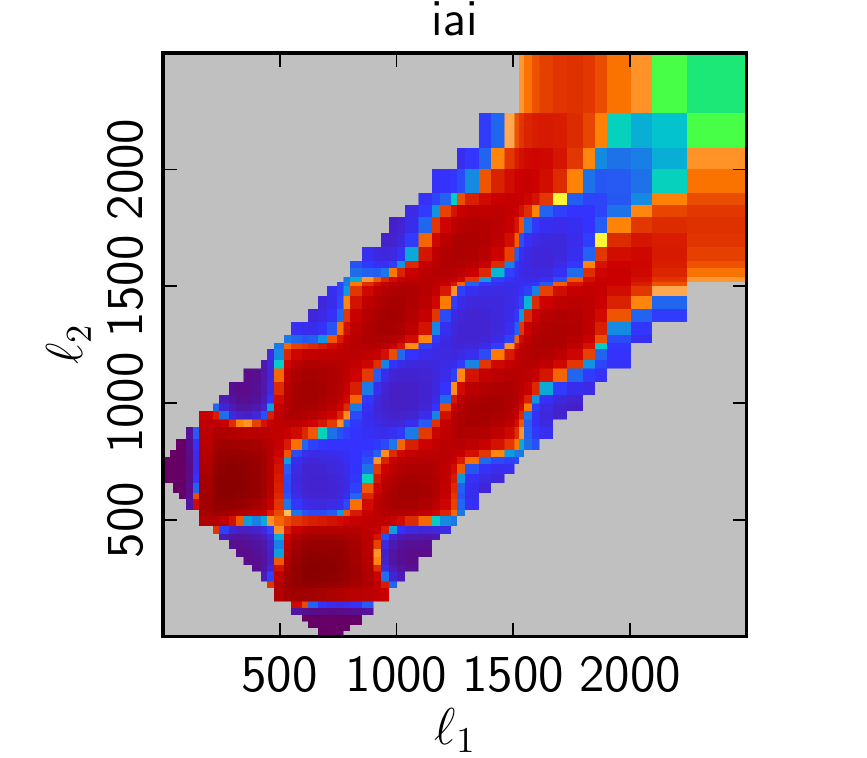}
\includegraphics[trim =3mm 0mm 12mm 0mm,clip,width=0.19\columnwidth]{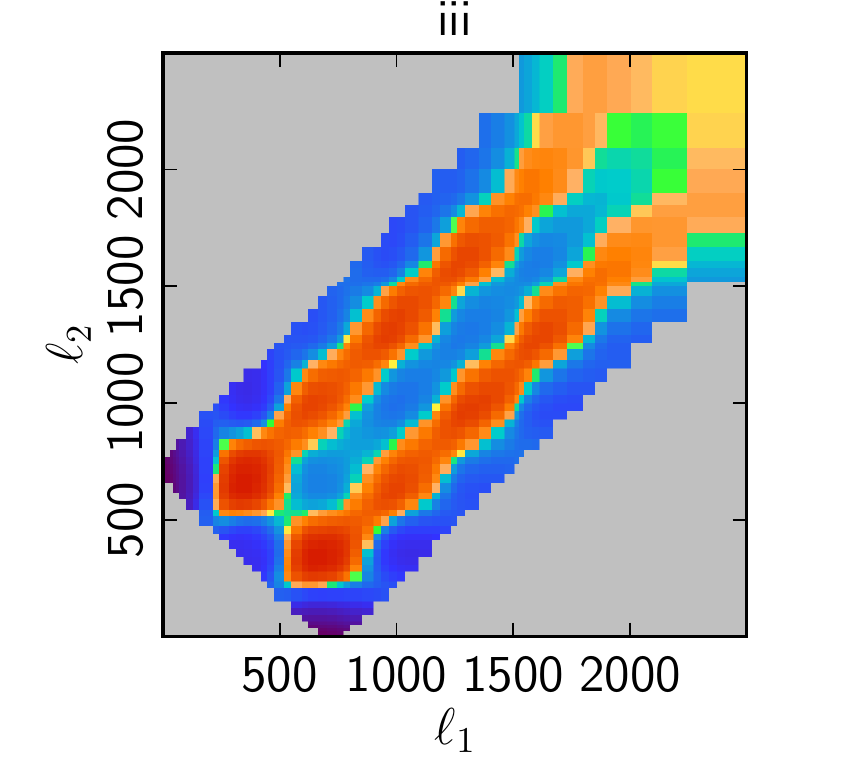}

\includegraphics[trim =3mm 0mm 12mm 0mm,clip,width=0.19\columnwidth]{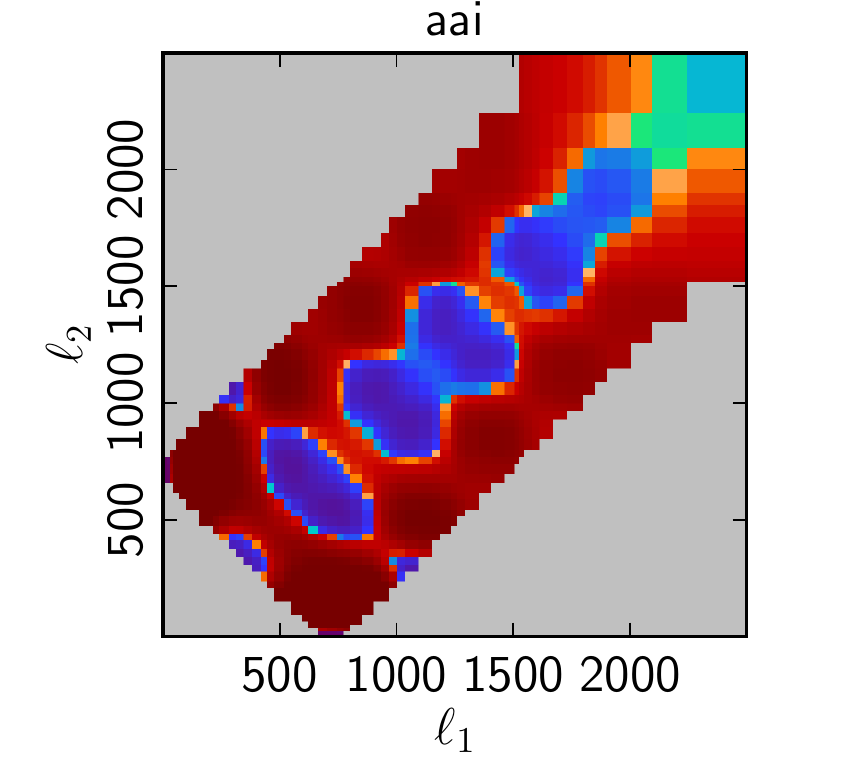}
\includegraphics[trim =3mm 0mm 12mm 0mm,clip,width=0.19\columnwidth]{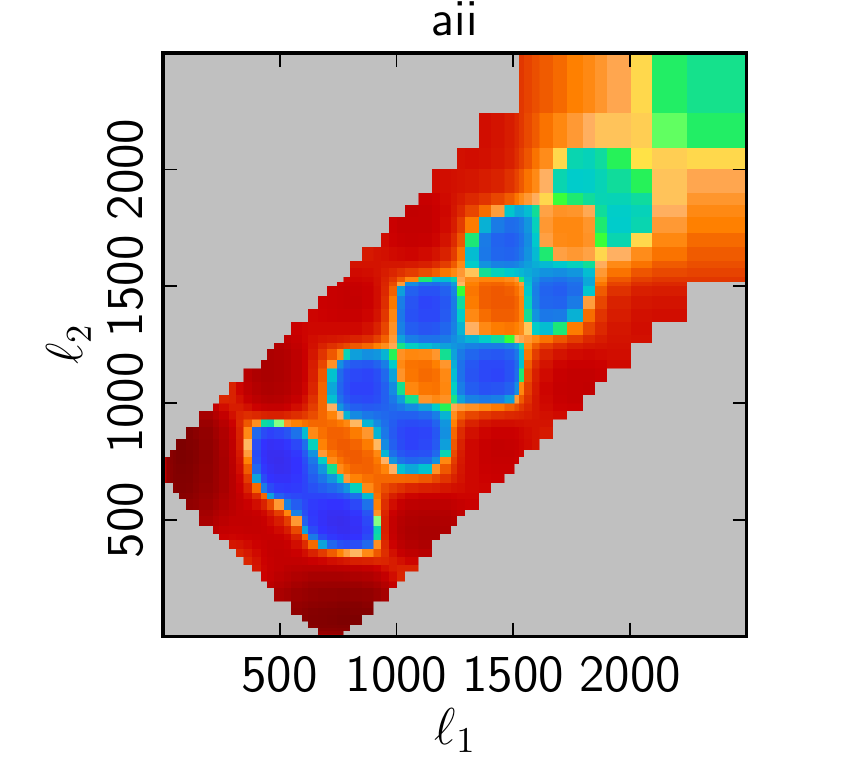}
\includegraphics[trim =3mm 0mm 12mm 0mm,clip,width=0.19\columnwidth]{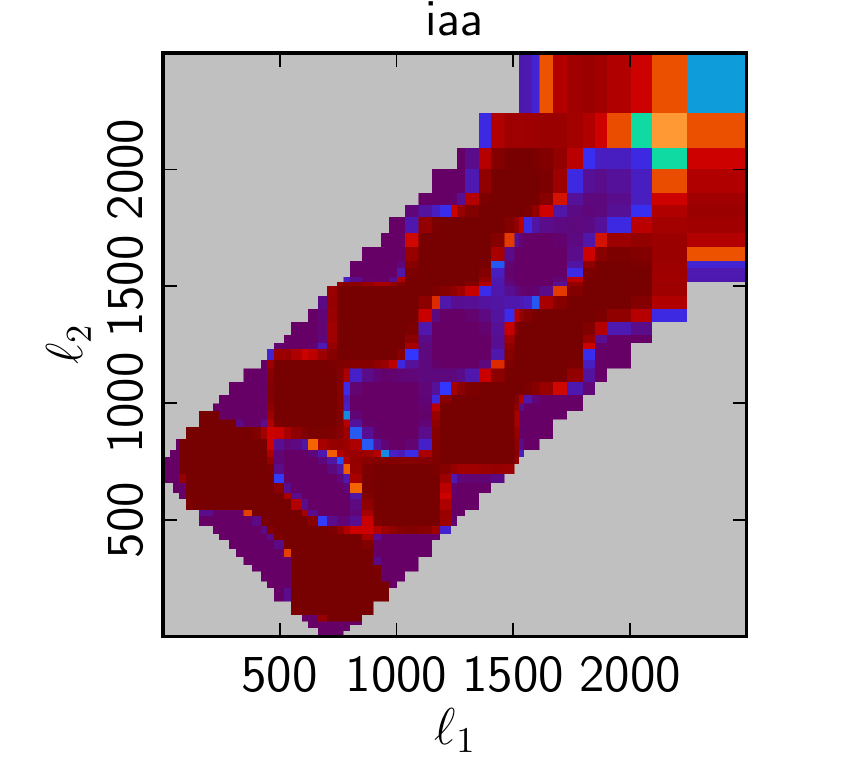}
\includegraphics[trim =3mm 0mm 12mm 0mm,clip,width=0.19\columnwidth]{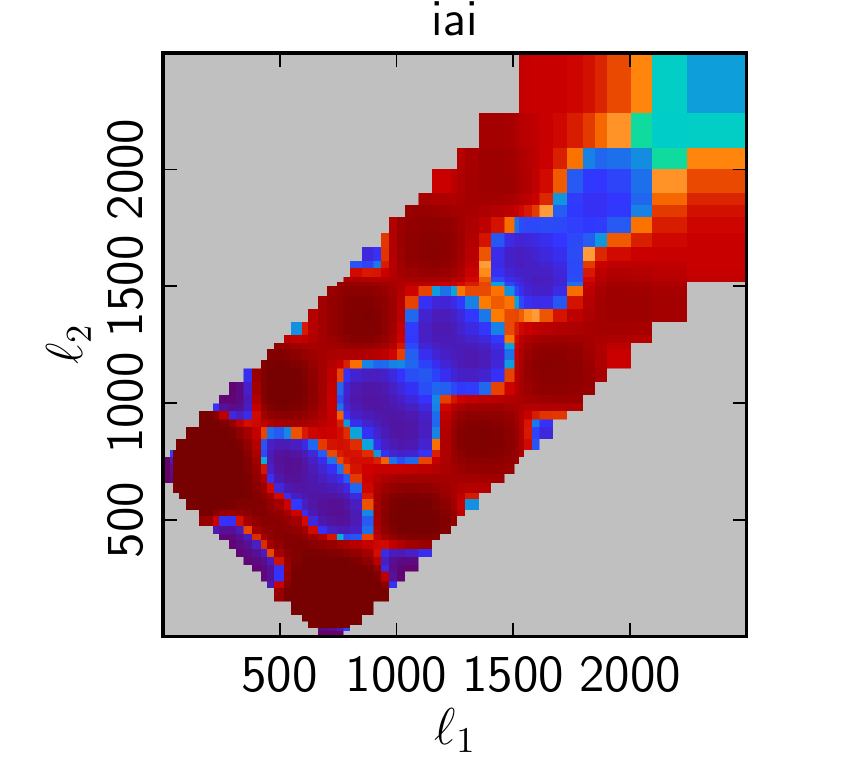}
\includegraphics[trim =3mm 0mm 12mm 0mm,clip,width=0.19\columnwidth]{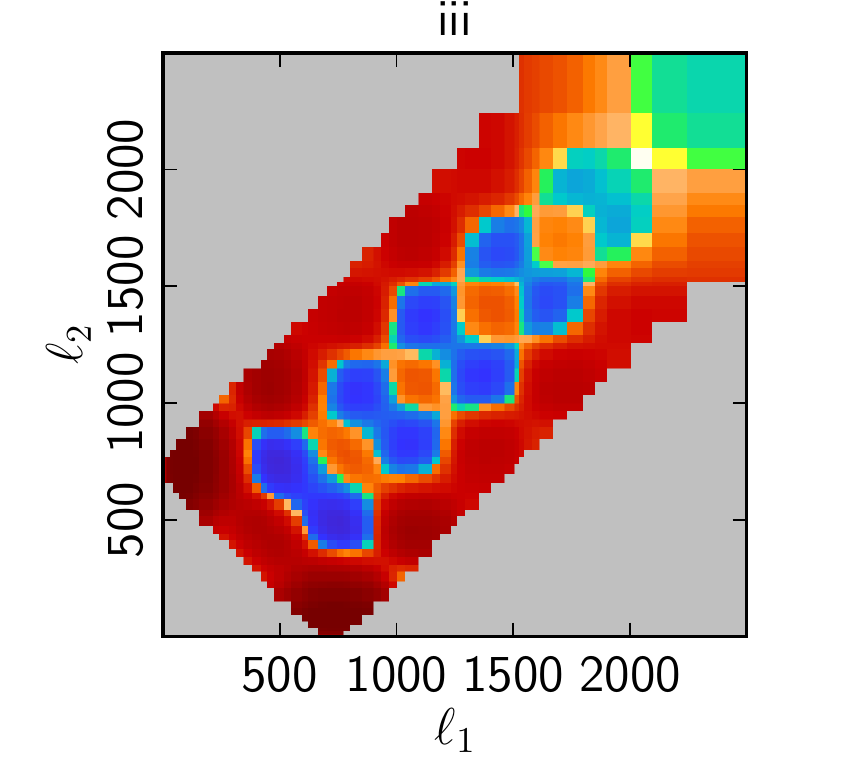}

\includegraphics[trim =3mm 0mm 12mm 0mm,clip,width=0.19\columnwidth]{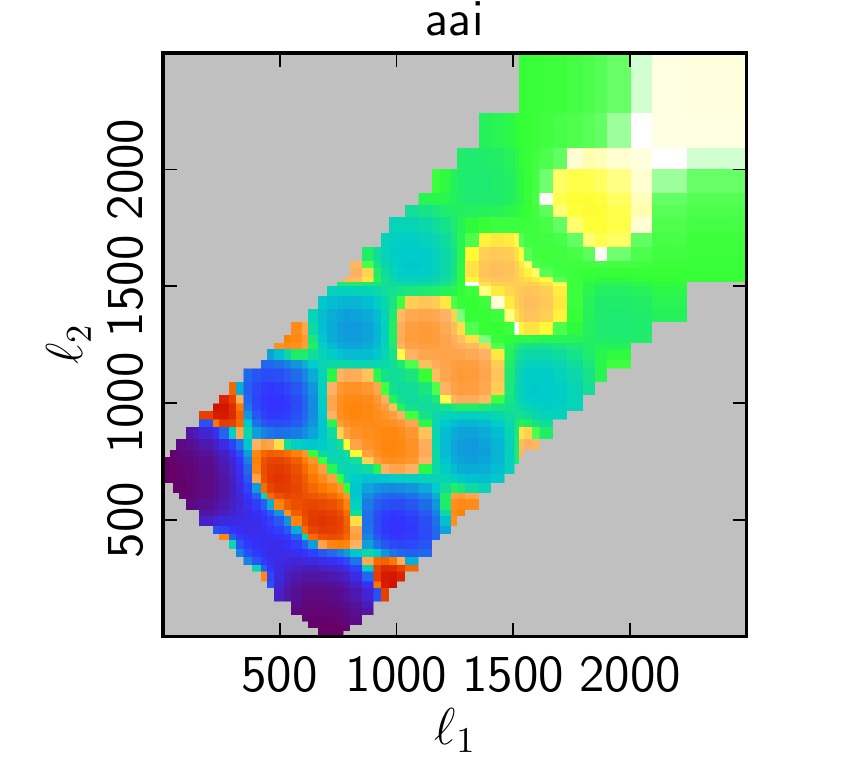}
\includegraphics[trim =3mm 0mm 12mm 0mm,clip,width=0.19\columnwidth]{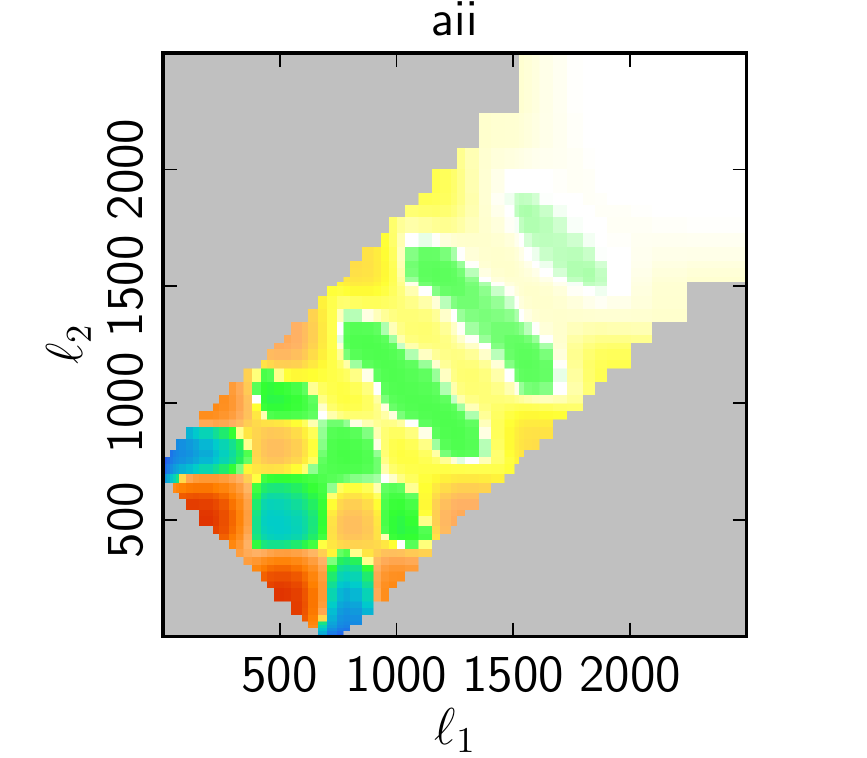}
\includegraphics[trim =3mm 0mm 12mm 0mm,clip,width=0.19\columnwidth]{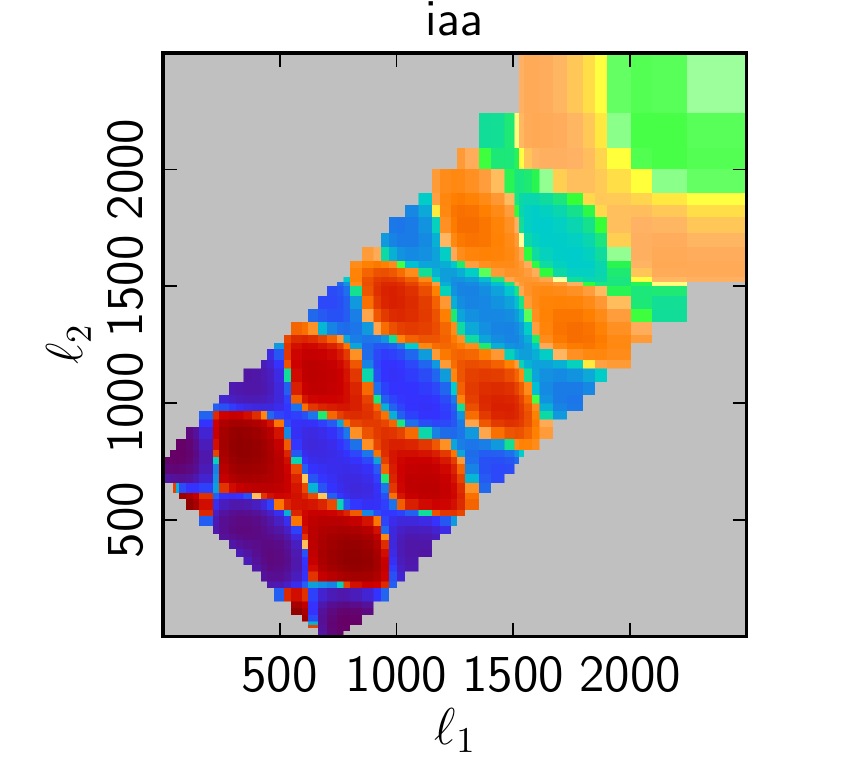}
\includegraphics[trim =3mm 0mm 12mm 0mm,clip,width=0.19\columnwidth]{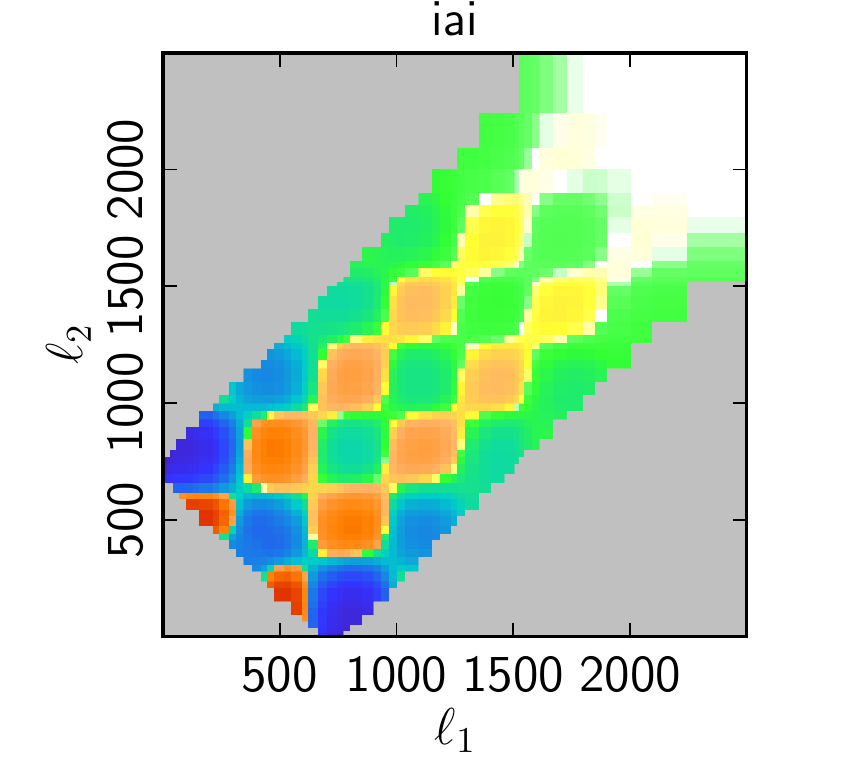}
\includegraphics[trim =3mm 0mm 12mm 0mm,clip,width=0.19\columnwidth]{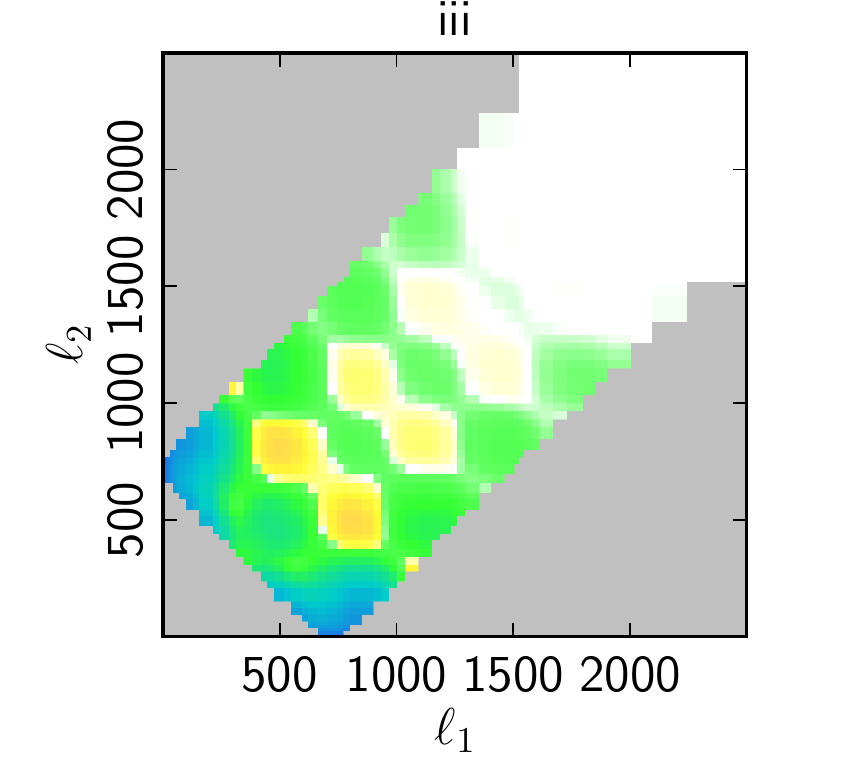}

\includegraphics[trim =3mm 0mm 12mm 0mm,clip,width=0.19\columnwidth]{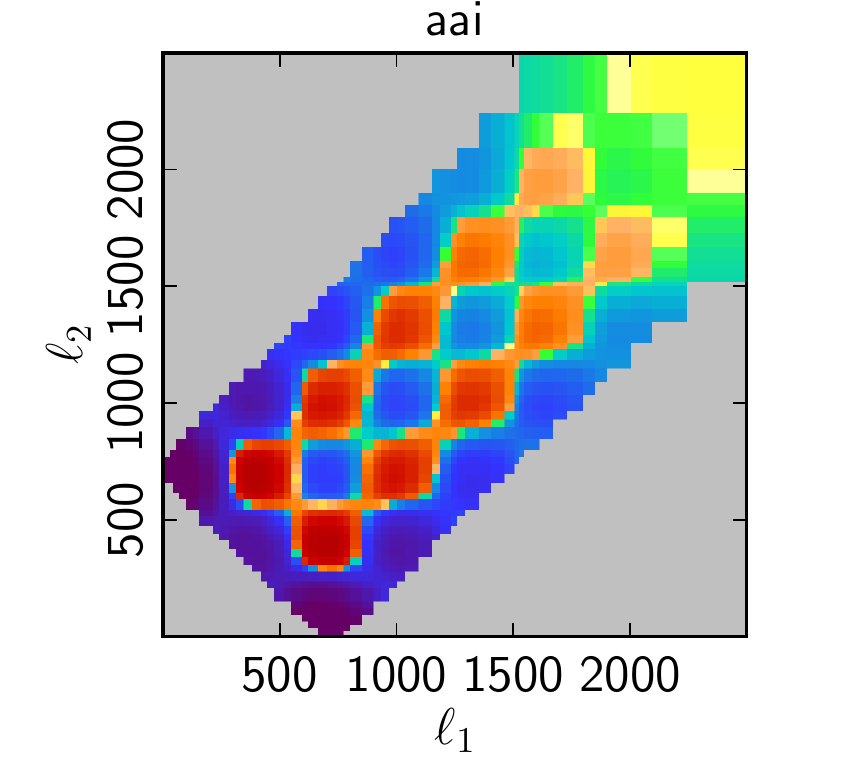}
\includegraphics[trim =3mm 0mm 12mm 0mm,clip,width=0.19\columnwidth]{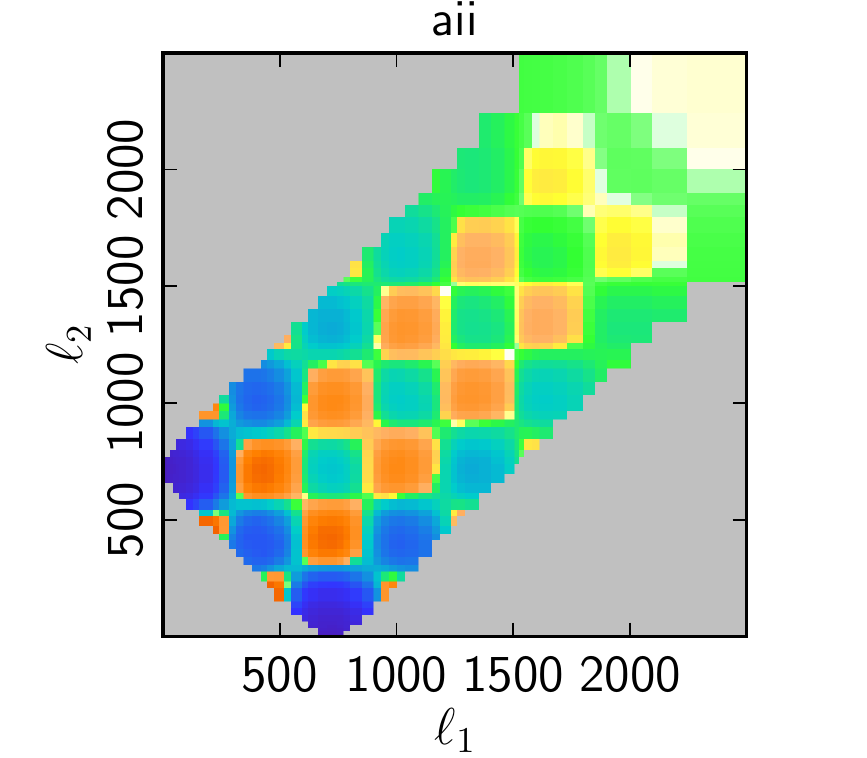}
\includegraphics[trim =3mm 0mm 12mm 0mm,clip,width=0.19\columnwidth]{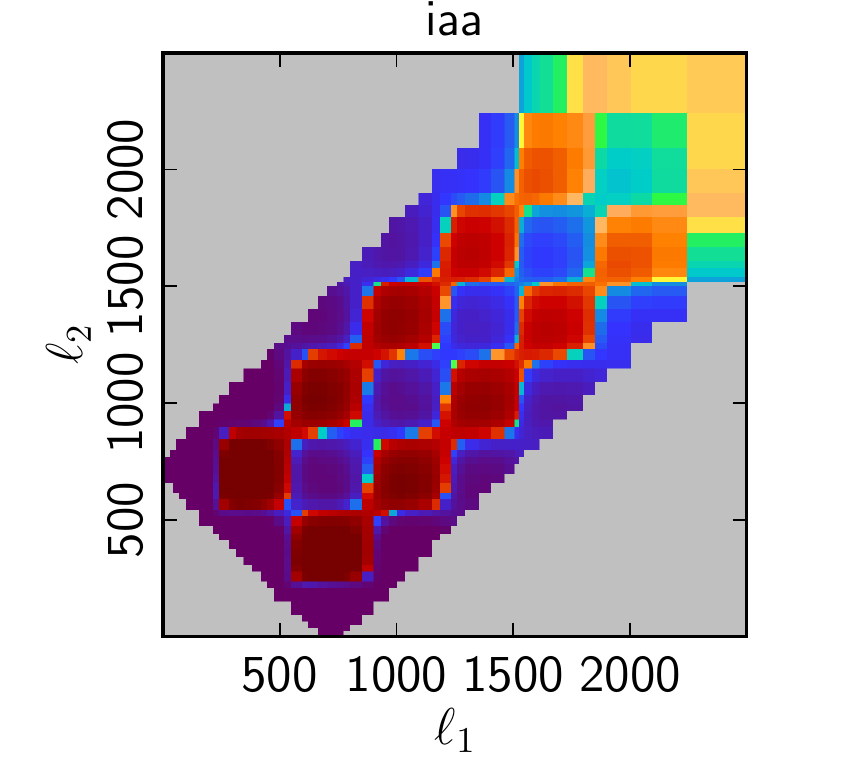}
\includegraphics[trim =3mm 0mm 12mm 0mm,clip,width=0.19\columnwidth]{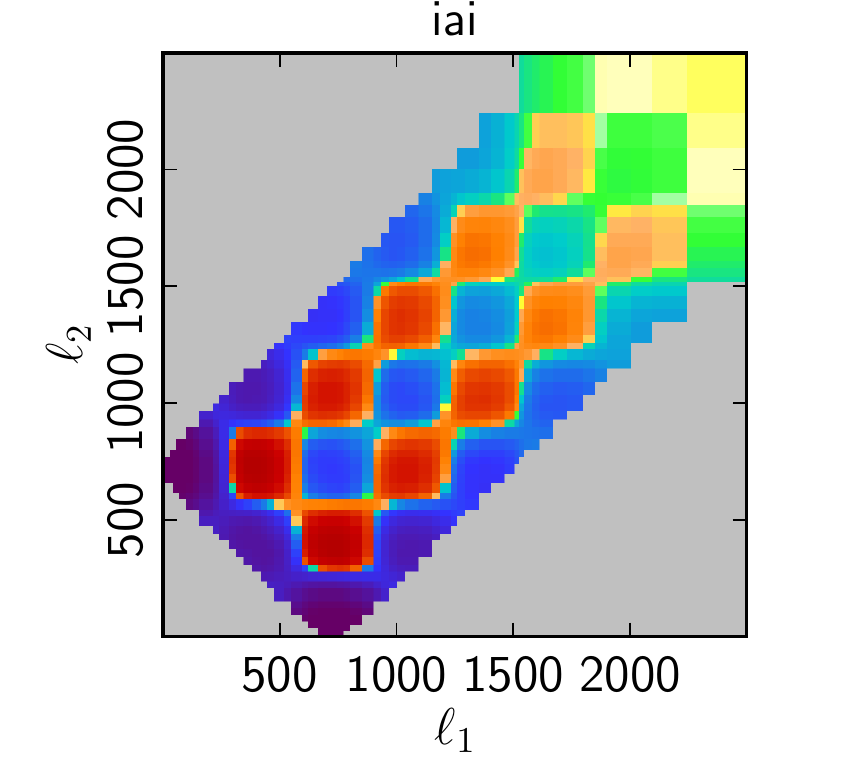}
\includegraphics[trim =3mm 0mm 12mm 0mm,clip,width=0.19\columnwidth]{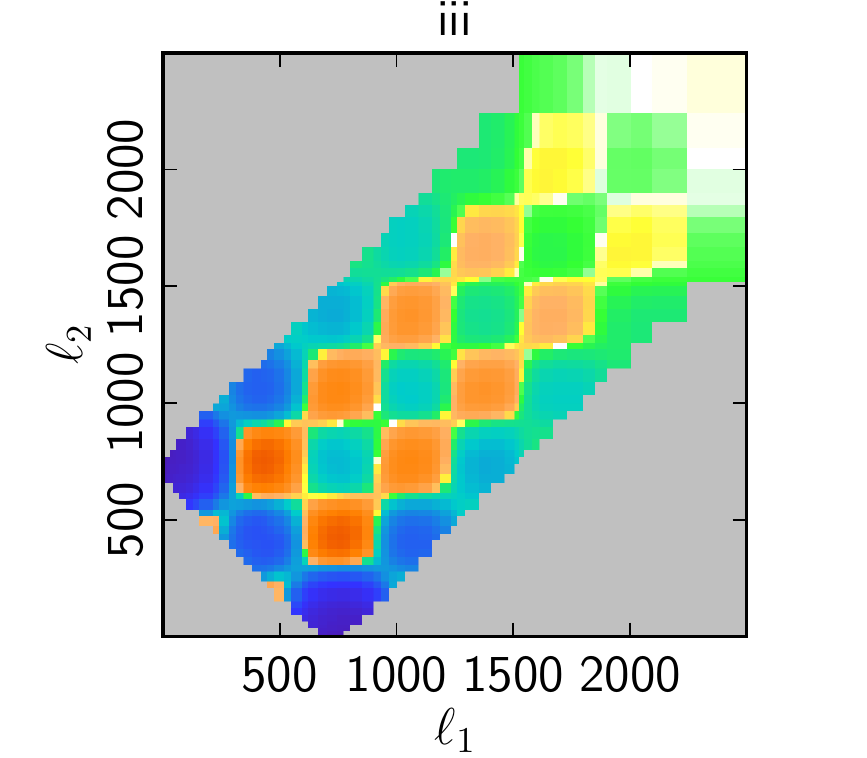}

\includegraphics[trim =3mm 0mm 12mm 0mm,clip,width=0.19\columnwidth]{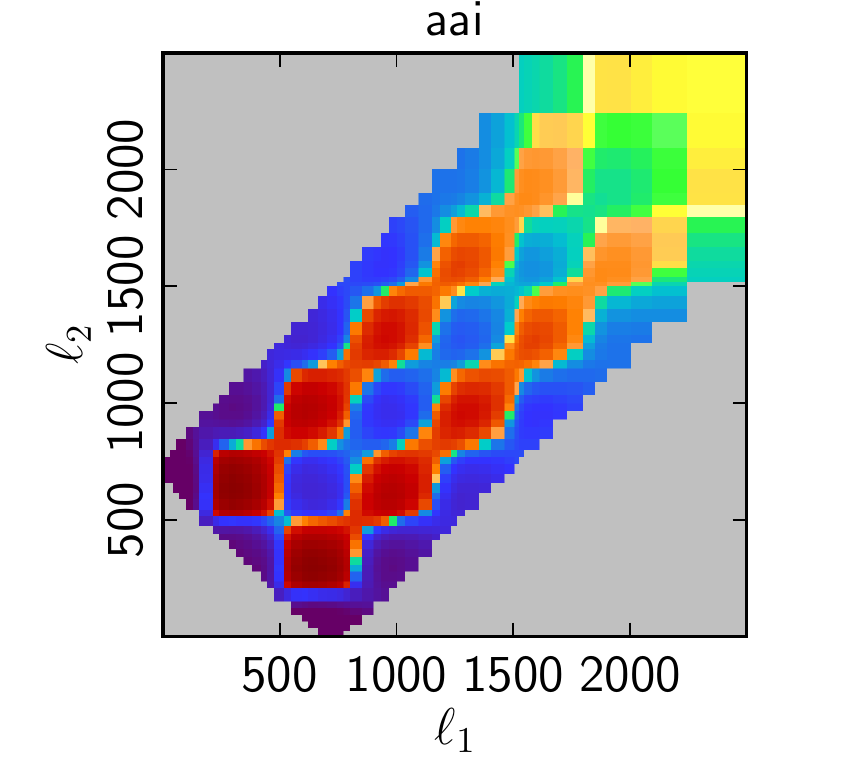}
\includegraphics[trim =3mm 0mm 12mm 0mm,clip,width=0.19\columnwidth]{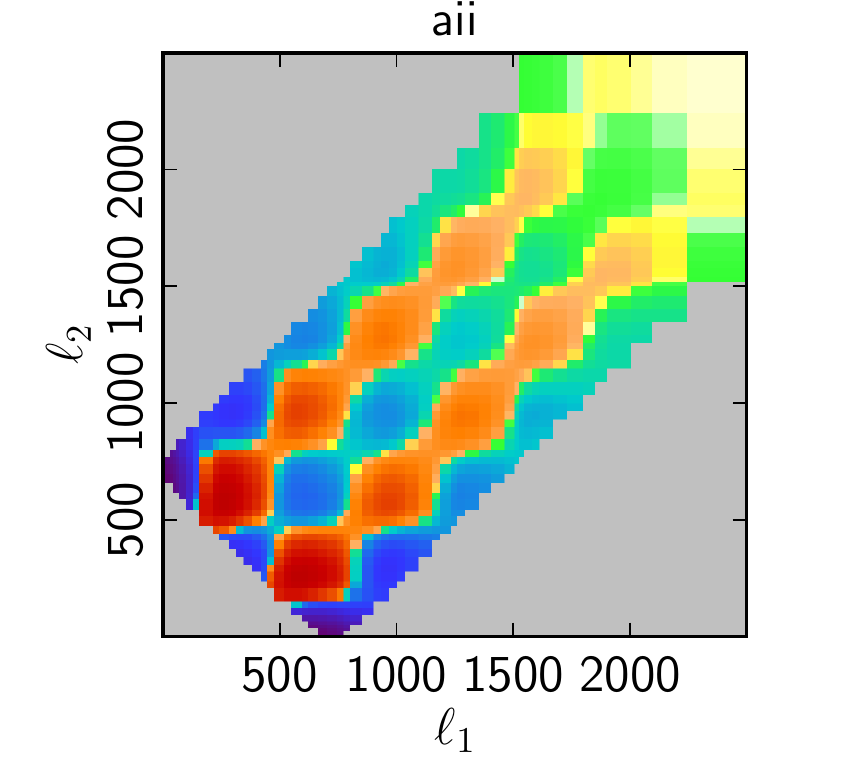}
\includegraphics[trim =3mm 0mm 12mm 0mm,clip,width=0.19\columnwidth]{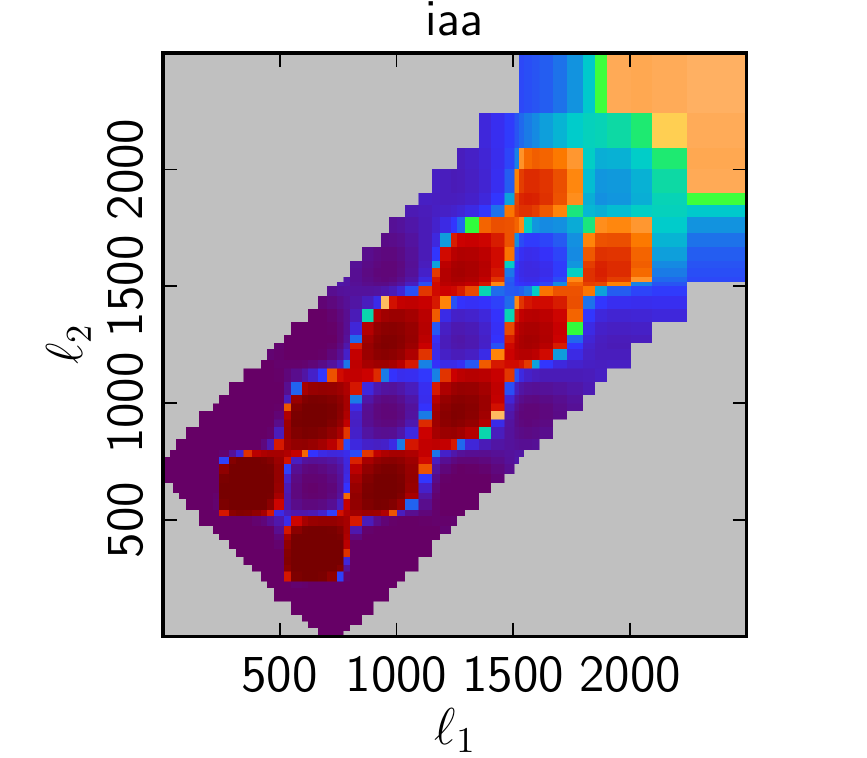}
\includegraphics[trim =3mm 0mm 12mm 0mm,clip,width=0.19\columnwidth]{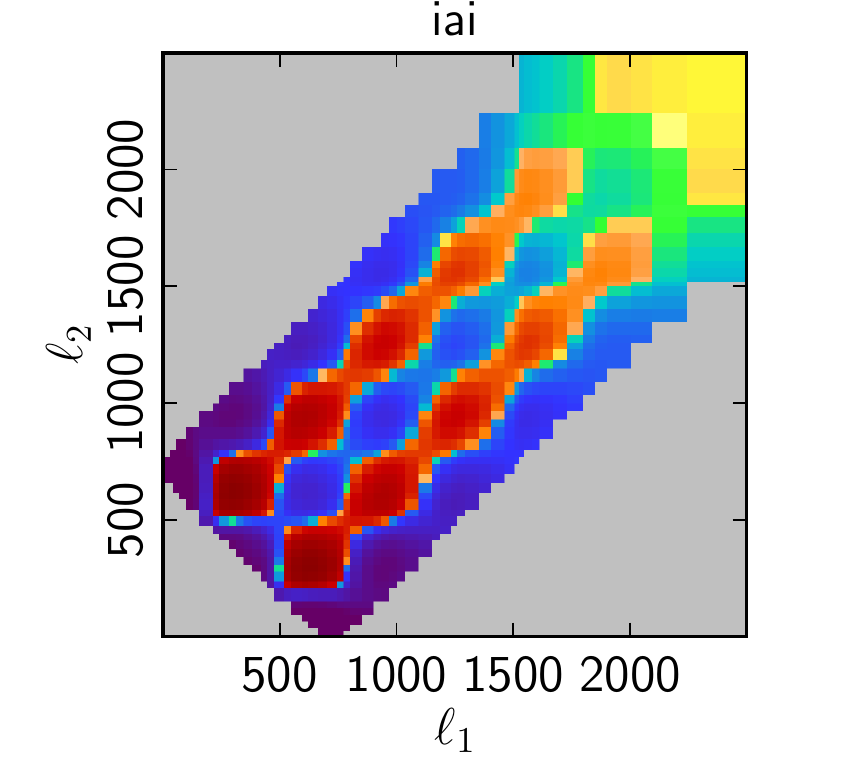}
\includegraphics[trim =3mm 0mm 12mm 0mm,clip,width=0.19\columnwidth]{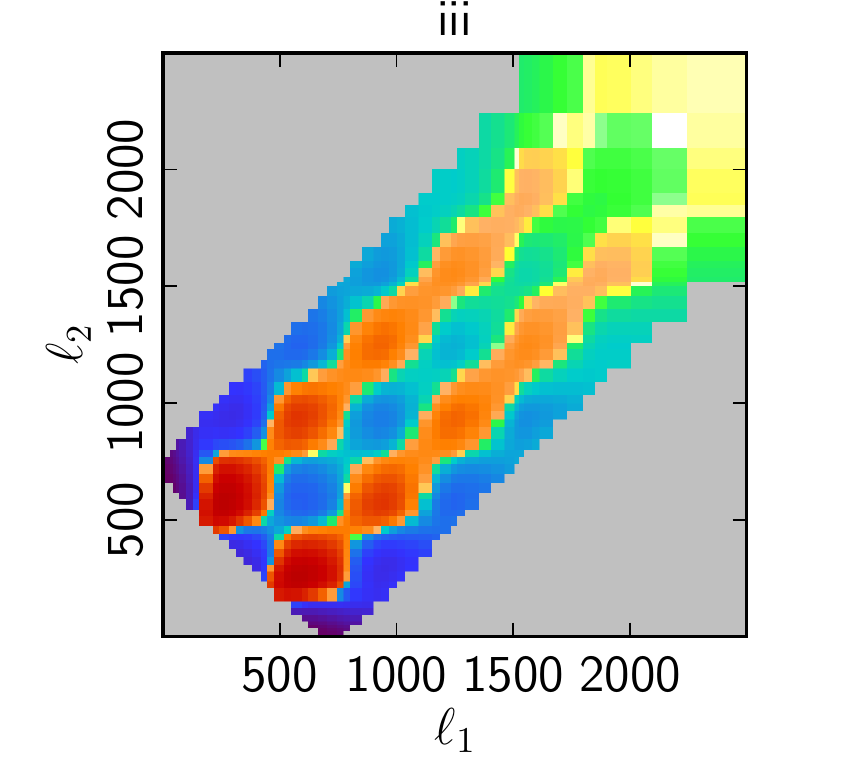}
\includegraphics[trim = 0mm 0mm 25mm 10mm,clip,width=0.9\columnwidth]{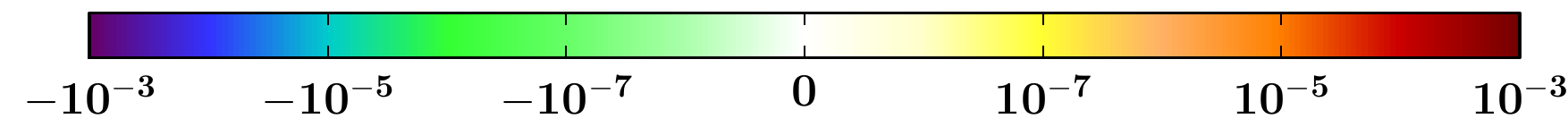}
\caption{Smoothed theoretical isocurvature bispectra for $\ell_3\in$ [700, 741].
The first row presents the 5 different $TTT$ bispectra one can have when 
considering just the adiabatic and the cold dark matter density isocurvature 
mode, excluding the purely adiabatic bispectrum. The second and third
row are the same, but taking instead the neutrino density and neutrino velocity
isocurvature mode, respectively. Finally the last three rows are similar to
the first three, but for $EEE$.
The bispectra, all with $f_\mathrm{NL}=1$, have been normalized by the expected 
standard deviation assuming the beam and noise characteristics of 
Section~\ref{masking}, so that a dimensionless signal-to-noise bispectrum 
results. The colour scale is logarithmic (with the white in the middle 
corresponding to values smaller than $10^{-9}$ in absolute value).}
\label{fig:theor_iso1}
\end{figure}

\begin{figure}
\centering
\includegraphics[trim =3mm 0mm 12mm 0mm,clip,width=0.19\columnwidth]{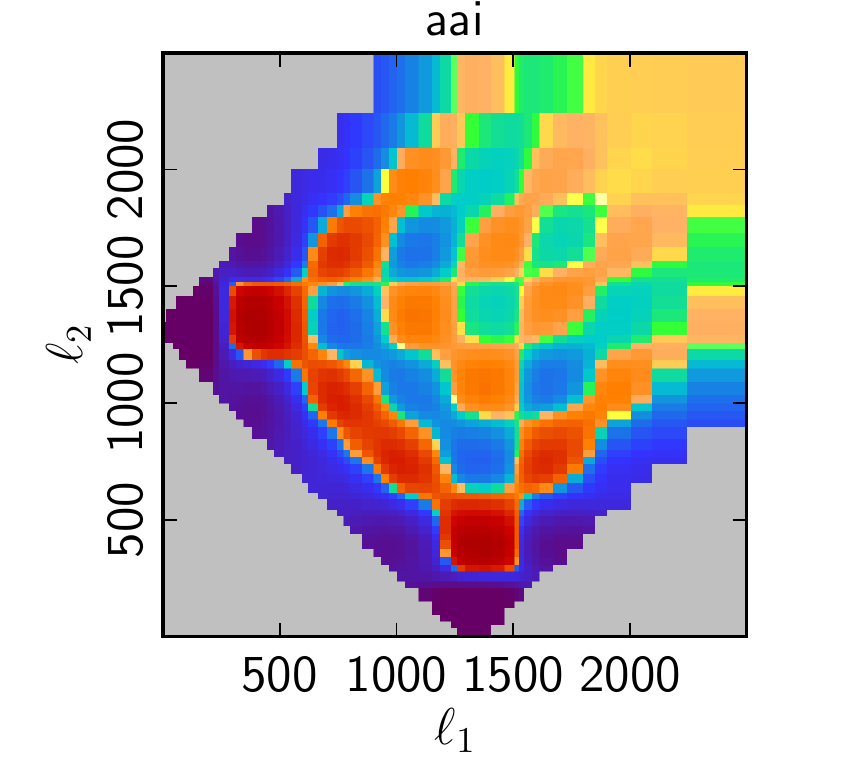}
\includegraphics[trim =3mm 0mm 12mm 0mm,clip,width=0.19\columnwidth]{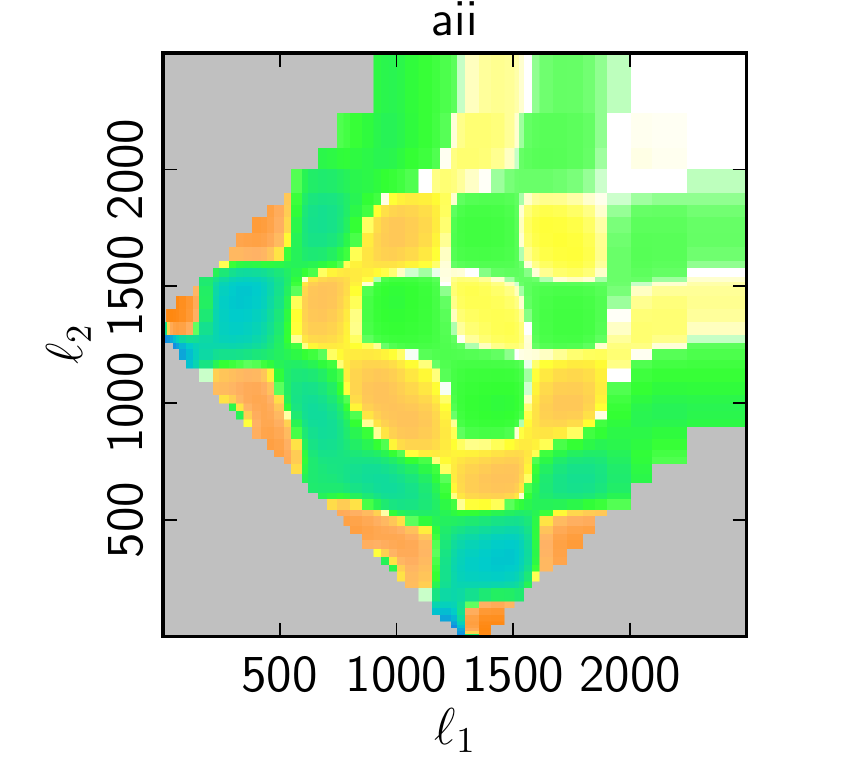}
\includegraphics[trim =3mm 0mm 12mm 0mm,clip,width=0.19\columnwidth]{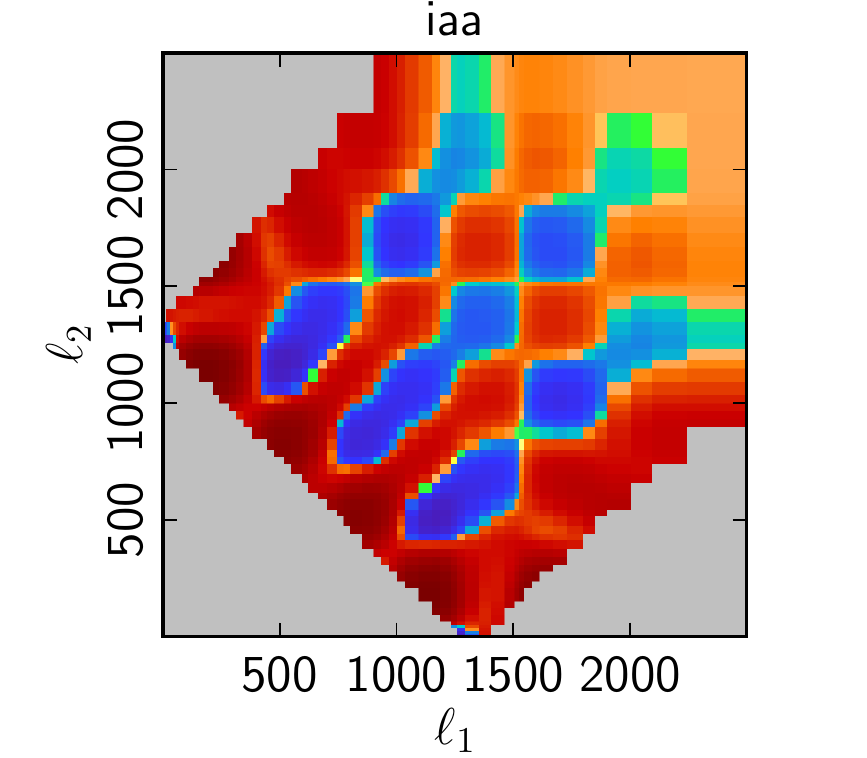}
\includegraphics[trim =3mm 0mm 12mm 0mm,clip,width=0.19\columnwidth]{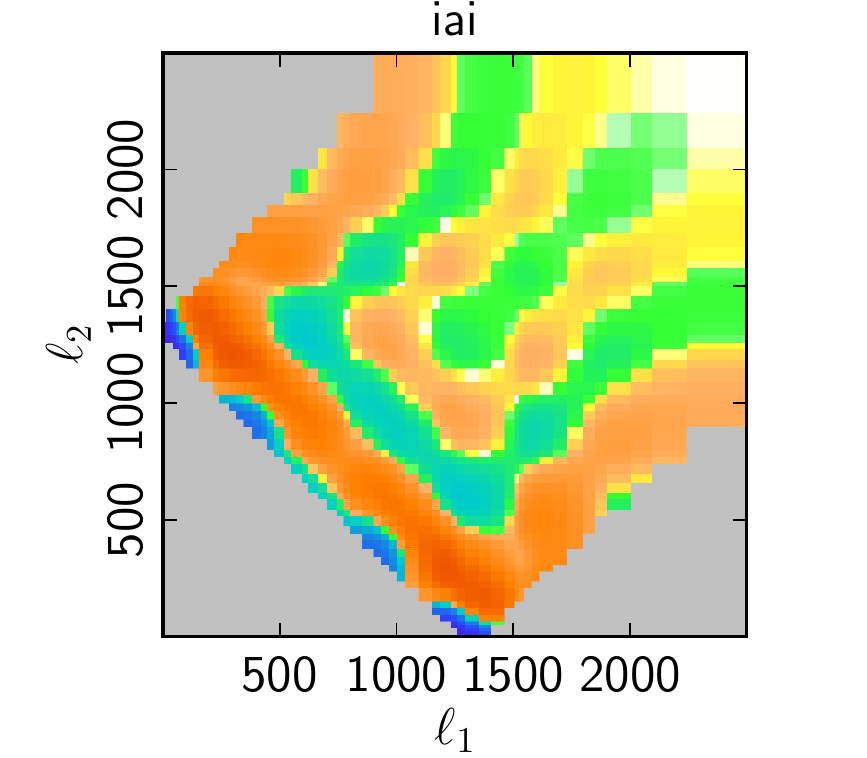}
\includegraphics[trim =3mm 0mm 12mm 0mm,clip,width=0.19\columnwidth]{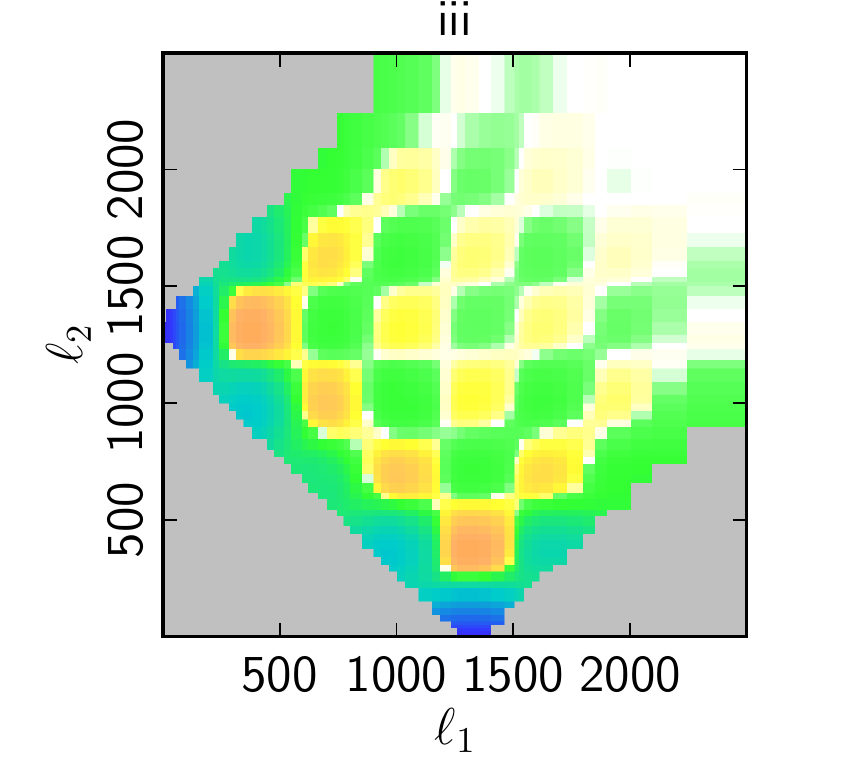}

\includegraphics[trim =3mm 0mm 12mm 0mm,clip,width=0.19\columnwidth]{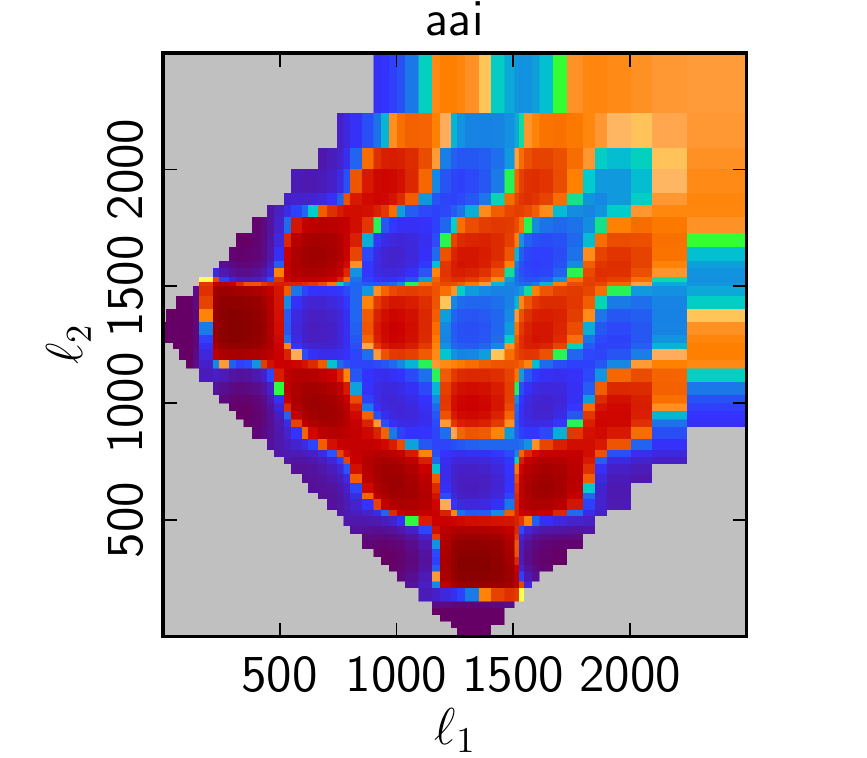}
\includegraphics[trim =3mm 0mm 12mm 0mm,clip,width=0.19\columnwidth]{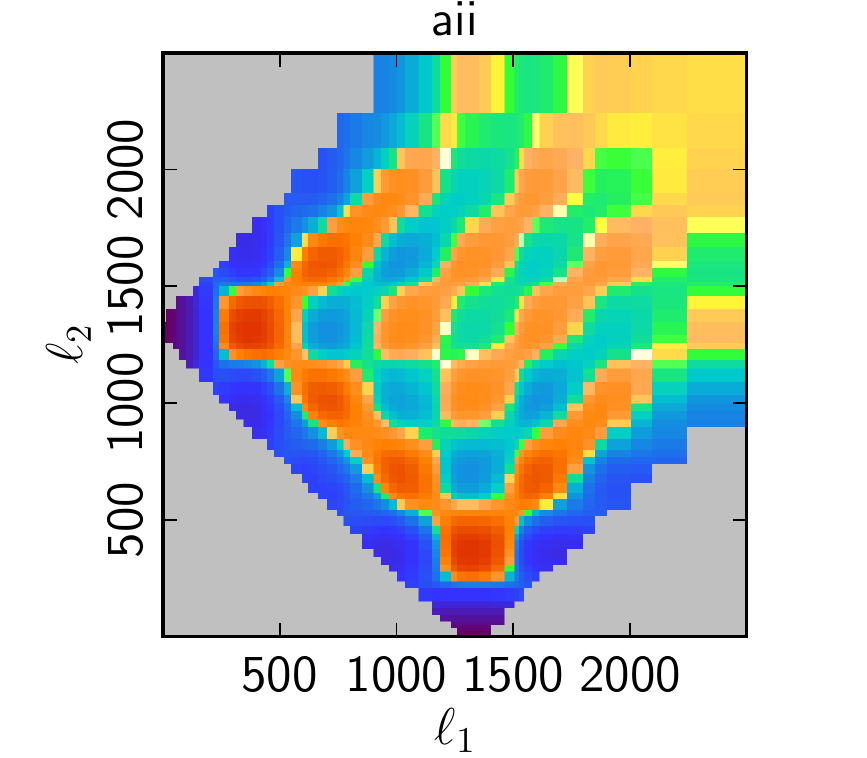}
\includegraphics[trim =3mm 0mm 12mm 0mm,clip,width=0.19\columnwidth]{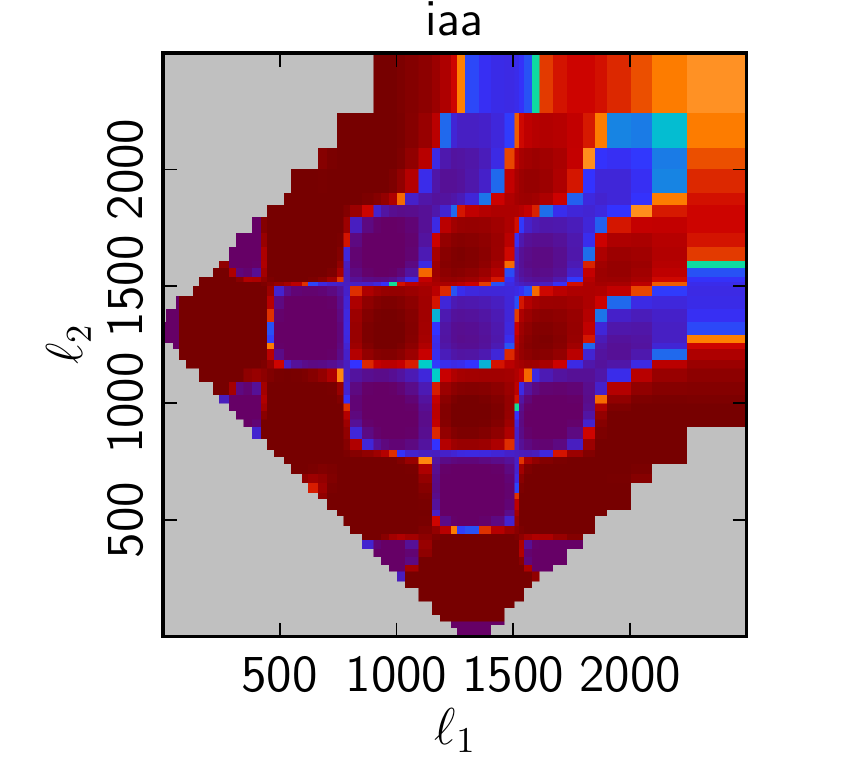}
\includegraphics[trim =3mm 0mm 12mm 0mm,clip,width=0.19\columnwidth]{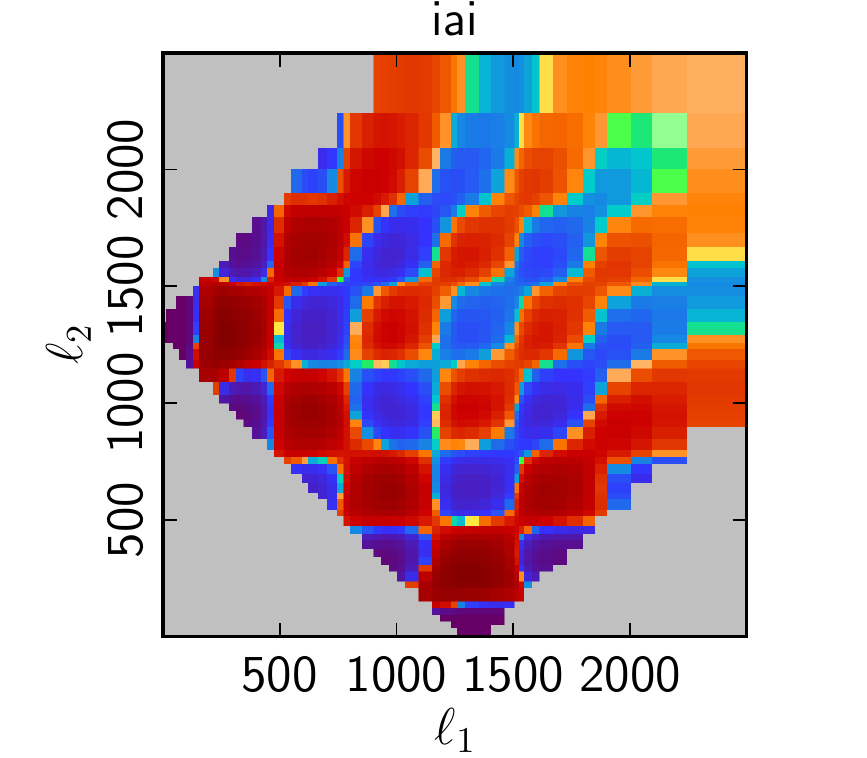}
\includegraphics[trim =3mm 0mm 12mm 0mm,clip,width=0.19\columnwidth]{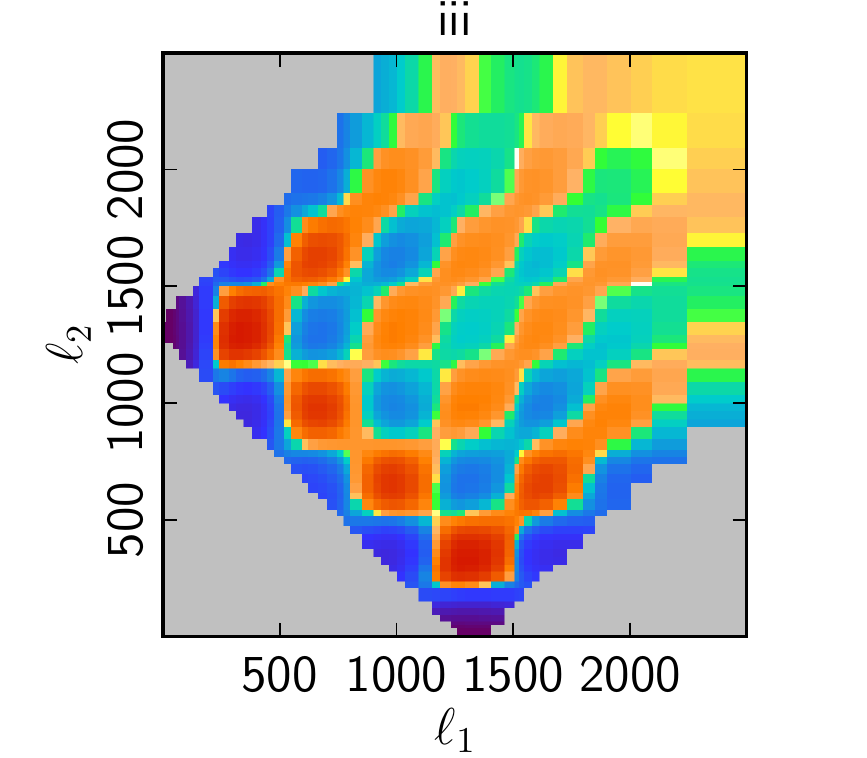}

\includegraphics[trim =3mm 0mm 12mm 0mm,clip,width=0.19\columnwidth]{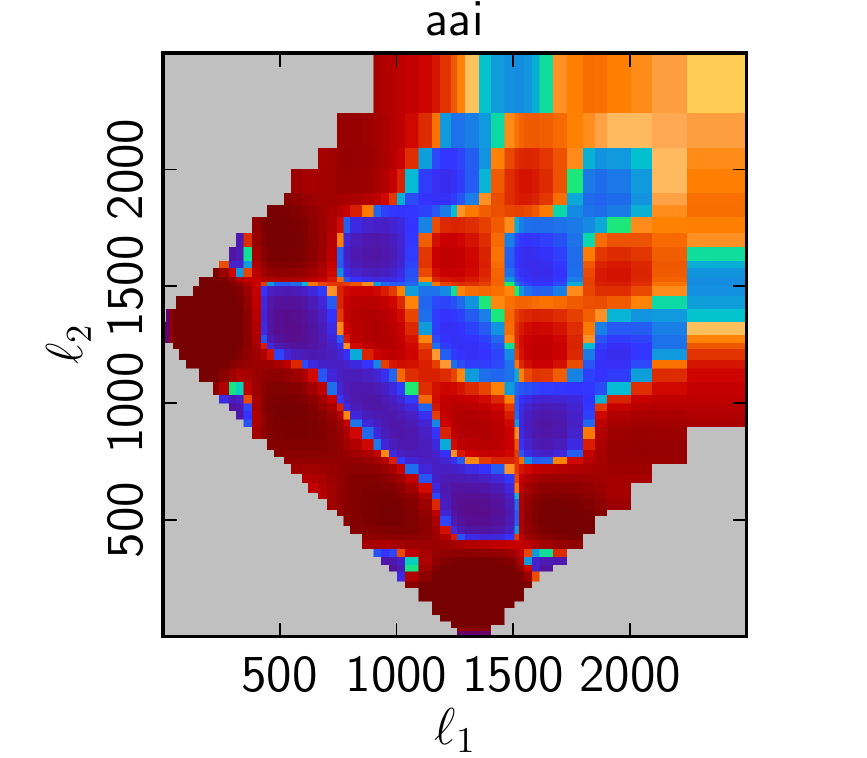}
\includegraphics[trim =3mm 0mm 12mm 0mm,clip,width=0.19\columnwidth]{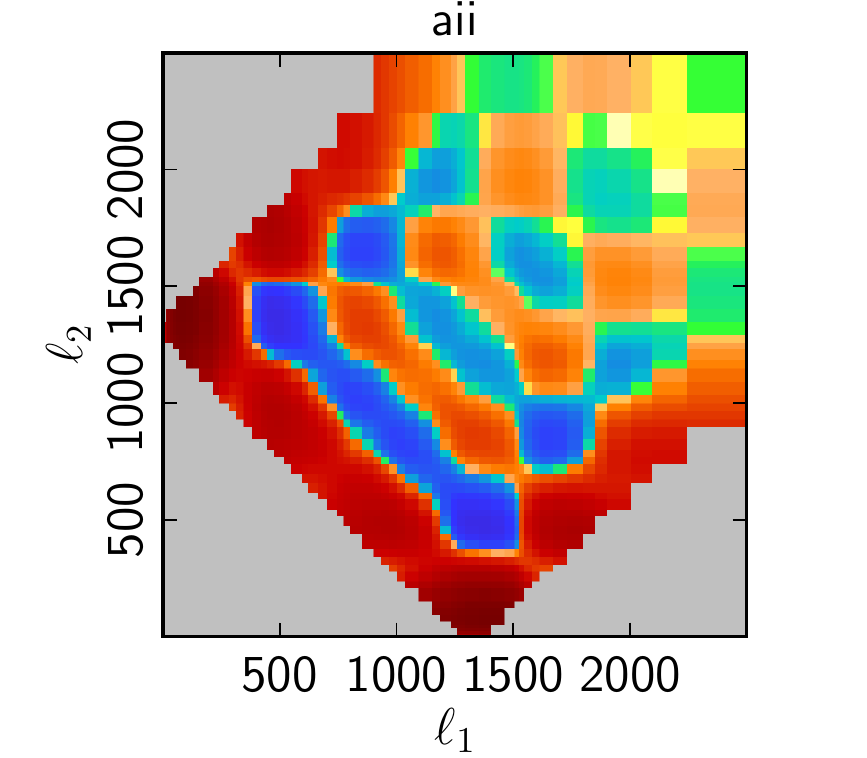}
\includegraphics[trim =3mm 0mm 12mm 0mm,clip,width=0.19\columnwidth]{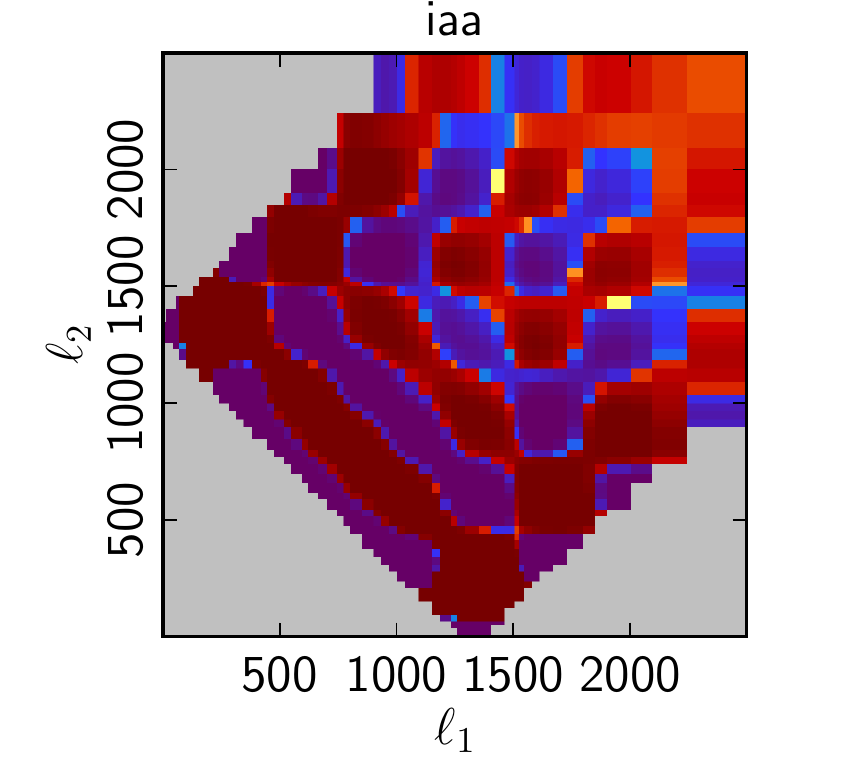}
\includegraphics[trim =3mm 0mm 12mm 0mm,clip,width=0.19\columnwidth]{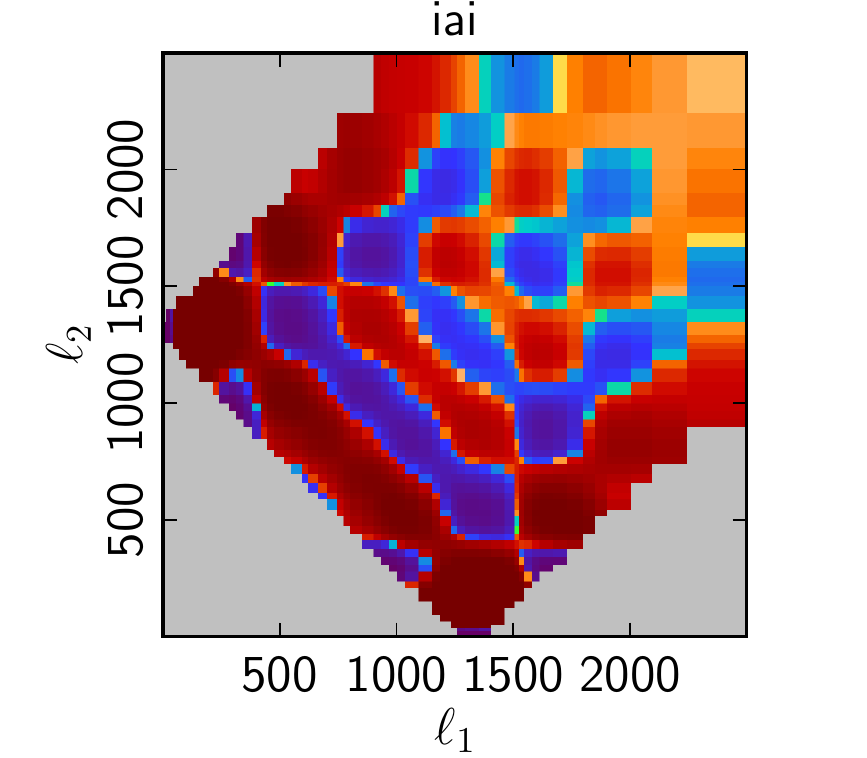}
\includegraphics[trim =3mm 0mm 12mm 0mm,clip,width=0.19\columnwidth]{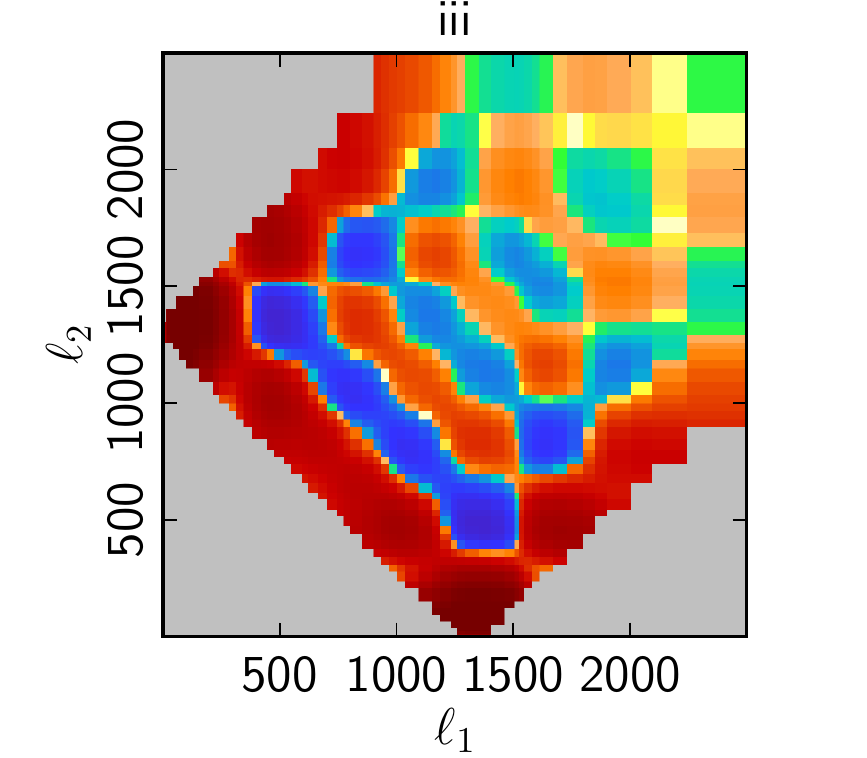}

\includegraphics[trim =3mm 0mm 12mm 0mm,clip,width=0.19\columnwidth]{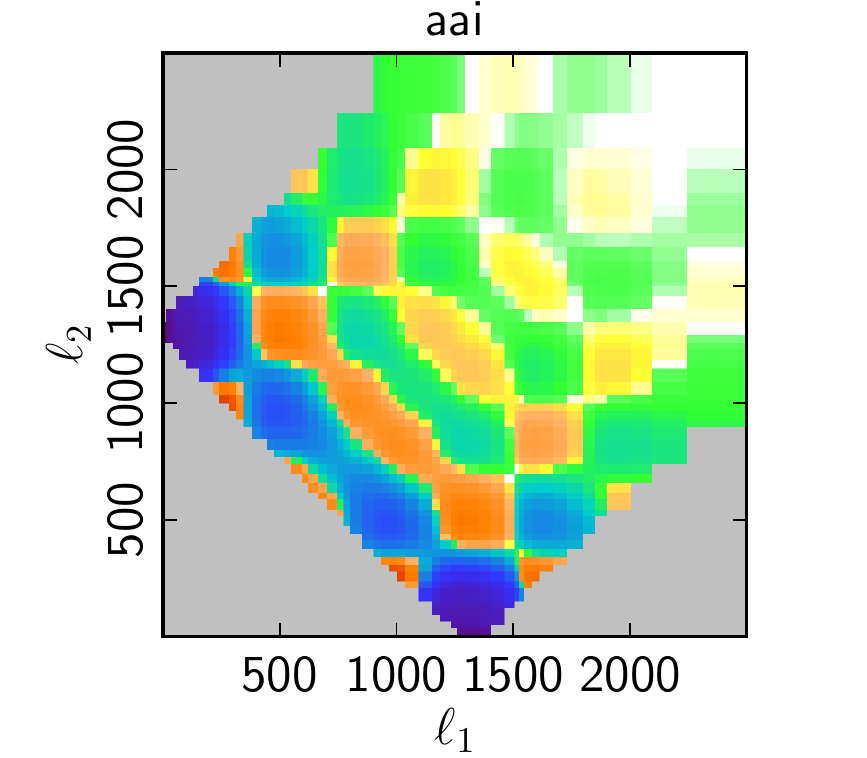}
\includegraphics[trim =3mm 0mm 12mm 0mm,clip,width=0.19\columnwidth]{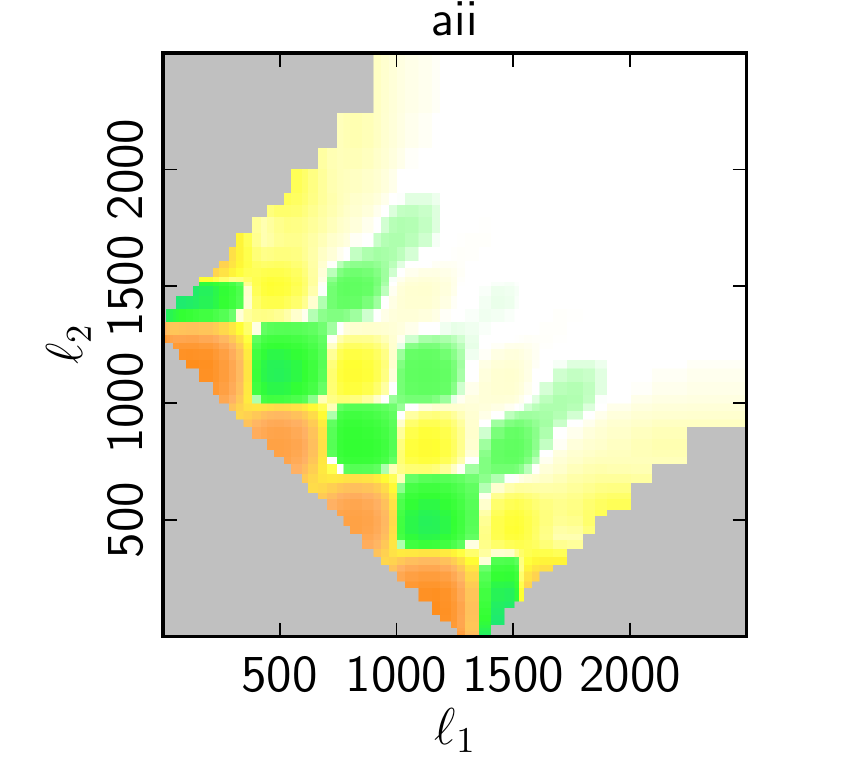}
\includegraphics[trim =3mm 0mm 12mm 0mm,clip,width=0.19\columnwidth]{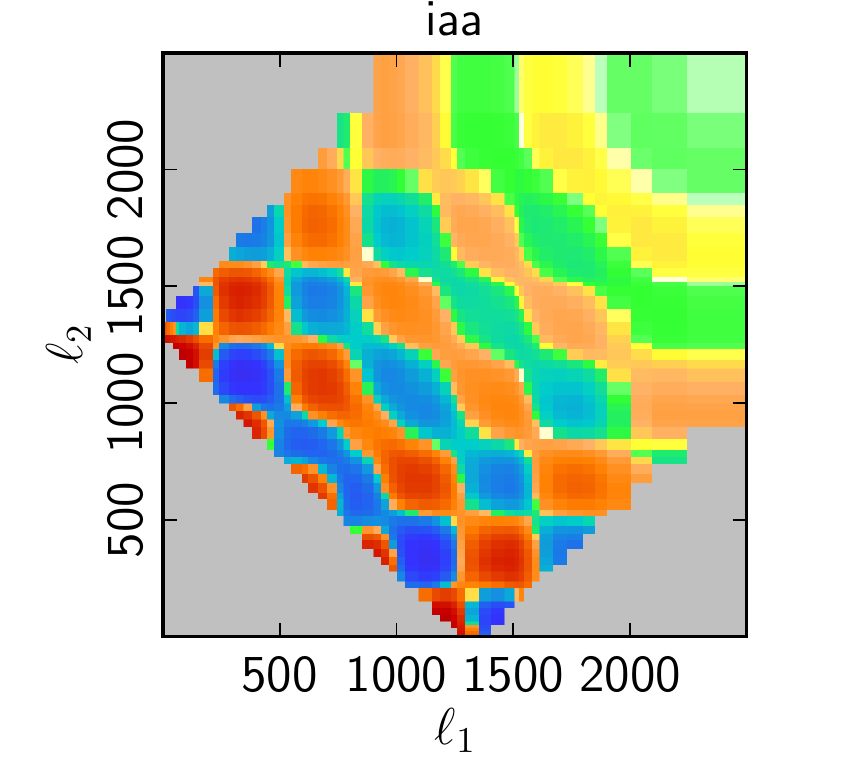}
\includegraphics[trim =3mm 0mm 12mm 0mm,clip,width=0.19\columnwidth]{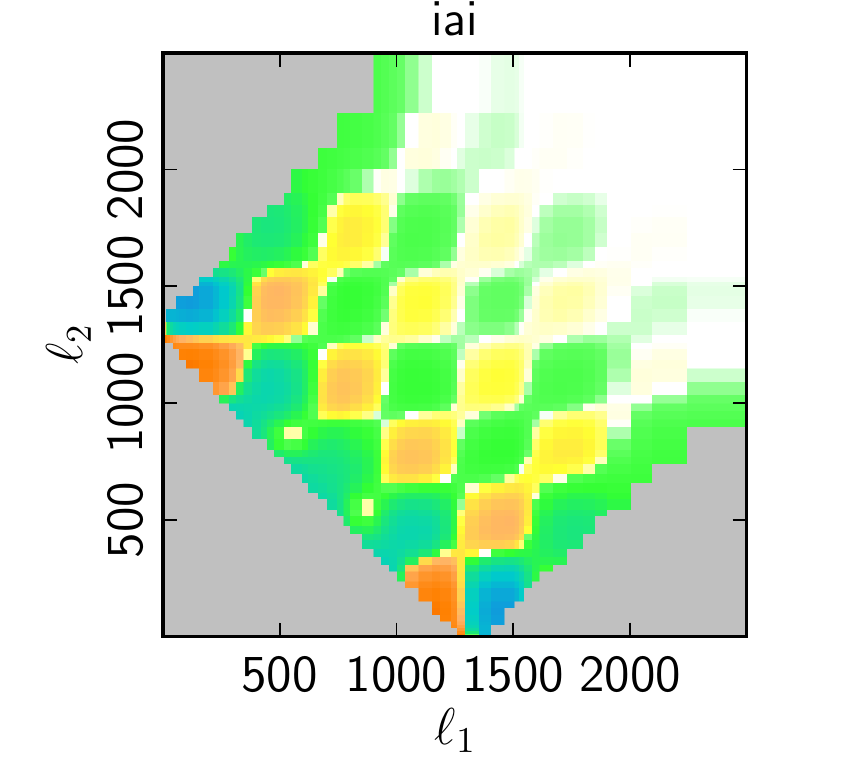}
\includegraphics[trim =3mm 0mm 12mm 0mm,clip,width=0.19\columnwidth]{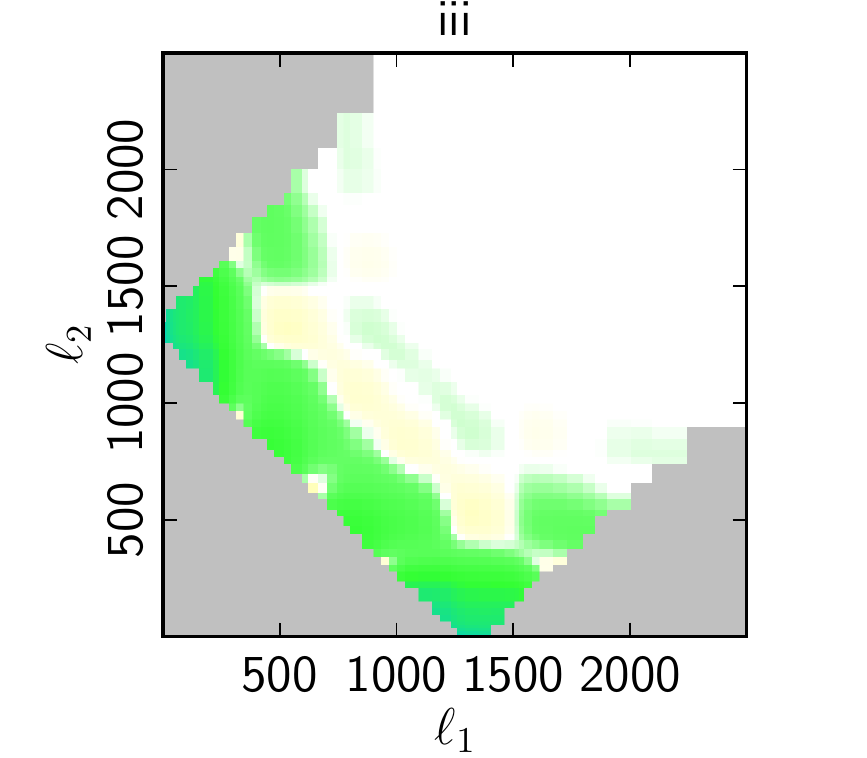}

\includegraphics[trim =3mm 0mm 12mm 0mm,clip,width=0.19\columnwidth]{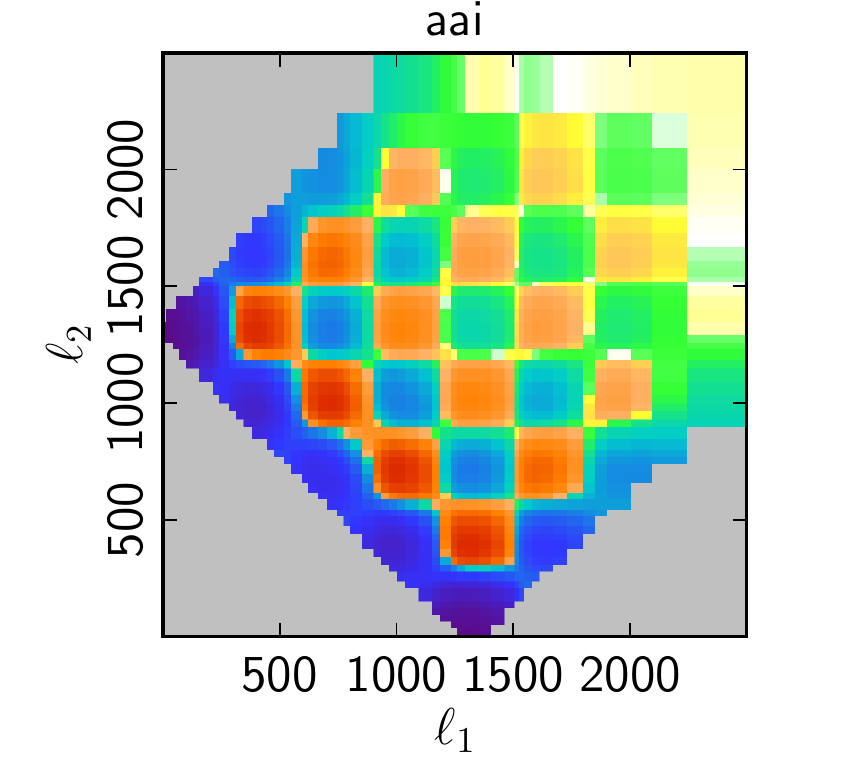}
\includegraphics[trim =3mm 0mm 12mm 0mm,clip,width=0.19\columnwidth]{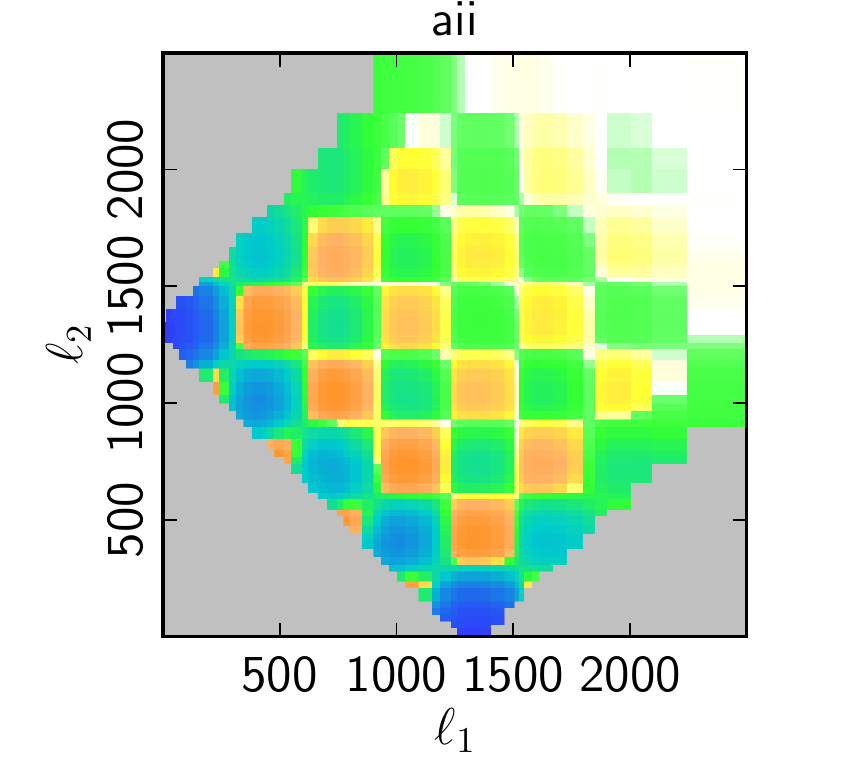}
\includegraphics[trim =3mm 0mm 12mm 0mm,clip,width=0.19\columnwidth]{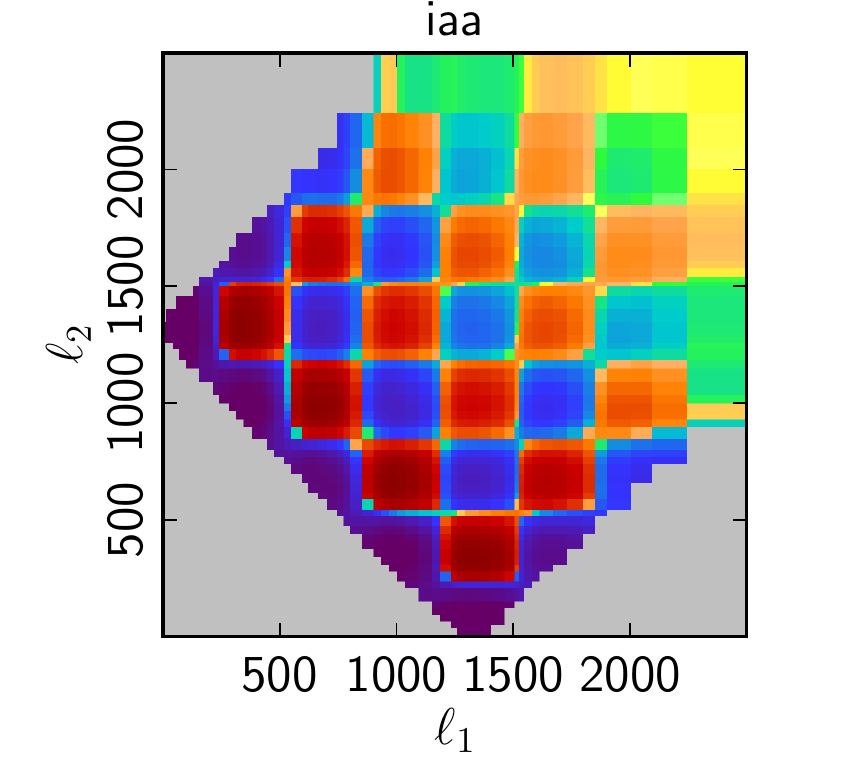}
\includegraphics[trim =3mm 0mm 12mm 0mm,clip,width=0.19\columnwidth]{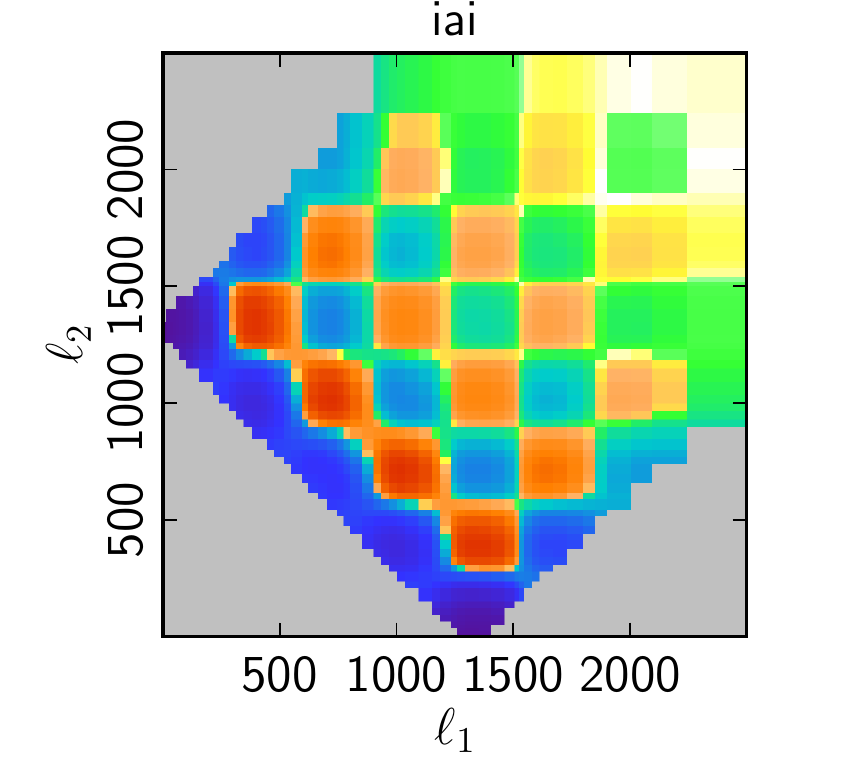}
\includegraphics[trim =3mm 0mm 12mm 0mm,clip,width=0.19\columnwidth]{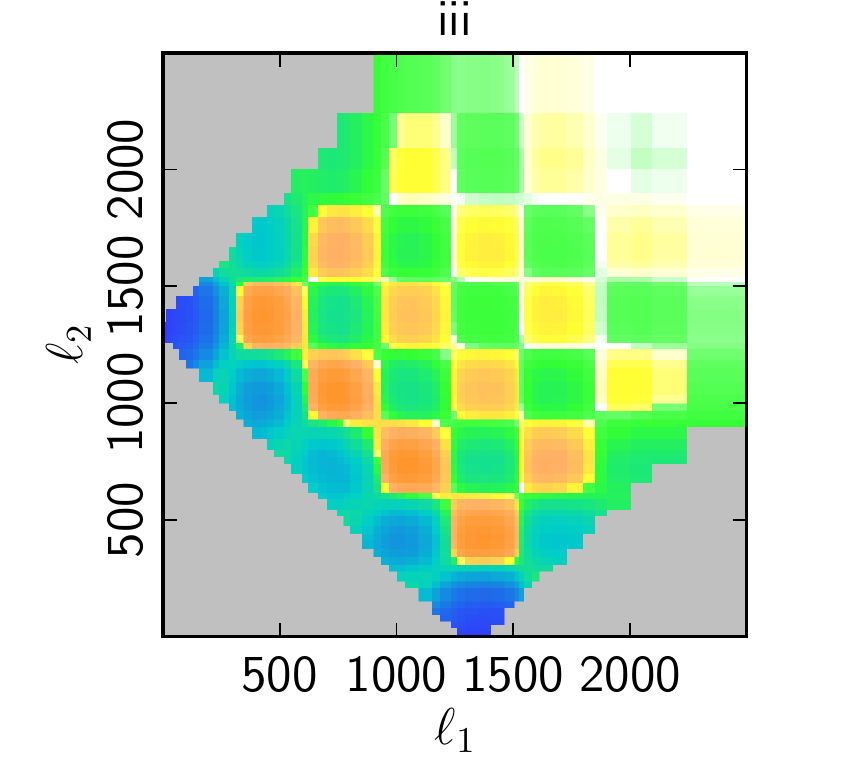}

\includegraphics[trim =3mm 0mm 12mm 0mm,clip,width=0.19\columnwidth]{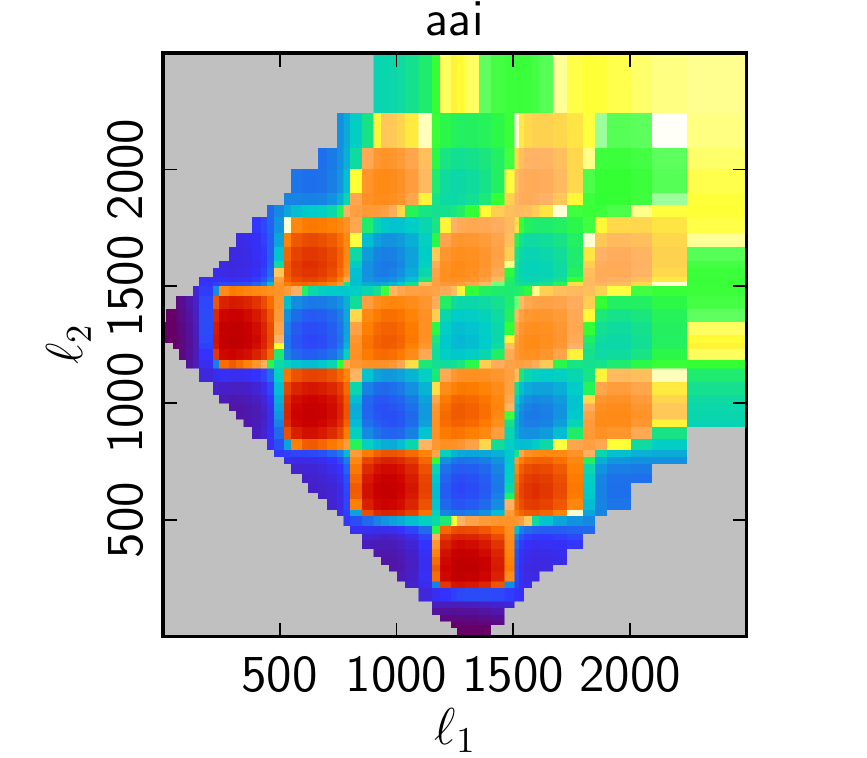}
\includegraphics[trim =3mm 0mm 12mm 0mm,clip,width=0.19\columnwidth]{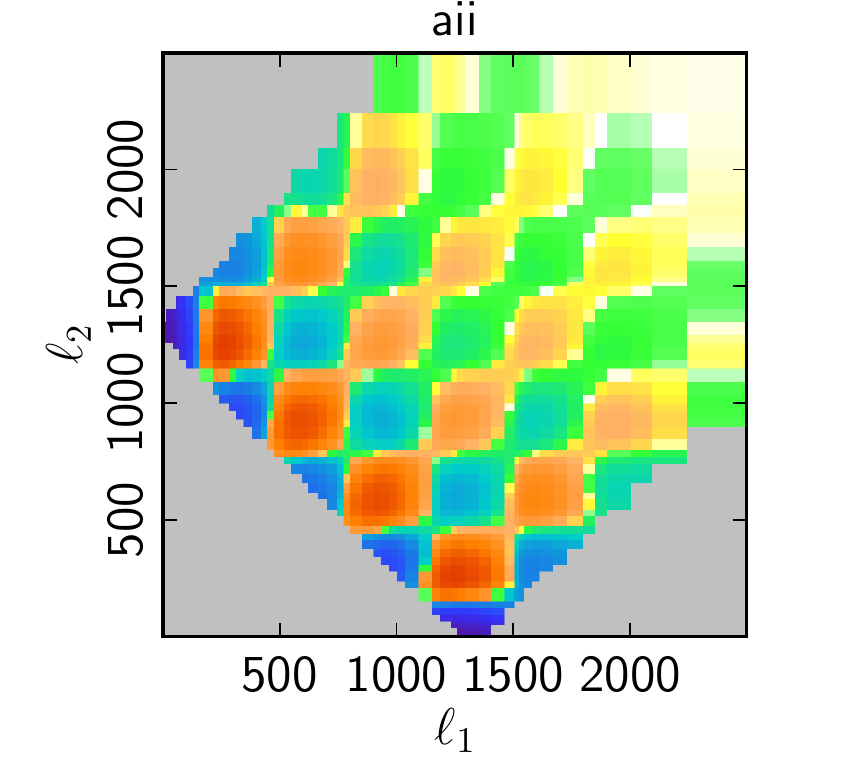}
\includegraphics[trim =3mm 0mm 12mm 0mm,clip,width=0.19\columnwidth]{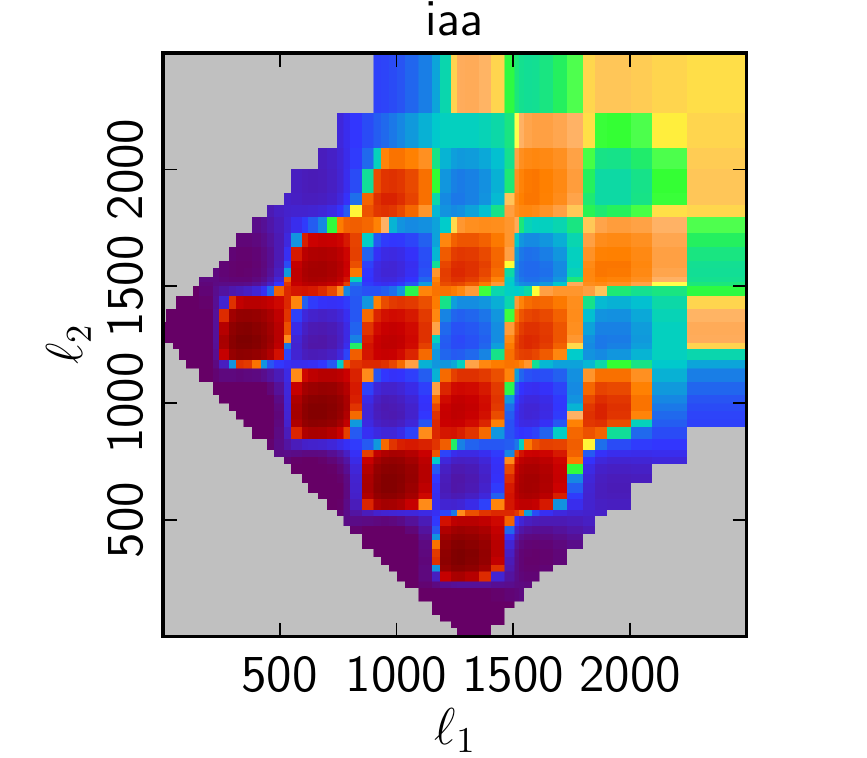}
\includegraphics[trim =3mm 0mm 12mm 0mm,clip,width=0.19\columnwidth]{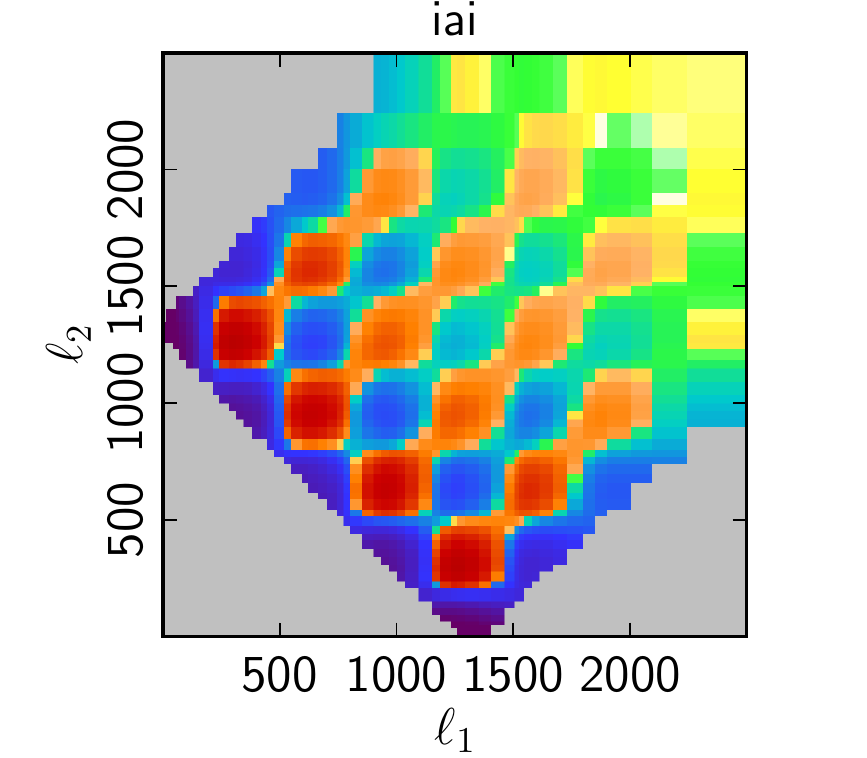}
\includegraphics[trim =3mm 0mm 12mm 0mm,clip,width=0.19\columnwidth]{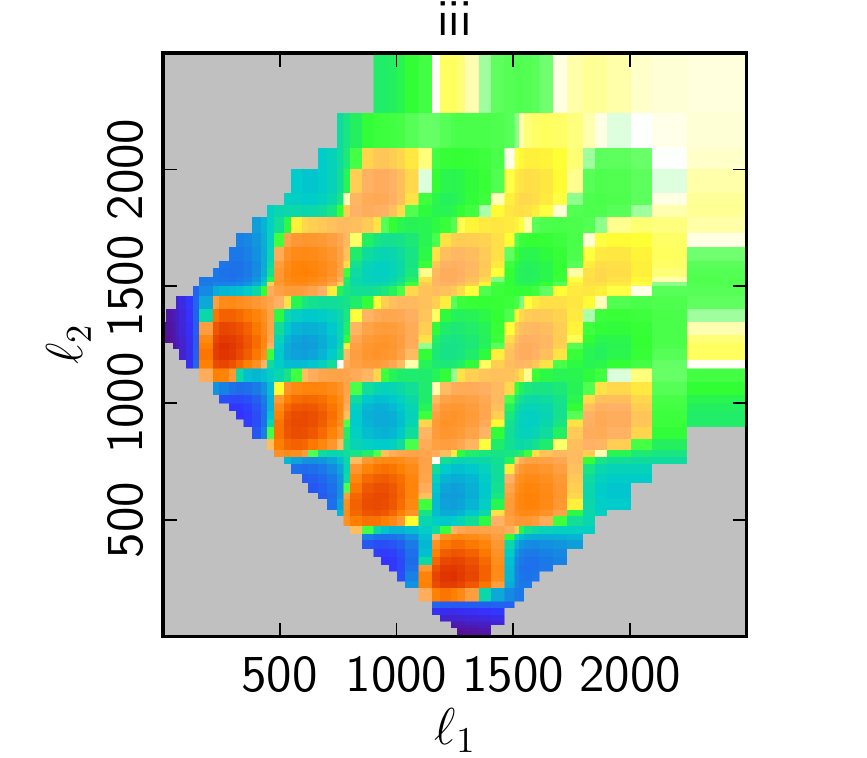}
\includegraphics[trim = 0mm 0mm 25mm 10mm,clip,width=0.9\columnwidth]{figures/colorbars_lin_iso_monobar_signifrenorm_martincolor.pdf}
\caption{Similar to Fig.~\ref{fig:theor_iso1}, but for $\ell_3\in$ 
[1291, 1345].} 
\label{fig:theor_iso3}
\end{figure}

Figs.~\ref{fig:theor_iso1} and~\ref{fig:theor_iso3} show the isocurvature 
templates discussed in Section~\ref{sec_isocurv}. Considering only the 
adiabatic mode
together with one isocurvature mode (for which we take consecutively
cold dark matter density isocurvature, neutrino density isocurvature, and
neutrino velocity isocurvature) gives six independent bispectrum 
templates. These are denoted respectively by $aaa$, $aai$, $aii$, $iaa$, $iai$,
and $iii$ with $a$ for adiabatic and $i$ for isocurvature (see 
Section~\ref{sec_isocurv} for more details). Since the pure adiabatic
$aaa$ template is just the standard local template from 
Fig.~\ref{fig:theor_smoo}, it is not shown here again. The other five
templates correspond to one row in the Figures. The different rows then
correspond to the different isocurvature modes and different polarization
components, as explained in the captions. The difference between the two
Figures is the choice of bin for $\ell_3$.

Since we want to show the relative size of these templates, we use a 
normalization different from the first Figure. The smoothed
bispectra are still normalized with the same expected standard deviation as 
before, but they all have $f_\mathrm{NL}=1$. Hence their signal-to-noise 
ratios are all very small and vary a lot from one template to another, so we 
have chosen a logarithmic colour scale over a large range. 
Since the isocurvature templates are all based
on the local shape, they are peaked in the squeezed
limit. The cold dark matter isocurvature case falls faster as a function
of $\ell$ than the others (its power spectrum is proportional to $\ell^{-4}$
instead of $\ell^{-2}$). 
The colour scale was chosen to make apparent which regions are positive
and which are negative. We see that, as is the case for the power 
spectrum, the acoustic oscillations are not in phase for different 
isocurvature sources.

\clearpage

\bibliographystyle{JHEP}
\bibliography{NG_binned_refs}

\end{document}